\documentclass[twocolumn,trackchanges]{aastex631}

\usepackage{natbib}
\usepackage{graphicx}
\usepackage{epstopdf}
\usepackage{gensymb}
\usepackage{amsmath}
\usepackage{hyperref}
\usepackage{subfigure}
\usepackage{threeparttable}  
\hypersetup{hidelinks,
	colorlinks=true,
	allcolors=blue,
	pdfstartview=Fit,
	breaklinks=true}
\let\oldAA\AA
\renewcommand{\AA}{\text{\normalfont\oldAA}}

\newcommand{\xmm}{\hbox{\hbox{XMM-Newton }\/}}

\newcommand{\fluxd}{{erg~cm$^{-2}$~Hz$^{-1}$~s$^{-1}~$}}
\newcommand{\flux}{{erg~cm$^{-2}$~s$^{-1}$}}
\newcommand{\lum}{{erg~s$^{-1}$}}

\begin{document}

\title{On the Extremely \hbox{X-ray} Variable Active Galactic Nuclei in the XMM-LSS Field}


\author[0000-0002-2420-5022]{Zijian Zhang}
\affiliation{Department of Astronomy, School of Physics, Peking University, Beijing 100871, China}
\affiliation{Kavli Institute for Astronomy and Astrophysics, Peking University, Beijing 100871, China}
\affiliation{School of Astronomy and Space Science, Nanjing University, Nanjing 210093, China}

\author[0000-0002-9036-0063]{Bin Luo}
\affiliation{School of Astronomy and Space Science, Nanjing University, Nanjing 210093, China}
\affiliation{Key Laboratory of Modern Astronomy and Astrophysics (Nanjing University), Ministry of Education, China}

\author[0000-0003-4176-6486]{Linhua Jiang}
\affiliation{Department of Astronomy, School of Physics, Peking University, Beijing 100871, China}
\affiliation{Kavli Institute for Astronomy and Astrophysics, Peking University, Beijing 100871, China}

\author[0000-0002-0167-2453]{W. N. Brandt}
\affiliation{Department of Astronomy \& Astrophysics, 525 Davey Lab, The Pennsylvania State University, University Park, PA 16802, USA}
\affiliation{Institute for Gravitation and the Cosmos, The Pennsylvania State University, University Park, PA 16802, USA}
\affiliation{Department of Physics, 104 Davey Lab, The Pennsylvania State University, University Park, PA 16802, USA}

\author[0000-0002-9036-0063]{Jian Huang}
\affiliation{School of Astronomy and Space Science, Nanjing University, Nanjing 210093, China}
\affiliation{Key Laboratory of Modern Astronomy and Astrophysics (Nanjing University), Ministry of Education, China}

\author[0000-0002-8577-2717]{Qingling Ni}
\affiliation{Max-Planck-Institut f\"{u}r extraterrestrische Physik (MPE), Gie{\ss}enbachstra{\ss}e 1, D-85748 Garching bei M\"unchen, Germany}

\begin{abstract}
We present a systematic investigation of extremely \hbox{X-ray} variable active galactic nuclei (AGNs) in the \hbox{$\approx 5.3~{\rm deg}^2$} \hbox{XMM-SERVS} \hbox{XMM-LSS} region. Eight variable AGNs are identified with rest-frame 2 keV flux density variability amplitudes around 6--12. We comprehensively analyze the \hbox{X-ray} and multiwavelength data to probe the origin of their extreme \hbox{X-ray} variability. It is found that their extreme \hbox{X-ray} variability can be ascribed to changing accretion state or changing obscuration from \hbox{dust-free} absorbers. For five AGNs, their \hbox{X-ray} variability is attributed to changing accretion state, supported by contemporaneous multiwavelength variability and the absence of \hbox{X-ray} absorption in the low-state spectra. With new Multiple Mirror Telescope (MMT) spectra for four of these sources, we confirm one changing-look AGN. One MMT AGN lacks multi-epoch spectroscopic observations, while the other two AGNs do not exhibit changing-look behavior, likely because the MMT observations did not capture their high states. The \hbox{X-ray} variability of the other three AGNs is explained by changing obscuration, and they show only mild long-term optical/IR variability. The absorbers of these sources are likely clumpy accretion-disk winds, with variable column densities and covering factors along the lines of sight.

\end{abstract}


\section{Introduction}
\label{sec:intro}
Active galactic nuclei (AGNs) are powered by accretion onto supermassive black holes (SMBHs) in the centers of massive galaxies. Luminous \hbox{X-ray} emission is a ubiquitous property of AGNs, which is believed to mainly originate from the ``corona'' in the vicinity of the SMBH via Comptonization of optical/ultraviolet (UV) seed photons \citep[e.g.,][]{2010arXiv1008.2287D,2014SSRv..183..121G,2017AN....338..269F,2023arXiv230210930G}. The typical size of the corona is constrained to be around 6--10$~r_{\rm g}$ \citep[e.g.,][]{2009ApJ...693..174C,2010ApJ...709..278D,2013ApJ...769L...7R,2014ApJ...783..116S}, where $r_{\rm g} = GM_{\rm BH}/c^2$ is the gravitational radius of the SMBH. Observations of radio-quiet AGNs have revealed a significant correlation between the coronal \hbox{X-ray} emission and the \hbox{accretion-disk} optical/UV emission, which indicates a physical connection between the accretion disk and the corona. This correlation is typically expressed as the relation between the \hbox{X-ray}-to-optical \hbox{power-law} slope parameter $\alpha_{\rm OX}$ and the 2500 $\rm \AA$ monochromatic luminosity $L_{\rm 2500 \AA}$ over $\approx 5$ orders of magnitude in $L_{\rm 2500 \AA}$ \citep[e.g.,][]{2006AJ....131.2826S,2007ApJ...665.1004J,2017A&A...602A..79L}.
Here $\alpha_{\rm OX}$ is defined as $\alpha_{\rm OX}=-0.3838 \log(f_{\rm 2500 \AA}/f_{\rm 2 keV})$, where $f_{\rm 2500 \AA}$ and $f_{\rm 2 keV}$ are the rest-frame 2500 $\rm \AA$ and 2 keV flux densities, respectively.

An important feature of AGN \hbox{X-ray} emission is its variability, which is a useful probe of the underlying nature of the coronal region. Analyses of \hbox{long-term} (year timescales) \hbox{multi-epoch} \hbox{X-ray} observations of large AGN samples have revealed that typical AGNs exhibit \hbox{X-ray} variability amplitudes of about $20\%$--$50\%$ \citep[e.g.,][]{2004ApJ...611...93P,2012ApJ...746...54G,2016ApJ...831..145Y,2020MNRAS.498.4033T}. Such AGN \hbox{X-ray} variability is generally believed to be related to instability of the corona or fluctuations in the accretion flow \citep[e.g., energy dissipation from magnetic flares or variation of the optical depth;][]{1997ARA&A..35..445U}. AGNs that show strong (e.g., amplitude $\gtrsim 2$) or extreme (e.g., amplitude $\gtrsim$ 6) \hbox{X-ray} variability are rare \citep[e.g.,][]{2020MNRAS.498.4033T}, and they need to be interpreted with additional mechanisms. 

Strong and extreme \hbox{X-ray} variability of AGNs may be caused by changes of intrinsic \hbox{X-ray} emission intensity and/or external effects. Possible scenarios include changing accretion state \cite[][]{2015ApJ...800..144L,2019ApJ...883...94T,2023NatAs...7.1282R}, changing obscuration \citep[][]{2003MNRAS.342..422M,2004PASJ...56L...9T,2022ApJ...930...53L,2023NatAs...7.1282R}, and tidal disruption events \citep[TDEs; e.g.,][]{2021ARA&A..59...21G}. We note that TDEs are considered one possible mechanism for changes of accretion states in AGNs \citep[e.g.,][]{2023NatAs...7.1282R}, though it might not be the main driving mechanism \citep[e.g.,][]{2024ApJ...966...85Z}. Also, optical/UV selected TDEs do not necessarily show corresponding \hbox{X-ray} flares. In this study, the TDE scenario mainly refers to \hbox{X-ray} selected TDEs in inactive galaxies that display significant \hbox{X-ray} variability when compared to quiescent states.

As AGN radiation is powered by accretion, a large change in the accretion rate naturally leads to strong optical/UV and \hbox{X-ray} continuum variability. Broad emission lines (e.g., Balmer lines and the Mg II line) will respond to the increase/decrease of the continuum flux and show a similar variability trend. An extreme example of such a case is \hbox{changing-look} AGNs, which show strong continuum variability and disappearance or appearance of broad emission lines \cite[e.g.,][]{2015ApJ...800..144L,2016ApJ...826..188R,2024ApJS..270...26G}. The timescale of such changing-look behavior is of the order of years, considerably shorter than the viscous timescale of AGN accretion disk \citep[e.g.,][]{2018ApJ...864...27S,2023NatAs...7.1282R}. Some studies suggest that changing-look AGNs have preferentially relatively low Eddington ratios; however, these results may be subject to various observational biases \citep{2019ApJ...874....8M,2022ApJ...933..180G,2023NatAs...7.1282R}. Variable AGNs induced by changing accretion state will show contemporaneous \hbox{multiwavelength} variability. The \hbox{X-ray} spectra of such AGNs generally show no signs of obscuration in either the high or low states \cite[e.g.,][]{2015ApJ...800..144L,2023ApJ...953...61Y}.

Another possible scenario is changing obscuration of the \hbox{X-ray} emission. In the AGN unification model \citep[e.g.,][]{2015ARA&A..53..365N}, Type II AGNs are considered to be obscured by the so-called ``torus'' as they are observed at higher inclinations (inclination is measured relative to the direction perpendicular to the central disk). Some Type II AGNs were found to show strong \hbox{X-ray} variability caused by variations in the absorbing column densities on timescales of \hbox{months-to-years}, where the variable absorbing material is likely gas clumps located in the dusty torus or the \hbox{broad-line} region \citep[e.g.,][]{2002MNRAS.329L..13G,2002ApJ...571..234R,2011ApJ...742L..29R,2014MNRAS.439.1403M}. Type I AGNs may also show obscuration-induced extreme variability. For example, some \hbox{narrow-line} Seyfert 1 galaxies (NLS1s) have strong and sometimes rapid (down to minutes) \hbox{X-ray} variability, possibly caused by variable \hbox{X-ray} obscuration from clumpy accretion-disk winds \citep[e.g.,][]{2007ApJ...668L.111G}. NLS1s generally have small BH masses ($\lesssim 10^7 M_{\odot}$) and large Eddington ratios \citep[$\lambda_{\rm Edd} \gtrsim 0.1$; e.g.,][]{1996A&A...305...53B}, and powerful disk winds launched via radiation pressure are expected in such systems \citep[e.g.,][]{2014MNRAS.439..503S,2014ApJ...796..106J,2019ApJ...880...67J}. Strong \hbox{X-ray} variability events have also been observed in more luminous Type I quasars that have high accretion rates, and they were also attributed to variable wind obscuration \cite[e.g.,][]{2012MNRAS.425.1718M,2020ApJ...889L..37N,2022MNRAS.511.5251N,2022ApJ...930...53L,2023ApJ...950...18H}. Such \hbox{X-ray} absorbers may vary in covering factor, column density, and/or ionization parameter \citep[e.g.,][]{2004PASJ...56L...9T,2009MNRAS.399..750B,2022ApJ...930...53L}. AGNs with obscuration-induced extreme \hbox{X-ray} variability tend to show substantial negative deviations from the expected $\alpha_{\rm OX} \textrm{--} L_{\rm 2500 \AA}$ relation, accompanied with absorption features in their low-state \hbox{X-ray} spectra. In the changing obscuration scenario, as the SMBH's accretion rate does not change significantly, there is no contemporaneous strong optical/UV continuum or emission-line variability.

TDEs may also cause strong \hbox{X-ray} variability in both AGNs and inactive galaxies \citep[e.g.,][]{2021ARA&A..59...21G,2021SSRv..217...54Z}. Typical TDEs are generally observed in inactive galaxies, and only a few AGN TDE candidates were proposed \citep[e.g.,][]{1995MNRAS.273L..47B,2015MNRAS.452...69M,2017ApJ...843..106B,2019MNRAS.490L..81Z,2020ApJ...894...93L,2020ApJ...898L...1R}. It is difficult to distinguish real TDEs from strong \hbox{X-ray} variability of pre-existing AGNs \citep[e.g.,][]{2020SSRv..216...85S}. A significant difference is that the \hbox{X-ray} flux of a typical X-ray selected TDE will rise sharply (weeks-to-month timescales) and then gradually decline within a few years to even decades, and there should be no recurrence of such transient behavior, while the variability of AGNs could be more stochastic \citep[e.g.,][]{2021SSRv..217...54Z}. There are some other features that favor a TDE over variable AGN activity, including a soft \hbox{X-ray} spectrum (photon index $\Gamma \geq 3$), weak or even absent Mg II $\lambda 2800$ line emission, and luminous He II $\lambda 4686$ line emission \citep[e.g.,][]{2021SSRv..217...54Z}.

Besides the \hbox{X-ray} variability itself, the X-ray emission strength evaluated by the $\alpha_{\rm OX} \textrm{--} L_{\rm 2500 \AA}$ relation also provides useful constraints on the underlying physics. For example, for X-ray variability induced by changing obscuration, the AGN will vary between X-ray normal ($\Delta \alpha_{\rm OX} \approx 0$) and X-ray weak states \citep[$\Delta \alpha_{\rm OX} < 0$; e.g.,][]{2012MNRAS.425.1718M,2015ApJ...805..122L,2022MNRAS.511.5251N,2022ApJ...930...53L,2023ApJ...954..159Z}. Here the $\Delta \alpha_{\rm OX}$ parameter is defined as the difference between the observed and expected $\alpha_{\rm OX}$ values: $\Delta \alpha_{\rm OX}=\alpha_{\rm OX} - \alpha_{\rm OX,exp}$. For \hbox{X-ray} variability induced by changing accretion state, whether the low-state and high-state \hbox{X-ray} emission strengths follow the $\alpha_{\rm OX} \textrm{--} L_{\rm 2500 \AA}$ relation is still uncertain, as some changing-state AGNs follow the $\alpha_{\rm OX} \textrm{--} L_{\rm 2500 \AA}$ relation while some do not \citep[e.g., see Table 8 of][]{2023ApJ...953...61Y}.

The XMM-Spitzer Extragalactic Representative Volume Survey \citep[XMM-SERVS;][]{2018MNRAS.478.2132C,2021ApJS..256...21N} is one of the major X-ray surveys with \hbox{medium-deep} X-ray coverage. XMM-SERVS encompasses three fields (XMM-LSS, W-CDF-S, and ELAIS-S1) with a total area of $\approx 13~\rm deg^2$. All of these fields have \hbox{multi-epoch} X-ray observations and rich multiwavelength coverage, which make them promising for studying AGN X-ray variability. In this paper, we present a systematic investigation of \hbox{X-ray} selected AGNs that show extreme \hbox{X-ray} variability in a XMM-SERVS field that has the largest area ($\approx 5.3~\rm deg^2$), the XMM-Large Scale Structure (XMM-LSS) field. We select eight extremely variable AGNs, perform comprehensive analyses of their multiwavelength properties, and investigate the possible origin of their extreme \hbox{X-ray} variability. The paper is organized as follows. In Section \ref{sec:sample_multidata}, we describe our selection of extremely variable \hbox{X-ray} AGNs in the XMM-LSS field, their archival \hbox{X-ray} data, and new optical spectroscopic observations. The \hbox{X-ray} and multiwavelength properties of the selected sources are presented in Section \ref{sec:Xray and multiwavelength data}. In Section \ref{sec:Discussion}, we discuss the physical origin of the extreme \hbox{X-ray} variability of these AGNs. Conclusions are given in Section \ref{sec:Conclusion}. Throughout this paper, we use a cosmology with $H_{0}=67.4~ \rm km~s^{-1}~Mpc^{-1}$, $\Omega_{\rm M} = 0.315$, and $\Omega_{\rm \Lambda} = 0.686$ \citep{2020A&A...641A...6P}. By default, measurement uncertainties are quoted at a 1$\sigma$ confidence level, while upper limits are quoted at a 90\% confidence level.

\section{Sample Selection and Multiwavelength Data}
\label{sec:sample_multidata}

\subsection{Sample Selection}
\label{subsec:sample}
The XMM-SERVS XMM-LSS is a \hbox{5.3-square-degree} region with a total of 2.7 Ms of \xmm \hbox{flare-filtered} exposure, supplemented with superb multiwavelength surveys (see, e.g., \citealt{2022ApJS..262...15Z} for details). This field builds upon the archival observations from previous surveys (e.g., XXL and SXDS) and uses a potential multi-epoch observation approach to minimize the effects of background flaring. As a result, most of the sources in this region have \hbox{multi-epoch} (3--8) \hbox{X-ray} observations with typical cadences of \hbox{$\approx$ 5--15 years}; these data serve as a good dataset to study AGN \hbox{X-ray} variability. \cite{2018MNRAS.478.2132C} presented an \hbox{X-ray} \hbox{point-source} catalog containing 5242 \hbox{X-ray} sources detected in this field, and 5071 of these sources were identified as AGNs. The source detection and photometry in \cite{2018MNRAS.478.2132C} were performed on the co-added image and cannot be used for variability analyses. Therefore, we used the measurements in the \xmm Serendipitous Source Catalog \citep[4XMM DR13;][]{2020A&A...641A.136W,2020A&A...641A.137T}. The 4XMM catalog includes a ``slim'' source catalog with unique sources and a ``full'' detection catalog with individual detections. We first selected sources from the 4XMM slim source catalog in the sky-region with $34.2\degree \leq$ RA $\leq 37.125\degree$ and $-5.72\degree \leq$ DEC $\leq -3.87 \degree$ (the same as the \citealt{2018MNRAS.478.2132C} catalog), resulting in 4648 sources. All these sources have at least one detection in the 4XMM \hbox{full-detection} catalog within a matching radius of 10$\rm \arcsec$. Some \hbox{X-ray} sources identified by \cite{2018MNRAS.478.2132C} are not present in the 4XMM catalog. We verified that they tend to exhibit relatively low \hbox{X-ray} fluxes (the \hbox{0.5--10} keV band fluxes of most of them are less than $1 \times 10^{-14}$ \flux), and they can only be detected in the co-added images.

Our investigation aimed to identify AGNs exhibiting extreme \hbox{X-ray} variability in this field. In the \hbox{sample-selection} step, we computed the \hbox{X-ray} variability amplitude using the observed \hbox{full-band} (0.2--12 keV) count rates measured by the pn detector. The count-rate values provided by the 4XMM catalog are corrected to the on-axis position \citep{2016A&A...590A...1R}. The flux values in the 4XMM catalog are proportional to the count-rate values, as a constant conversion factor was used to convert count rates to fluxes for each band \citep{2016A&A...590A...1R}. We used the count rates of individual detections in the 4XMM detection catalog, with a summary flag threshold of $>0$ to avoid spurious or problematic detections as suggested in the 4XMM user guide.\footnote{\url{http://xmmssc.irap.omp.eu/Catalogue/4XMM-DR10/4XMM-DR10_Catalogue_User_Guide.html\#Catalogue}.} The error of the variability amplitude is propagated from the count rate errors provided by the 4XMM catalog following the method described in Section 1.7.3 of \citet{1991pgda.book.....L}. From the parent sample, we first required sources to have multi-epoch detections in 4XMM catalog and calculated the \hbox{X-ray} variability amplitude between every two detections. Then we selected sources with an \hbox{X-ray} variability amplitude $> 10$ between any two detections.\footnote{We chose a conservative variability amplitude threshold of 10 in the selection here, as the 4XMM fluxes might have additional systematic uncertainties. A lower threshold value might introduce sources that are not extremely variable from the spectral analysis below (Section \ref{3subsec:xray_spec}). There would also be more sources without spectroscopic AGN classification. Thus a conservative threshold was chosen here to reduce the workload.} The lower relative error of the variability amplitude was also limited to $< 40\%$ to avoid the influence of count rates with large errors. We found seven sources that meet the above criteria, but one of them was excluded as it is a star. The other six sources are classified as galaxies/AGNs based on their optical spectra.

X-ray variable sources may have \hbox{non-detections} in some of the individual observations. To incorporate count-rate upper limits from \hbox{non-detection} observations in our selection procedure, we used the RapidXMM upper limit server \citep{2022MNRAS.511.4265R}, which provides upper limits on the XMM-Newton coverage within HEALPix \citep[Hierarchical Equal Area isoLatitude Pixelization; ][]{2011ascl.soft07018G} cells of size $\approx 3$$\rm \arcsec$. For parent-sample sources with upper-limit constrains, we compared their pn full-band 3$\sigma$ count-rate upper limit to count-rate measurements with $\leq$ 20\% relative error. We found three additional sources with a maximum variability amplitude $> 10$. One of them was excluded by visual inspection of its \hbox{X-ray} images, as the high-state detection is caused by problematic pixels. The other two sources are spectroscopically confirmed as AGNs.

\begin{figure}
\hspace{-0.4cm}
 \includegraphics[width=0.46\textwidth]{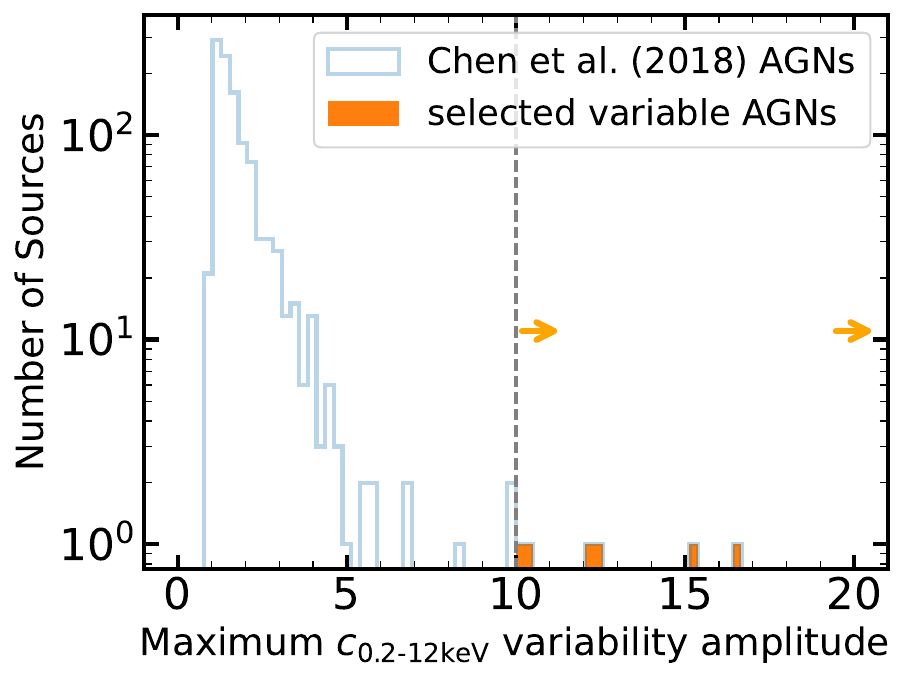}
 \centering
 \caption{Maximum full-band count rate variability amplitude distributions of the X-ray AGNs in the XMM-LSS field and the selected extreme X-ray variable AGNs. The two arrows represent the lower limits on the maximum variability amplitudes for VIDs 6 and 7. The dashed vertical line corresponds to a variability amplitude of 10.}
 \label{fig:xvar_amp_ratio}
\end{figure}

We selected the above eight (6+2) sources as our final sample and adopted their optical counterparts following \citet{2018MNRAS.478.2132C}. All these optical counterparts are considered to be reliable according to the flags in the \citet{2018MNRAS.478.2132C} catalog. A comparison of their maximum variability amplitudes and those of the other \citet{2018MNRAS.478.2132C} AGNs in the 4XMM detection catalog is shown in Figure \ref{fig:xvar_amp_ratio}, demonstrating the extreme variability behavior of the selected objects. The basic object properties and the \hbox{X-ray} observations of these eight sources are listed in Table \ref{tab:xray_variability_object_info}. In the following, we refer to each source using the abbreviations in parentheses (VIDs 1--8) in Table \ref{tab:xray_variability_object_info}.

VIDs 1--4, VID 6, and VID 8 have SDSS spectra and we adopted their spectroscopic redshifts (spec-$z$s) from the SDSS DR16. The spec-$z$ of VID 7 is from the Dark Energy Spectroscopic Instrument Early Data Release \citep[DESI EDR;][]{2024AJ....168...58D}. For VID 5, its spec-$z$ is measured using the new Multiple Mirror Telescope (MMT) spectrum (see Section \ref{subsec:MMT Data} below) with the { \sc specpro} tool \citep{2011PASP..123..638M}. The single-epoch virial SMBH masses and Eddington-ratios estimates for VID 1, VID 3, and VID 4 in Table \ref{tab:xray_variability_object_info} are from the SDSS DR16 quasar catalog \citep{2022ApJS..263...42W}. The other five sources are not in this catalog, and their SMBH masses and Eddington ratios are estimated as described in Section \ref{subsec:BHmass_estimation} below.

Figure \ref{fig:z_M2500_plt} shows the distribution of our sample in the redshift versus rest-frame 2500 $\AA$ absolute magnitude space. We used the rest-frame 2500 $\AA$ absolute magnitude as host-galaxy contamination is generally minimized at this wavelength. We derived the 2500 $\AA$ flux density of each source via interpolation of the SDSS $u$-band flux densities, assuming a power-law index $\alpha_{\nu}$ of $-0.44$ \citep{2001AJ....122..549V}. Overall, these eight \hbox{X-ray} variable AGNs have relatively low 2500 $\AA$ luminosities compared to the SDSS DR16 quasars. Since our selected sources are detected in all/most of the XMM-SERVS individual observations, they are generally brighter compared to the other \cite{2018MNRAS.478.2132C} X-ray AGNs which were identified from the co-added image.

\begin{figure}
\hspace{-0.4cm}
 \includegraphics[width=0.49\textwidth]{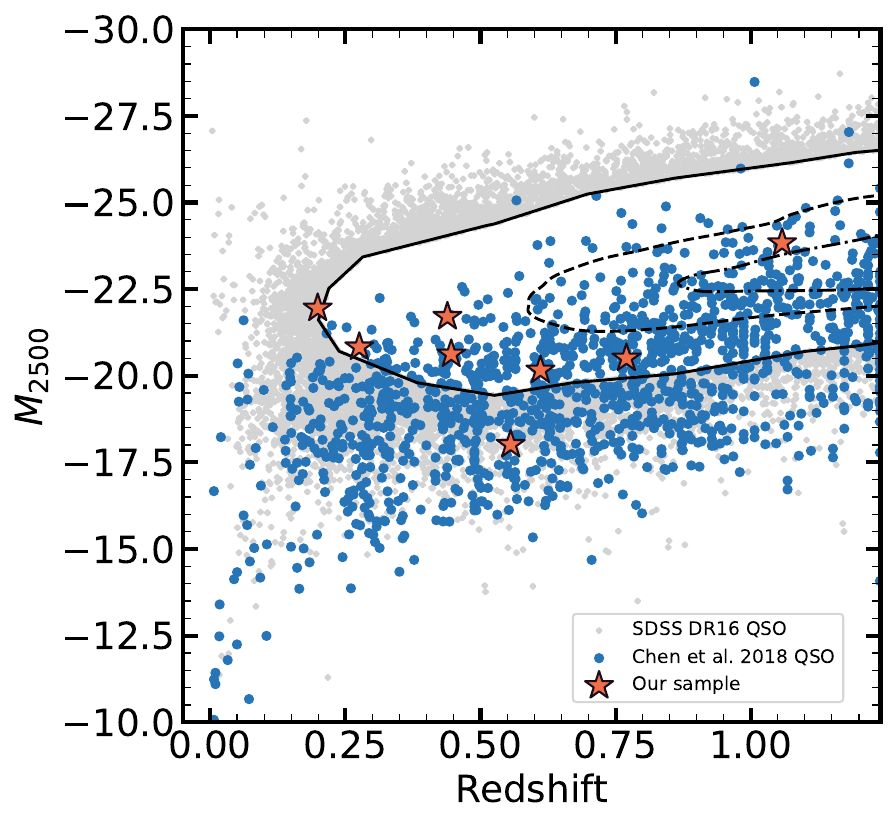}
 \centering
 \caption{Distribution of our sample in the redshift and rest-frame 2500 $\AA$ absolute-luminosity ($M_{2500}$) space. Orange stars represent the eight selected AGNs that show extreme \hbox{X-ray} variability. The blue dots are all $z<1.25$ \hbox{X-ray} selected AGNs with spec-$z$s or phot-$z$s from \cite{2018MNRAS.478.2132C}, complemented with $\sim 400$ new spec-$z$s from our MMT observations (Zhang Z.J. et al. in prep.). The black contours and gray dots are all $z<1.25$ quasars in the SDSS DR16 quasar catalog. The contours enclose 35, 68, and 95 percent of the SDSS DR16 quasars, respectively. The eight \hbox{X-ray} variable AGNs have relatively low $M_{2500}$ compared to the SDSS DR16 quasar sample, but they are generally brighter compared to the other \cite{2018MNRAS.478.2132C} X-ray AGNs.}
 \label{fig:z_M2500_plt}
\end{figure}

\subsection{Archival \hbox{X-ray} Observations and Data Analysis}
\label{subsec:X-ray data}

\begin{table*}
\caption{Basic Object Properties and Their \hbox{X-ray} Observations}
\label{tab:xray_variability_object_info}
\hspace*{-3.1cm}
\setlength\tabcolsep{3pt}
\begin{threeparttable}  
\begin{tabular}{lccccccccc}
\hline
\hline
Object&$z$&$m_{\rm i}$&$\log M_{\rm BH}$&$L/L_{\rm Edd}$&$N_{\rm H,Gal}$&Observation&Observation&Exposure&Comments\\
(SDSS J)&&&$(M_{\odot})$&&($10^{20}~\rm cm^{-2}$)&ID&Date&(ks)& \\
(1)&(2) &(3)  &(4) &(5)  &(6) &(7) &(8)&(9)&(10)  \\
\hline
021832.06$-$041345.1&0.439&18.8&8.78 ($9.20$)&0.014&1.99&0404966901&2007 Jan 07&8.2&pn, M1, M2\\
(VID 1)&&&&&&0404967301&2007 Jan 08&13.5&M1, M2\\
&&&&&&0404967401&2007 Jan 08&11.5&pn, M1, M2\\
&&&&&&0553911501&2009 Jan 01&10.4&pn, M1, M2\\
&&&&&&0785100101&2016 Jul 01&17.2&pn, M1, M2\\
&&&&&&0785100201&2016 Jul 01&17.8&pn, M1, M2\\
&&&&&&0785100301&2016 Jul 02&14.9&pn, M2\\
\hline
021952.37$-$042448.8&0.556&19.8&($8.78$)&$0.031$&2.00&0037982301&2003 Jan 26&7.2&pn, M1, M2\\
(VID 2)&&&&&&0785100301&2016 Jul 02& 17.1&pn, M1, M2\\
 &&&&&&0785100601&2016 Jul 03& 21.5&pn, M2\\
 &&&&&&0793580401&2017 Jan 02& 5.0&pn, M1, M2\\
\hline
022105.62$-$044101.6&0.199&16.8&8.33 ($7.93$)&0.034&1.92&0037982001&2002 Aug 14&11.8&pn, M1, M2\\
(VID 3)&&&&&&0037982101&2002 Aug 14&10.8&pn, M1, M2\\
 &&&&&&0785100501&2016 Jul 03&15.7&pn, M1, M2\\
 &&&&&&0785101301&2016 Jul 06&13.3&pn, M1, M2\\
 &&&&&&0785101401&2016 Jul 07&16.3&pn\\
 &&&&&&18257&2016 Oct 08& 10.0&ACIS-S\\ 
 &&&&&&0793580301&2017 Jan 01&5.0&pn, M1\\
 &&&&&&0793580601&2017 Jan 03&10.4&pn, M1, M2\\
 &&&&&&23741&2020 Dec 06& 2.0&ACIS-S\\
\hline

022202.76$-$050944.5&0.276&18.7&9.05 ($8.83$)&0.004&2.08&0111110501&2001 Jul 04& 17.1 &pn, M1, M2\\
(VID 4)& &&&& &0147111301&2003 Jul 24& 9.9&pn, M1, M2\\
 & &&&& &0785101101&2016 Jul 05& 12.2 &pn, M1, M2\\
 & &&&& &0785101701&2016 Jul 08&4.3&pn, M1  \\
 & &&&& &0793581201&2017 Jan 01&27.4&M1, M2  \\
 & &&&& &0793580701&2017 Jan 03&11.1&pn, M1, M2\\

\hline
022240.36$-$050700.4&0.770&21.3&7.92 ($7.89$)&0.049&2.10&0111110501&2001 Jan 04&17.2&pn, M1, M2\\
(VID 5)& &&&& &0785101701&2016 Jul 08&4.3&pn, M2 \\
 & &&&& &0780450101&2016 Aug 13&18.9&pn, M1, M2\\
 & &&&& &0793581201&2017 Jan 01&16.3&pn, M2 \\
\hline

022541.66$-$043417.7 &0.611&20.6&8.26 ($8.04$)&0.049 &2.20&0112681001&2002 Jan 30&19.4&no detection in pn  \\

(VID 6) &&&&&&6864&2006 Nov 12& 29.7&ACIS-S\\
&&&&&&0780450701&2016 Aug 14& 13.0&pn, M2\\
  &&&&&&0780451001&2017 Jan 07&7.1&pn \\
   &&&&&&0780451401&2017 Jan 09&7.6&pn, M2 \\

\hline
022558.05$-$045720.9&1.058&20.3&8.20 ($9.85$)&0.132&2.14&0109520301&2002 Feb 02& 16.0&no detection in pn \\
(VID 7) &&&&&&18264 &2016 Sept 27& 22.8&ACIS-S\\
 &&&&&&0780450901&2017 Jan 06&8.7&pn, M1, M2\\
 &&&&&&0780451301&2017 Jan 08&6.5&pn, M1, M2\\
 &&&&&&0780452601&2017 Feb 10&9.5&pn, M1, M2\\

\hline
022739.76$-$050047.0 &0.446&19.1&($9.17$)&$0.006$&2.00&0111110101&2001 Jul 06&15.6&pn, M1, M2\\ 
(VID 8)&&&&&&0109520201&2002 Jan 29&18.6&pn, M1, M2\\
 &&&&&&0780451901&2017 Jan 11& 13.9&pn, M2\\
 &&&&&&0780452001&2017 Jan 11&17.5&pn, M1, M2\\
 
\hline
\end{tabular}

\end{threeparttable}      
\tablecomments{
Cols. (1) and (2): object name and spec-$z$.
Col. (3): HSC/CFHT $i$-band magnitude.
Cols. (4) and (5): \hbox{single-epoch} virial SMBH mass and Eddington ratio adopted from \citet{2022ApJS..263...42W} or estimated as described in Section \ref{subsec:BHmass_estimation}. The black-hole masses in parentheses are estimated using the $M_{*} \textrm{--} M_{\rm BH}$ relation. 
Col. (6): Galactic neutral hydrogen column density. 
Col. (7): XMM-Newton/Chandra observation ID.
Col. (8): observation start date.
Col. (9): cleaned exposure time. For XMM-Newton observations, the pn exposure time is shown when available, otherwise the MOS1/MOS2 exposure time is shown.
Col. (10): detectors used in each observation: pn, M1, and M2 for XMM-Newton observations and ACIS-S for Chandra observations.
}
\end{table*}

\begin{figure*}
\centering
\begin{minipage}{0.81\columnwidth}
      \includegraphics[clip,trim=0 0.9cm 0 0cm, width=1.02\columnwidth]{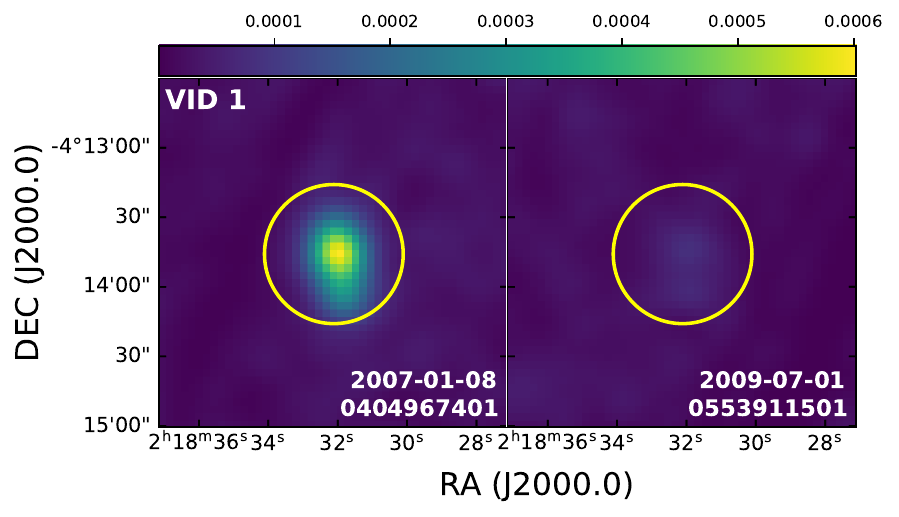}

     \vspace{0.2cm}
     \includegraphics[clip,trim=0 0.8cm 0 0cm, width=1.0\columnwidth]{figure/Ximg/xrayimgVID1.pdf}

    \includegraphics[clip,trim=0 0.8cm 0 0cm, width=1.0\columnwidth]{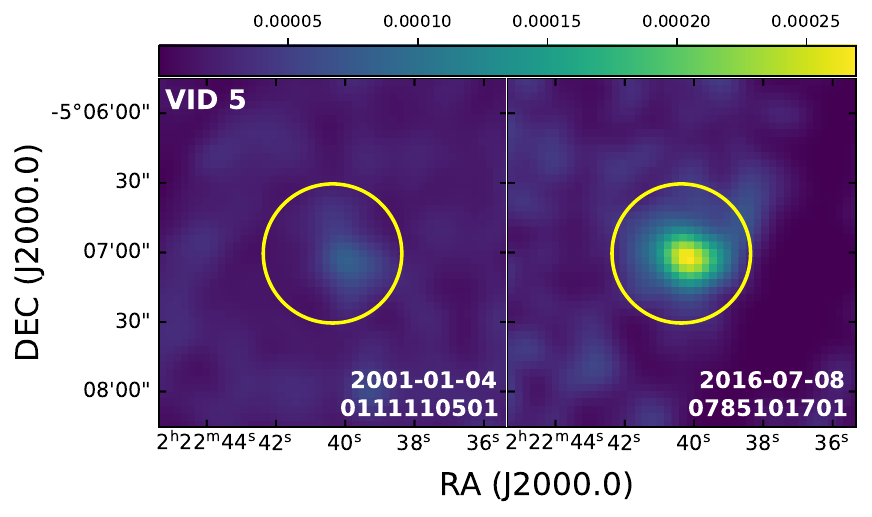}

     \includegraphics[clip,trim=0 0cm 0 0cm, width=1.0\columnwidth]{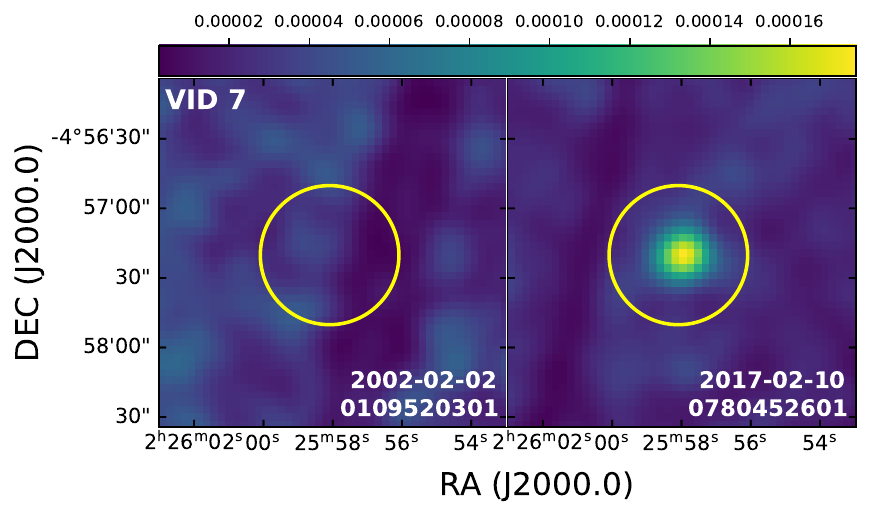}

\end{minipage}
\begin{minipage}{0.81\columnwidth}
      \includegraphics[clip,trim=0 0.8cm 0 0cm, width=1.0\columnwidth]{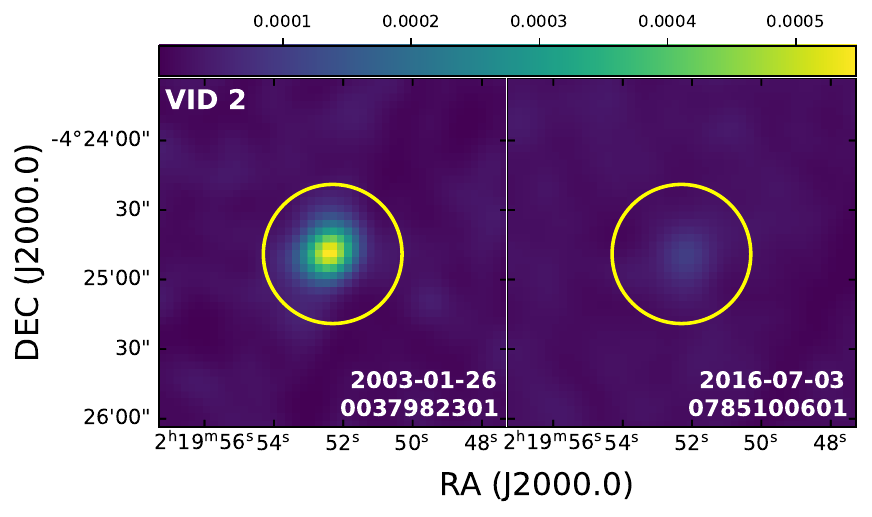}

       \includegraphics[clip,trim=0 0.8cm 0 0cm, width=1.04\columnwidth]{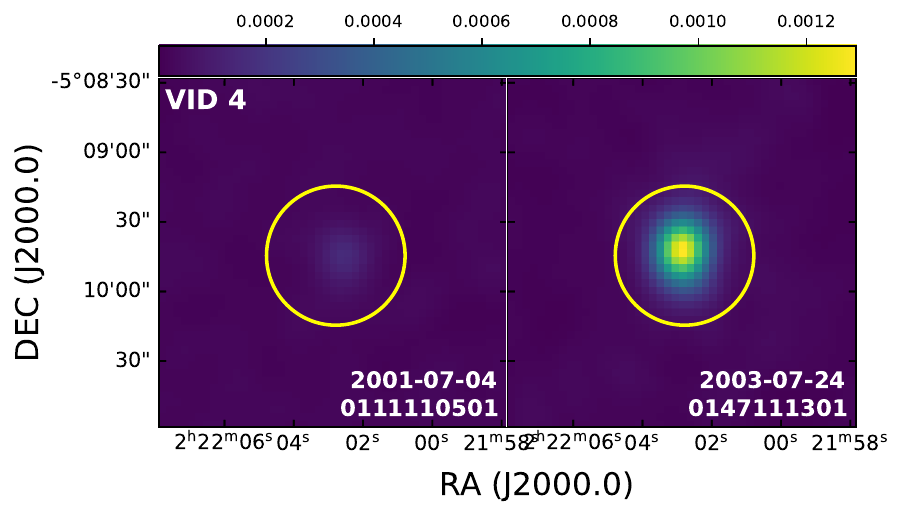}
      
      \includegraphics[clip,trim=0 0.8cm 0 0cm, width=1.005\columnwidth]{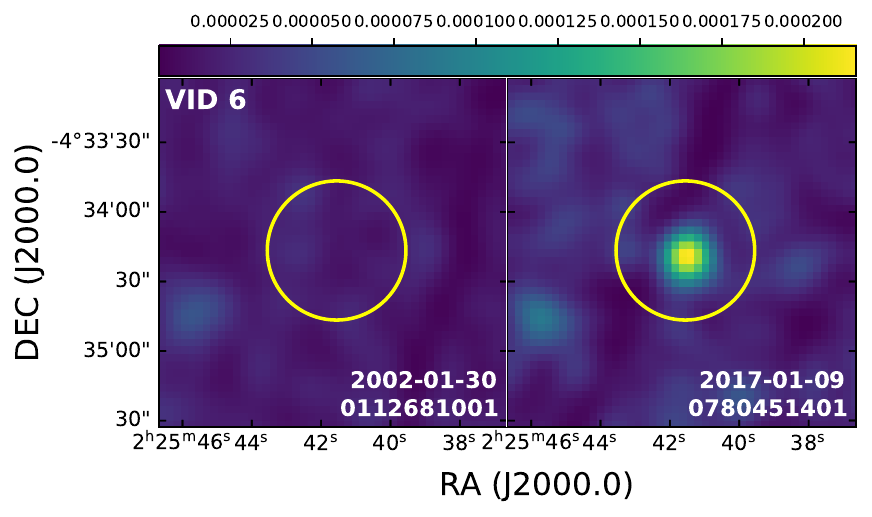}

     \includegraphics[clip,trim=0 0cm 0 0cm, width=1.0\columnwidth]{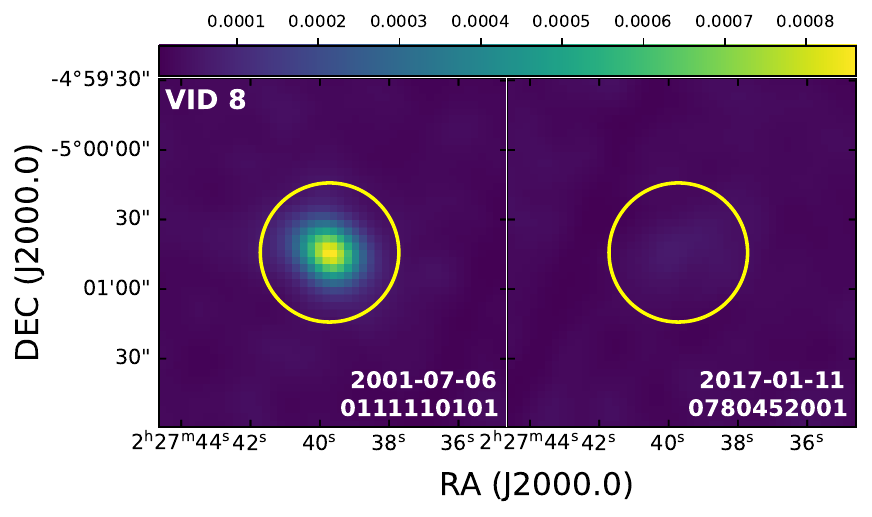}

\end{minipage}
\caption{Highest-state and \hbox{lowest-state} XMM-Newton 0.3--10.0 keV exposure-corrected images arranged in chronological order for each source. The images are smoothed with a symmetric 2D Gaussian kernel with a standard deviation of 2 pixels. The units of the colorbars are counts per second. For comparison, each pair of the highest-state image and \hbox{lowest-state} image use the same scale. The name of each source is noted on the upper-left of their left panels. The observation date and observation ID of each image are noted at the bottom-right corner. The yellow circle in each image is the circular region with a radius of $30''$ centered on the optical source position used to extract the source spectrum. It is apparent that all the sources exhibit significant variability.}
\label{fig:xray_image}
\end{figure*}

VIDs 1--8 each have 3--7 multi-epoch detections in the 4XMM detection catalog. For VID 6 and VID 7, they each have one upper-limit constraint. We retrieved their XMM-Newton observations from the \xmm Science Archive.\footnote{\url{https://www.cosmos.esa.int/web/xmm-newton/xsa}.} We also searched the Chandra archive\footnote{\url{https://cxc.cfa.harvard.edu/cda/}.} and found that VID 3, VID 6, and VID 7 have archival Chandra observations. The \xmm and Chandra observations of the eight AGNs are listed in Table \ref{tab:xray_variability_object_info}. We reduced the data of these observations. For \xmm data reduction, we processed the \hbox{X-ray} data using the \xmm Science Analysis System \citep[SAS v.20.0.0;][]{2004ASPC..314..759G} and the latest calibration files. All the EPIC pn, MOS1, and MOS2 data were used in our study when available. We reduced the pn and MOS data following the standard procedure described in the SAS Data Analysis Threads.\footnote{\url{https://www.cosmos.esa.int/web/XMM-Newton/sas-threads}.} Background flares were filtered to generate cleaned event files. The cleaned pn exposure times of each observation are listed in Table \ref{tab:xray_variability_object_info}, and the cleaned MOS1/MOS2 exposure times are shown when pn data are not available. For most of the observations, our AGNs are significantly detected in both the pn and MOS images. In a few observations, the target may be outside the field of view of one or two detectors. All the sources have pn coverage in both their highest and lowest states. Figure \ref{fig:xray_image} shows the highest- and \hbox{lowest-state} pn images of each source. It is apparent that all the sources exhibit significant variability.

We then extracted the X-ray spectra of each source. For each observation of each source, a source spectrum was extracted using a circular region with a radius of 30$\rm \arcsec$ centered on the optical source position. A corresponding background spectrum was extracted from a few nearby circular source-free regions on the same CCD chip with a total area of about four times the area of the source region. Spectral response files were generated using the {\sc rmfgen} and {\sc arfgen} tasks. For XMM-Newton observations made on the same day or within several days, if there is no significant variation in their spectra, we combined them using the {\sc epicspeccombine} script to improve the \hbox{signal-to-noise} ratio. VID 4 shows significant variability in the spectra observed on 2016 Jul 05 and 2016 Jul 08. Therefore, we did not combine the spectra of these two observations. The spectra of VID 6 observed on 2017 Jan 07 and 2017 Jan 09 were not combined either due to variability. Finally, if the total spectral counts number in the 0.3--10 keV band combining the pn and MOS spectra is larger than 500, we grouped the source spectra with at least 25 counts per bin for spectral fitting. Otherwise, we grouped the source spectra with at least one count per bin. There are 10 (16) spectra sets that are grouped with at least 25 (1) counts per bin. For the XMM-Newton observations where VID 6 and VID 7 were not detected, we retrieved the 3$\sigma$ 0.2--12.0 keV flux upper limits from the XMM-Newton Science Archive and converted them to the rest-frame \hbox{2 keV} flux density and 2--10 keV luminosity upper limits assuming a photon index of 2, which is the average value for luminous AGNs \citep[e.g.,][]{2011MNRAS.417..992S}.

To reduce the Chandra data, we used the Chandra Interactive Analysis of Observations (CIAO; v4.15)\footnote{\url{http://cxc.harvard.edu/ciao/}.} tool. We first used the CIAO {\sc  chandra\_repro} script to create new bad-pixel files and new level 2 event files. Background flares were removed using the {\sc deflare} script with an iterative 3$\sigma$ clipping algorithm. Then a source spectrum was extracted using the {\sc specextract} tool from a circular region centered on the
optical position. The radius of the circle was chosen to enclose $90\%$ of the point spread function (PSF) at 1 keV using the {\sc psf} command. A background spectrum was extracted from an annular region centered on the source position, with the inner and outer radii chosen to be the source radius plus 4$\rm \arcsec$ and 8$\rm \arcsec$, respectively. We have visually inspected the background-extraction regions and veriﬁed that they do not contain any \hbox{X-ray} sources.

For the \xmm observations of each source, we also searched for corresponding Optical Monitor (OM) photometric detections in the \xmm Science Archive. Some XMM-Newton observations of VID 1, VID 3, and VIDs 5--7 have simultaneous OM measurements in certain bands, while the other three sources have no OM coverage.

Five sources with relatively smaller right ascensions (VIDs 1--5) are covered by the SRG/eROSITA \hbox{X-ray} Survey of the UKIDSS Ultra Deep Survey Field \citep[eUDS;][]{2024MNRAS.528.1264K}, providing the most recent \hbox{X-ray} measurements of these sources. We estimated the \hbox{power-law} effective photon index from the eROSITA 0.3--0.6 keV and 0.6--2.3 keV band fluxes using the PIMMS tool.\footnote{\url{https://heasarc.gsfc.nasa.gov/cgi-bin/Tools/w3pimms/w3pimms.pl}.} We then converted the \hbox{0.6--2.3 keV} band flux to the rest-frame 2 keV flux density and the \hbox{X-ray} luminosity. The errors of the 2 keV flux densities were propagated from the flux errors in the \hbox{0.6--2.3 keV} band. These photon indices and $f_{\rm 2keV}$ values are shown in Table \ref{tab:xray_variability_X_spec_info}.

\subsection{MMT Observations and Data Analysis}

\label{subsec:MMT Data}
Except for VID 5 and VID 7, the other six sources each have at least one archival SDSS spectrum. VIDs 1--4 and VID 7 each have one spectrum from DESI \citep[][]{2022AJ....164..207D,2023arXiv230606308D}. We obtained new optical spectra for VIDs 1--2, VIDs 4--5, and VID 8 using Hectospec on the 6.5 m MMT at the Fred Lawrence Whipple Observatory \citep{2005PASP..117.1411F}. Hectospec is a multi-fiber optical spectrograph. It has 300 fibers with size of $1\farcs{5}$ in diameter over a field of view of $\sim1 \degree$ in diameter. These MMT observations belong to a campaign that targets \hbox{X-ray} sources and other interesting objects \citep[e.g.,][]{2024arXiv241206923H} in the XMM-LSS field. The details of the MMT observations relevant to our objects are summarized in Table \ref{tab:MMT_obs}. Combining all the spectra, 6/8 of the sources have multi-epoch spectroscopic measurements, while VID 5 and VID 6 have only one spectrum each.

\begin{table}
\centering
\caption{Relevant MMT observation log}
\label{tab:MMT_obs}
\hspace*{-2.1cm}
\begin{tabular}{lcccccc}
\hline
\hline
Object&Observation Date&Exposure&Airmass\\
\hline
VIDs 1--2&2021-12-08&3.67 hours 	&1.35\\
VIDs 4--5&2023-10-16&2.3 hours 	&1.34\\
VID 8&2022-11-27&2 hours 	&1.27\\
\hline
\end{tabular}
\end{table}

We used Hectospec's 270 gpm grating, which provides a wavelength coverage of 3700--9200 $\AA$ with a resolving power of $\sim1000$. The MMT data were reduced using the official IDL pipeline HSRED (v2.1).\footnote{\url{https://github.com/MMTObservatory/hsred}.} We followed the standard processing process of HSRED. We first used the pipeline to debias and flat-field the raw data and remove cosmic rays. We then used domed flat fields to remove CCD fringing and the high-frequency inhomogeneity of the pixel response. The pipeline constructed a sky model for each pointing based on the interpolation of the sky spectra obtained by the sky fibers, and then the model was used to compute the sky spectrum for each fiber. The modeled sky spectrum is scaled according to the strength of skylines in the individual spectrum and then subtracted. Finally, the spectrum of each source was extracted, and wavelength correction was performed by cross-correlating the observed spectra with the calibration arc spectra. For the flux calibration, we used $\sim 5$ MMT fibers to observe standard stars for each observation. We compared the SDSS SEDs of the standard stars with the built-in stellar spectrum models to determine the spectral types of the standard stars and then obtained the flux-calibration curves. The average flux calibration curve was applied for flux calibration.

\subsection{Multiwavelength Multi-epoch Data}
\label{subsec:multi-wave_lc_construction}
We also collected multi-epoch data of these eight sources from the Zwicky Transient Facility \citep[ZTF;][]{2019PASP..131a8003M},  the Panoramic Survey Telescope and Rapid Response System \citep[Pan-STARRS;][]{2020ApJS..251....7F}, the Near-Earth Object Wide-field Infrared Survey Explorer Reactivation Mission \citep[NEOWISE;][]{2011ApJ...731...53M}, and the Catalina Real-Time Transient Survey \citep[CRTS;][]{2009ApJ...696..870D} catalogs. The images of the CRTS survey are taken unfiltered in the optical band to maximize throughput. 

The Galaxy Evolution Explorer \citep[GALEX;][]{2005ApJ...619L...1M} also has multi-epoch observations for both the NUV and FUV filters in the XMM-LSS field. We excluded GALEX detections marked as artifacts according to the official GALEX instruction \citep{2017ApJS..230...24B}. The High-quality Extragalactic Legacy-field Monitoring \citep[HELM;][]{2024arXiv240206052Z} survey performed by the Dark Energy Camera (DECam) covers the XMM-LSS field and provides 3.5 years of optical light curves during 2019--2022. The HELM survey overlaps with ZTF in both filter band and time span but achieves a deeper depth of observation. We adopted the ZTF data when available, as they have better time coverage; otherwise, we used the HELM data. VID 8 is an exception, for which the HELM data were adopted as they cover more epochs.

\section{Results}
\label{sec:Xray and multiwavelength data}

\begin{table*}
\caption{\hbox{X-ray} Spectral Fitting Results of the Power-law Model and Upper-Limit Constraints}
\label{tab:xray_variability_X_spec_info}
\hspace*{-1.0cm}
\begin{tabular}{lccccccc}
\hline
\hline
Target&Observation&$\Gamma$ &$\chi^2 /\rm dof$&$P_{\rm null}$&$f_{\rm 2keV}$&$L_{\rm X}$ \\
&Start Date& &or $W$-stat/dof&& \\
(1)&(2) &(3)  &(4) &(5)  &(6) &(7)  \\
\hline
VID 1&2007 Jan 07+2007 Jan 08&$ 1.87_{-0.06}^{+0.06}$&27.6/40&0.93&$1.21_{-0.05}^{+0.08}$&$7.49_{-0.62}^{+0.72}$\\

&2009 Jan 01&$ 2.60_{-0.32}^{+0.40}$&125.4/166&...&$0.14_{-0.04}^{+0.05}$&$0.49_{-0.18}^{+0.36}$  \\
&2016 Jul 01+2016 Jul 02&$1.86_{-0.08}^{+0.08}$&  43.0/33&0.1&$0.35_{-0.02}^{+0.02}$&$2.18_{-0.22}^{+0.25}$  \\

&2019 Sept 01$^a$&2.06&...&...&$1.05\pm0.08$&...  \\

\hline
VID 2&2003 Jan 26&$ 1.64_{-0.08}^{+0.08}$&$     342.7/372$&...&$0.88^{+0.04}_{-0.06}$&$11.78^{+1.53}_{-1.20}$\\
 &2016 Jul 02+2016 Jul 03&$ 1.20_{-0.17}^{+0.18}$& $432.5/401$&...&$0.12^{+0.04}_{-0.04}$&$1.85^{+0.32}_{-0.74}$ \\
 
 
 &2017 Jan 02&$ 1.44_{-0.11}^{+0.11}$ & $     340.1/314$&...&$0.59^{+0.07}_{-0.06}$& $9.46^{+1.40}_{-1.23}$\\

 &2019 Sept 01$^a$&0.81&...&...&$0.17\pm0.02$&...  \\
\hline
VID 3 &2002 Aug 14&$     1.96_{-0.02}^{+0.02}$ &$     271.7/228$&0.03&$7.70^{+0.11}_{-0.15}$&$7.10^{+0.24}_{-0.23}$  \\

 &2016 Jul 03$+$2016 Jul 06$+$2016 Jul 07&$     1.81_{-0.05}^{+0.05}$ &$     74.0/70$&0.35&$0.70^{+0.03}_{-0.03}$&$0.73^{+0.05}_{-0.06}$  \\
 &2016 Oct 08&$     1.39_{-0.20}^{+0.21}$ &$     65.7/73$&...&$0.76^{+0.07}_{-0.13}$&$1.16^{+0.16}_{-0.21}$  \\
 &2017 Jan 01$+$2017 Jan 03&$      1.85_{-0.07}^{+0.07}$ &$     25.8/31$&0.73&$0.91^{+0.04}_{-0.06}$&$0.92^{+0.07}_{-0.08}$  \\
  &2019 Sept 01$^a$&1.87&...&...&$3.60\pm0.12$&...  \\
 &2020 Dec 06&$     1.72_{-0.19}^{+0.20}$&$     75.9/81$&...&$5.75^{+0.35}_{-0.71}$&$6.48^{+0.61}_{-1.11}$  \\
\hline

VID 4 &2001 Jul 04&$1.47_{-0.11}^{+0.12}$ &$     346.2/358$&...&$0.34_{-0.03}^{+0.04}$&$0.99^{+0.17}_{-0.19}$\\
 &2003 Jul 24&$1.66_{-0.05}^{+0.05}$ &$     35.3/41$&0.72&$2.38_{-0.12}^{+0.09}$&$5.99^{+0.33}_{-0.46}$\\
 &2016 Jul 05&$1.87_{-0.04}^{+0.04}$ &$     53.7/65$&0.84&$2.37_{-0.09}^{+0.10}$&$4.94^{+0.37}_{-0.28}$\\
 &2016 Jul 08&$1.89_{-0.11}^{+0.11}$ &$      305.6/305$&...&$1.64_{-0.22}^{+0.28}$&$3.37^{+0.74}_{-0.33}$\\
 &2017 Jan 01$+$2017 Jan 03&$1.67_{-0.12}^{+0.13}$ &$     21.6/18$&0.25&$0.66_{-0.06}^{+0.10}$&$1.65^{+0.29}_{-0.33}$\\
  &2019 Sept 01$^a$&1.87&...&...&$0.29\pm0.21$&...  \\

\hline
VID 5 &2001 Jan 04&$2.01_{-0.25}^{+0.27}$ &$     254.3/267$&...&$0.08_{-0.01}^{+0.01}$&$1.77^{+0.38}_{-0.43}$\\
 &2016 Jul 08&$2.21_{-0.20}^{+0.21}$ &$     155.6/204$&...&$0.51_{-0.07}^{+0.10}$&$9.93^{+3.78}_{-2.44}$\\

 &2016 Aug 13&$1.77_{-0.10}^{+0.10}$ &$     421.7/445$&...&$0.41_{-0.03}^{+0.03}$&$11.57^{+1.09}_{-1.52}$\\

 &2017 Jan 01&$1.51_{-0.19}^{+0.20}$&$     209.9/239$&...&$0.19_{-0.02}^{+0.02}$&$6.83^{+1.43}_{-1.43}$\\

  &2019 Sept 01$^a$&2.26&...&...&$0.21\pm0.15$&...  \\ 
\hline

VID 6 &2002 Jan 30&...&...&...&$<0.04$&$<0.46$\\
&2006 Nov 12&$ 2.07_{-0.40}^{+0.43}$&$     22.0/29$&...&$0.08_{-0.02}^{+0.01}$&$0.95^{+0.32}_{-0.33}$\\

&2016 Aug 14&$ 2.90_{-0.25}^{+0.26}$&$     100.0/151$&...&$0.23_{-0.04}^{+0.05}$&$1.47^{+0.50}_{-0.41}$\\
&2017 Jan 07&$ 3.33_{-0.40}^{+0.43}$&$     75.5/87$&...&$0.12_{-0.04}^{+0.05}$&$0.59^{+0.57}_{-0.25}$\\
&2017 Jan 09&$ 2.63_{-0.24}^{+0.28}$ &$     150.8/131$&...&$0.40_{-0.07}^{+0.07}$&$3.09^{+1.29}_{-0.59}$\\
\hline
VID 7 &2002 Feb 02&...&...&...&$<0.04$&$<1.71$\\
&2016 Sept 27&$ 1.89_{-0.32}^{+0.33}$&$     32.8/48$&...&$0.19_{-0.05}^{+0.03}$&$10.28^{+1.50}_{-2.89}$\\
&2017 Jan 06+2017 Jan 08&$ 1.62_{-0.15}^{+0.15}$&$     377.8/361$&...&$0.18_{-0.03}^{+0.02}$&$12.07^{+2.18}_{-3.04}$\\
&2017 Feb 10&$ 1.70_{-0.15}^{+0.15}$&$     268.0/283$&...&$0.21_{-0.03}^{+0.02}$&$13.21^{+2.87}_{-3.45}$\\

\hline
VID 8 &2001 Jul 06&$     2.23_{-0.05}^{+0.05}$&$     55.6/55$&0.45&$1.45_{-0.08}^{+0.09}$&$7.02^{+0.54}_{-0.52}$\\

 &2002 Jan 29&$     2.18_{-0.07}^{+0.07}$&$     23.95/32$&0.68&$0.81_{-0.04}^{+0.06}$ &$4.11^{+0.36}_{-0.40}$\\
 
 &2017 Jan 11&$     1.72_{-0.22}^{+0.23}$&$      314.7/305$&...&$0.13^{+0.01}_{-0.02}$&$0.94^{+0.30}_{-0.22}$ \\
 
 
\hline
\end{tabular}
\tablecomments{
Col. (1): object name.
Col. (2): observation start date.
Col. (3): \hbox{power-law} photon index.
Col. (4): $\chi^2$ or $W$-stat value divided by the degrees of freedom.
Col. (5): null-hypothesis probability of the model. If the spectra group is fitted using $W$-stat, it has no corresponding $P_{\rm null}$.
Col. (6): rest-frame 2 keV flux density in units of $10^{-31}$ $\rm erg~cm^{-2}~s^{-1}~Hz^{-1}$. 
Col. (7): rest-frame 2–10 keV luminosity in units of $\rm 10^{43}~erg\,s^{-1}$.
}
\tablenotetext{a}{Estimations from the eROSITA measurements. The observation date is a mean value as described in Section \ref{subsec:lc}. The effective photon index and $f_{\rm 2keV}$ are estimated as described in Section \ref{subsec:X-ray data}.}
\end{table*}

\begin{figure*}
\label{fig:xspec_example}
\hspace{-0.2cm}
 \includegraphics[width=0.87\textwidth]{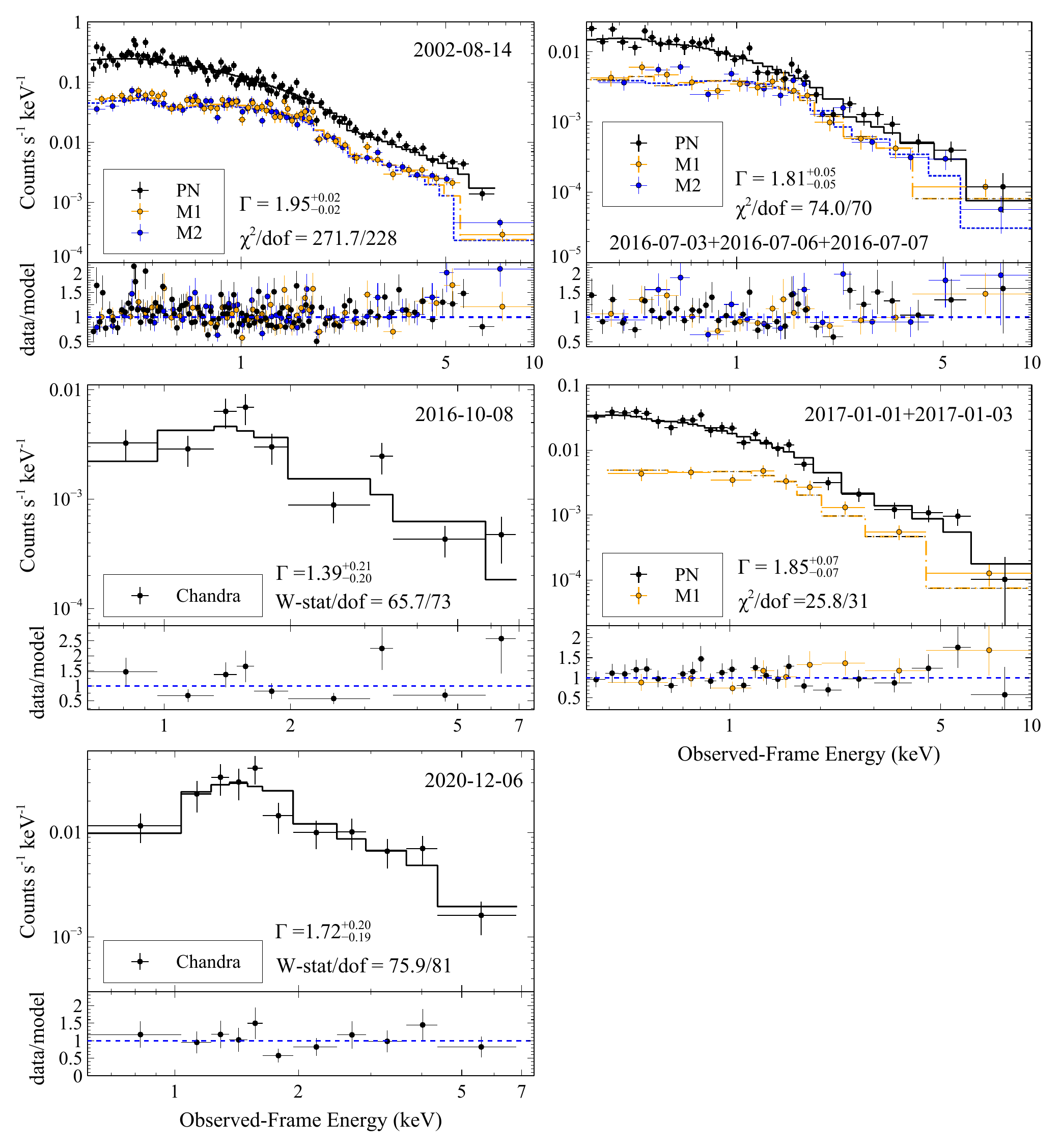}
 \centering
 \caption{\hbox{X-ray} spectra of VID 3 overlaid with the best-fit simple \hbox{power-law} models modified with Galactic absorption. For display purposes, we group the Chandra spectra so that each bin has at least $3\sigma$ significance. The bottom panels of each figure show the ratios of the spectral data to the best-fit models. For XMM-Newton observations, the EPIC pn (black), MOS1 (red), and MOS2 (green) spectra are jointly fitted if available. The observation date, best-fit power-law photon index, and $W$-stat/dof ($\chi^{2}$/dof) for each spectrum are also labeled in each panel.}
\end{figure*}

\subsection{\hbox{X-ray} Spectral Analyses}
\label{3subsec:xray_spec}

The \hbox{X-ray} spectra of each observation of each AGN are fitted using XSPEC (v12.12.1, \citealt{arnaud1996astronomical}). We used the $\chi^2$ statistic for the fitting of spectra that are grouped with at least 25 counts per bin. For spectra that are grouped with at least 1 count per bin, we used the $W$ statistic (\hbox{$W$-stat}).\footnote{See \url{https://heasarc.gsfc.nasa.gov/xanadu/xspec/manual/XSappendixStatistics.html} for details.} We first adopted a simple \hbox{power-law} model modified by Galactic absorption ({\sc zpowerlw*phabs}) to describe the 0.3--10 keV \xmm spectra or \hbox{0.5--7 keV} Chandra spectra. For each observation, we jointly fitted the available EPIC pn, MOS1, and MOS2 spectra. We also added normalization constants (\hbox{best-fit} values between 0.7 and 1.3) to the MOS spectra to account for small cross-calibration uncertainties. The simple \hbox{power-law} model can describe most of the spectra well, with small reduced $\chi^{2}$ values (\hbox{$W$-stat}/dof) and large null-hypothesis probabilities in cases of $\chi^{2}$ fitting. As an example, Figure \ref{fig:xspec_example} shows all the spectra of VID 3 and their best-fit power-law models. We consider spectra with best-fit $\Gamma \lesssim 1.2$ to be significantly affected by absorption \citep[e.g.,][]{2020ApJ...900..141P}. Only the \hbox{lowest-state} spectrum of VID 2 shows such a small photon index. The effective power-law photon index estimated from the eROSITA band ratio of VID 2 is $\sim 0.8$ (Section \ref{subsec:X-ray data}), also suggestive of \hbox{X-ray} absorption.

We also fitted the XMM-Newton spectra that have more than 300 counts and all the Chandra spectra with an additional intrinsic absorption component ({\sc zphabs}), with the $\Gamma$ and $N_{\rm H}$ both being the free parameters. Except for the \hbox{lowest-state} spectrum of VID 2, which has $N_{\rm H} = 6.26_{-3.03}^{+4.00} \times 10^{21} \rm ~cm^{-2}$, the derived upper bounds of $N_{\rm H}$ are small (below $8 \times 10^{20} \rm ~cm^{-2}$). We then compared the fitting statistics between the simple power-law model and the absorbed power-law model. For the lowest-state spectrum of VID 2, including the intrinsic absorption significantly improves the fitting with $\Delta W/\rm dof = 8.2/1$ and an $F$-test probability of 0.0056. For the other spectra, the $F$-test probabilities are all larger than 0.05, suggesting no significant improvements.

From the best-fit simple power-law models, we also computed the observed rest-frame 2 keV flux densities ($f_{\rm 2keV}$) and \hbox{2--10~keV} luminosities ($L_{\rm X}$) of each AGN in each epoch. The \hbox{best-fit} parameters and the derived fluxes/luminosities are shown in Table \ref{tab:xray_variability_X_spec_info}. The exposure-time averaged $L_{\rm X}$ values of our sources are consistent with (within $15\%$) the $L_{\rm X}$ values given by \citet{2018MNRAS.478.2132C}, which are measured in the co-add images. Based on the \hbox{best-fit} $f_{\rm 2keV}$, we calculated the maximum 2 keV flux density variability amplitude $A_{\rm var,2keV}$, defined as the ratio of the highest and lowest $f_{\rm 2keV}$. The maximum $A_{\rm var,2keV}$ values are listed in Table \ref{tab:properties_summary}, ranging from $\approx 6$ to $\approx 12$.

\begin{figure*}
 \hspace{-0.2cm}
 \includegraphics[width=0.99\textwidth]{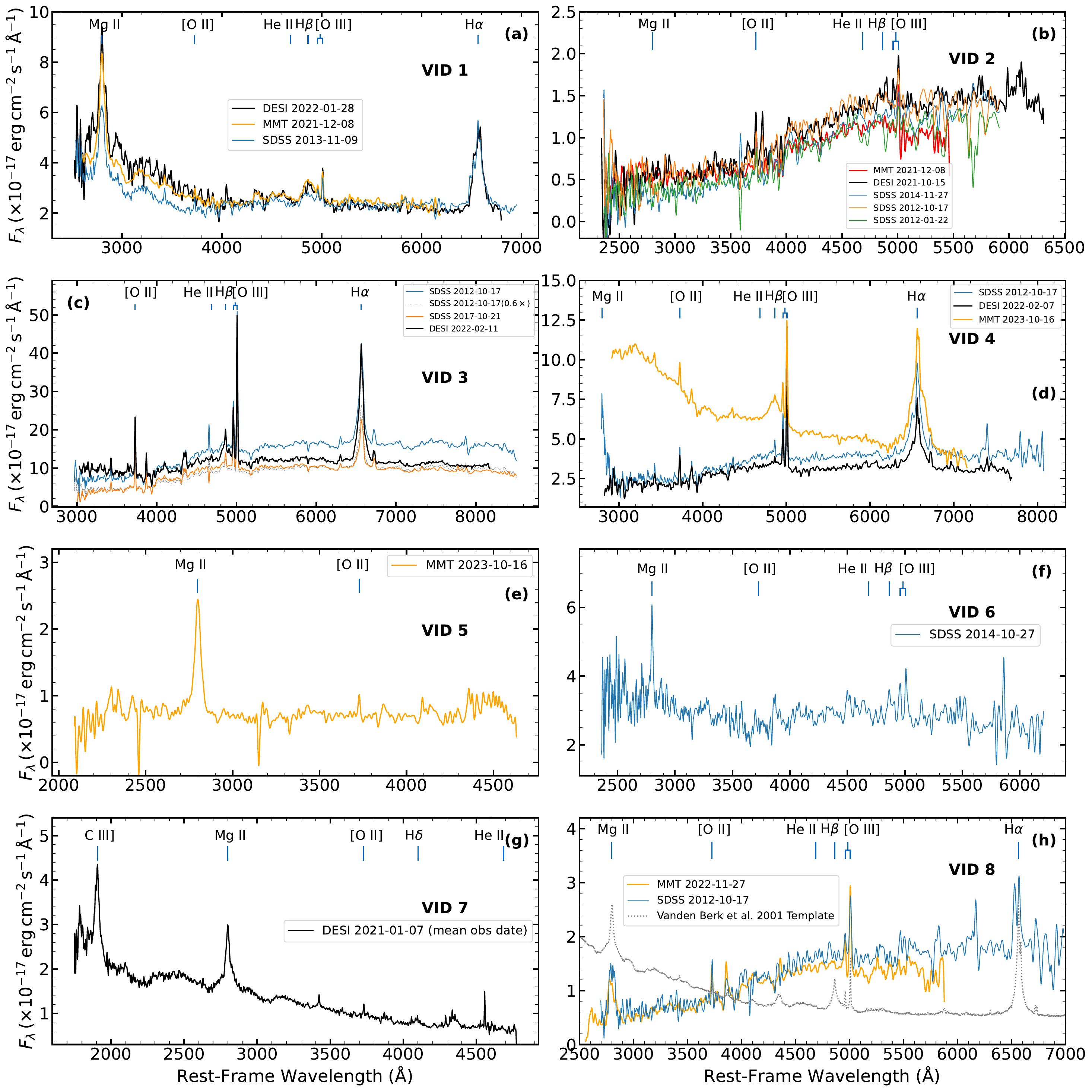}
 \centering
 \caption{Optical spectra of VIDs 1--8. The spectra are smoothed by a Gaussian kernel with a standard deviation of 5 pixels. The observation date of each spectrum is labeled. Some of the major emission lines are marked. The multi-epoch spectra of VID 3 and VID 4 show significant variability, while the spectra of the other six AGNs did not vary greatly or lack multi-epoch information. The gray curve in panel (h) shows the composite spectrum of typical SDSS quasars \citep{2001AJ....122..549V} for comparison, scaled to have a similar Mg II line intensity to VID 8.}
 \label{fig:all_opt_spec}
\end{figure*}

\subsection{Optical Spectral Properties}
\label{subsec:optical_spec}

Figure \ref{fig:all_opt_spec} shows the optical spectra of VIDs 1--8. VID 3 is a merging galaxy with two nuclei in the central region separated by $\sim 1''$, and only one nucleus shows AGN activity (source A, see the Appendix for more detail). The SDSS and DESI spectra of VID 3 were taken centered on source A. The SDSS spectrum of VID 6 shows a significantly lower flux level than its photometric measurements from the Hyper Suprime-Cam Deep Survey \citep[HSC;][]{2022PASJ...74..247A}, which were taken contemporaneously within one month. Therefore, we normalized the spectrum of VID 6 to the HSC $g$-band photometry. The optical spectra of all sources except VID 2 show significant broad emission lines, confirming their classification as Type I AGNs. For VID 2, VID 3, and VID 8, their continuum emission is dominated by their host galaxies, while the other five sources show relatively low galaxy contamination.

The MMT and SDSS spectra of VID 2 show no AGN signature and little variability. However, it shows multiple features that satisfy the \hbox{X-ray} AGN criteria \citep[e.g.,][]{2015A&ARv..23....1B,2017ApJS..228....2L}: (1) \hbox{0.5--10~keV} luminosity significantly larger than \hbox{$3 \times 10^{42}$ \lum}; (2) \hbox{X-ray}-to-optical ratio $\log f_{\rm X}/f_{r} > 1$ calculated from the HSC-SSP $r$-band photometry and mean \hbox{X-ray} flux; (3) \hbox{X-ray}-to-near-IR ﬂux ratio $\log f_{\rm X}/f_{Ks} > 1$ calculated from the VIDEO $Ks$- band photometry and mean \hbox{X-ray} flux; (4) extreme \hbox{X-ray} variability. Thus, VID 2 can be classified as an X-ray AGN.

Of the six sources with multiple spectroscopic observations, VID 3 and VID 4 exhibit significant variability, while the other sources show little variability overall. We note that the MMT, DESI, and SDSS/BOSS spectra have different fiber sizes ($1\farcs{5}$ to $3''$), and spurious variability might be introduced if different amounts of host-galaxy contribution are included in the spectra. The spectra of VID 4 show clearly the changing-look behavior, which is not caused by any host-galaxy contamination. For VID 3, the difference between the two SDSS spectra, which are both centered on source A, is unlikely caused by intrinsic variability of the object. In Figure \ref{fig:all_opt_spec}(c), we show that the SDSS spectrum observed on 2012 Oct 17 scaled by a factor of 0.6 (gray dotted line) agrees well with the SDSS spectrum observed on 2017 Oct 21. Thus, the difference between them is more likely caused by inaccurate flux calibration, as these two spectra are both from BOSS and they may have large calibration errors \citep[e.g.,][]{2016ApJ...831..157M}. The continuum flux level of its SDSS spectrum observed on 2017 Oct 21 is more consistent with its DESI spectrum and the HSC/SDSS photometry measurements, though enhancement of emission lines is present in the DESI spectrum (see Section \ref{subsec:accretion_rate_change} below).

For VID 4, the SDSS survey reported it as a broad-line AGN, but its new MMT spectrum obtained on 2023 Oct 16 reveals a significant enhancement in the quasar continuum emission and the broad Balmer emission compared to the SDSS spectrum. The variability of the spectra is consistent with the brightening trend of the ZTF $g$- and $r$-band light curves (see Section \ref{subsec:lc} and Figure \ref{fig:alllc1} below). Specifically, the broad H$\beta$ emission, which appears weak in the SDSS spectrum, exhibits a significant increase in the MMT spectrum. This result suggests that VID 4 is a changing-look AGN.

The optical spectra of VID 8 have a galaxy-dominated continuum and a significant Mg II broad line. For comparison, we also include the composite spectrum of typical SDSS quasars from \citet{2001AJ....122..549V} in Figure \ref{fig:all_opt_spec}(h), scaled to have a similar Mg II line intensity to VID 8. There appears to be no H$\beta$ broad line in the MMT and SDSS spectra, and the AGN continuum level and H$\alpha$ line intensity are also lower than that expected from the Mg II line. Such behavior is consistent with a rare population of Mg II emitters discovered in spectroscopically conﬁrmed massive galaxies from the SDSS \citep{2014ApJ...781...72R}. It can be interpreted as an intermediate stage of a changing-look AGN (see discussion in Section \ref{subsec:accretion_rate_change}).

\subsection{Optical Spectral Fitting and SMBH Mass Estimation}
\label{subsec:BHmass_estimation}

\subsubsection{Optical Spectral Fitting}
\label{subsubsec:optical_fit}

We estimate SMBH masses for the five AGNs that are not in the SDSS DR16 quasar catalog. For VIDs 5--7, we estimate their single-epoch virial SMBH masses and Eddington ratios following the method described in \citet{2022ApJS..263...42W}. We first perform the spectral fitting using QSOFITMORE \citep{2021zndo...5810042F}, which is a wrapper package based on PyQSOFit \citep{2018ascl.soft09008G}. The spectra are corrected for Galactic extinction using the $E_{B-V}$ value obtained from \cite{2011ApJ...737..103S} and the Milky Way extinction law ($R_V = 3.1$) from \citet{2019ApJ...886..108F}. 

For our AGNs with relatively low redshifts, their host galaxies may have significant contributions to the spectra. To extract the intrinsic AGN properties, we use PyQSOFit's built-in principal component analysis (PCA) method \citep{2004AJ....128.2603Y, 2004AJ....128..585Y} to decompose the spectrum into a host-galaxy component and an AGN component. For VID 6, the decomposition returned a host-galaxy spectrum with more than 100 pixels having negative flux. Thus, decomposition is not applied to this source. For VID 5 and VID 7, the decomposed host galaxies both contribute $\sim 25 \%$ of the total spectral flux at rest-frame 4500 $\AA$.

We then mask the emission lines and fit a pseudocontinuum for the isolated AGN spectrum of each source using a power-law component, a third-order polynomial component, and an Fe II component (see Section 3.2 of \citealt{2022ApJS..261...32F} for more details). The emission-line components are fitted with Gaussian profiles with the continuum component subtracted. For the ${\rm H}\beta$ and $\rm Mg~II$ lines, we fit both their broad and narrow components. The narrow-line component is fitted with one Gaussian profile with ${\rm FWHM} < 1200 {\rm~km~s^{-1}}$. The broad-line components of ${\rm H}\beta$ and $\rm Mg~II$ are fitted with up to three and two Gaussian profiles with ${\rm FWHM} > 1200 {\rm~km~s^{-1}}$, respectively, as suggested by \citet{2019ApJS..241...34S}. This fitting of the ${\rm H}\beta$ line yields results consistent with \citet{2006ApJ...641..689V}, as demonstrated by \citet{2011ApJS..194...45S}. Although the fitting of ${\rm FWHM}_{\rm MgII}$ in \citet{2011ApJS..194...45S} uses up to three Gaussian profiles for the broad component, two Gaussian profiles can already fit the broad Mg II components of our sources well. The Monte Carlo method is used to estimate the uncertainties of the measured quantities, incorporating 50 trials where random Gaussian noise ($\sigma$, the flux uncertainty) is added to the AGN spectrum for the continuum and emission line fits, with the resulting uncertainties calculated as standard deviations.


\subsubsection{SMBH Mass Estimation}
The single-epoch virial SMBH masses of VIDs 5--7 are then estimated using the continuum luminosities and line widths (FWHMs) of the broad components obtained from spectral fitting. For VID 5 and VID 7, the SMBH masses are estimated using the ${\rm FWHM}_{\rm MgII}-L_{3000}$ pair and the empirical equation from \citet{2011ApJS..194...45S}:
\begin{equation}
\begin{aligned}
&\log\left(\frac{M_{\rm BH,vir}}{M_{\odot}}\right) \\
    &= \log\left[\left(\frac{{\rm FWHM} ({\rm Mg II})}{\rm km~s^{-1}}\right)^{2}\left(\frac{L_{3000}}{10^{44} {\rm ~erg~s^{-1}}}\right)^{0.62}\right] + 0.74.
\end{aligned}
\end{equation}

For VID 6, we use the ${\rm FWHM}_{{\rm H} \beta}-L_{5100}$ pair and the empirical equation from \citet{2006ApJ...641..689V}:

\begin{equation}
\begin{aligned}
&\log\left(\frac{M_{\rm BH,vir}}{M_{\odot}}\right) \\
    &= \log\left[\left(\frac{{\rm FWHM} ({\rm H}\beta)}{\rm km~s^{-1}}\right)^2\left(\frac{L_{5100}}{10^{44} {\rm ~erg~s^{-1}}}\right)^{0.5}\right] + 0.91.
\end{aligned}
\end{equation}
We note that the black hole masses estimated using this single-epoch method have large systematic uncertainties, which may be $\sim $ 0.5--1 dex \citep[e.g.,][]{2009ApJ...692..246D}.

For VID 2, its spectra show no broad emission line. VID 8's spectra show broad Mg II but the continua are dominated by its host galaxy, and it is hard to subtract the galaxy component. Their SMBH masses cannot be estimated using the virial method. We thus estimate their SMBH masses using the $M_{*} \textrm{--} M_{\rm BH}$ relation. The stellar masses are taken from \citet{2022ApJS..262...15Z}, which are derived from SED fitting. The IR--UV SEDs of VID 2 and VID 8 are dominated by their hosts, and thus their $M_{*}$ values measured through SED fitting should be reliable. We use the $M_{*} \textrm{--} M_{\rm BH}$ relation in \citet{2023ApJ...954..173L}, calibrated using reverberation-mapping SMBH masses of quasars with similar redshifts to our sample. This relation has an intrinsic scatter of $0.47^{+0.24}_{-0.17}$ dex. We also use the same method to estimate the SMBH masses of the other six sources, and the results agree with their single-epoch masses within $\sim 0.5$ dex except VID 7.

We then estimate the bolometric luminosity $L_{\rm bol}$ of VIDs 5--7 using the measured continuum luminosity at rest-frame $3000\AA$ with a bolometric correction of 5.15, which is derived from the mean SED of quasars in \citet{2006ApJS..166..470R}. For VID 2 and VID 8, it is hard to estimate the bolometric luminosity directly from their optical spectra because of substantial galaxy contamination. We thus estimate the bolometric luminosity using their highest-state \hbox{X-ray} luminosity. We use the \hbox{luminosity-dependent} \hbox{X-ray} bolometric correction ($\kappa_{\rm X} \equiv L_{\rm bol}/L_{\rm X}$) following \citet{2020A&A...636A..73D}:
\begin{equation}
\begin{aligned}
\kappa_{\rm X}=12.76\left[ 1 + \frac{\log(L_{\rm bol}/L_{\odot})}{12.15}\right]^{18.78}.
\end{aligned}
\end{equation}
We first assume $\kappa_{\rm X} = 50$ and then calculate $L_{\rm bol}$ iteratively until it converged. The estimated single-epoch SMBH masses (for VID 2 and VID 8 the SMBH masses estimate by the $M_{*} \textrm{--} M_{\rm BH}$ relation are used) and bolometric luminosities are used to calculate the Eddington ratio $\lambda \equiv L_{\rm bol}/L_{\rm Edd}$, where $L_{\rm Edd} = 1.28\times10^{38}(M_{\rm BH}/M_{\odot})~\rm erg~s^{-1}$ is the Eddington luminosity. The Eddington ratios of these eight AGNs range from 0.004 to 0.132.

\begin{figure*}
\centering
\begin{minipage}{1.01\columnwidth}
      \includegraphics[clip,trim=0 0.cm 0 0cm, width=1.0\columnwidth]{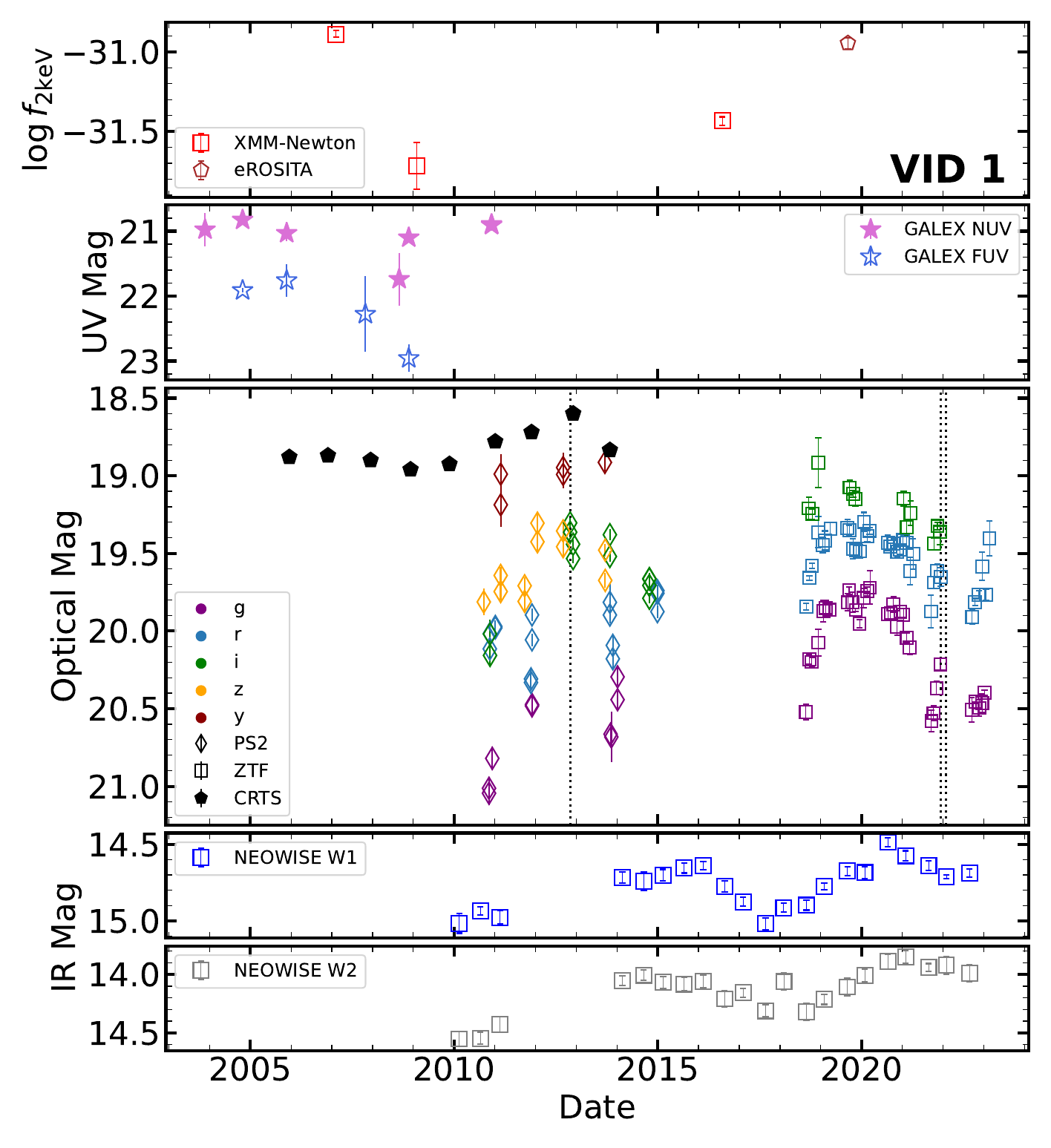}
      
     \vspace{0.2cm}
     \includegraphics[clip,trim=0 0.cm 0 0cm, width=1.0\columnwidth]{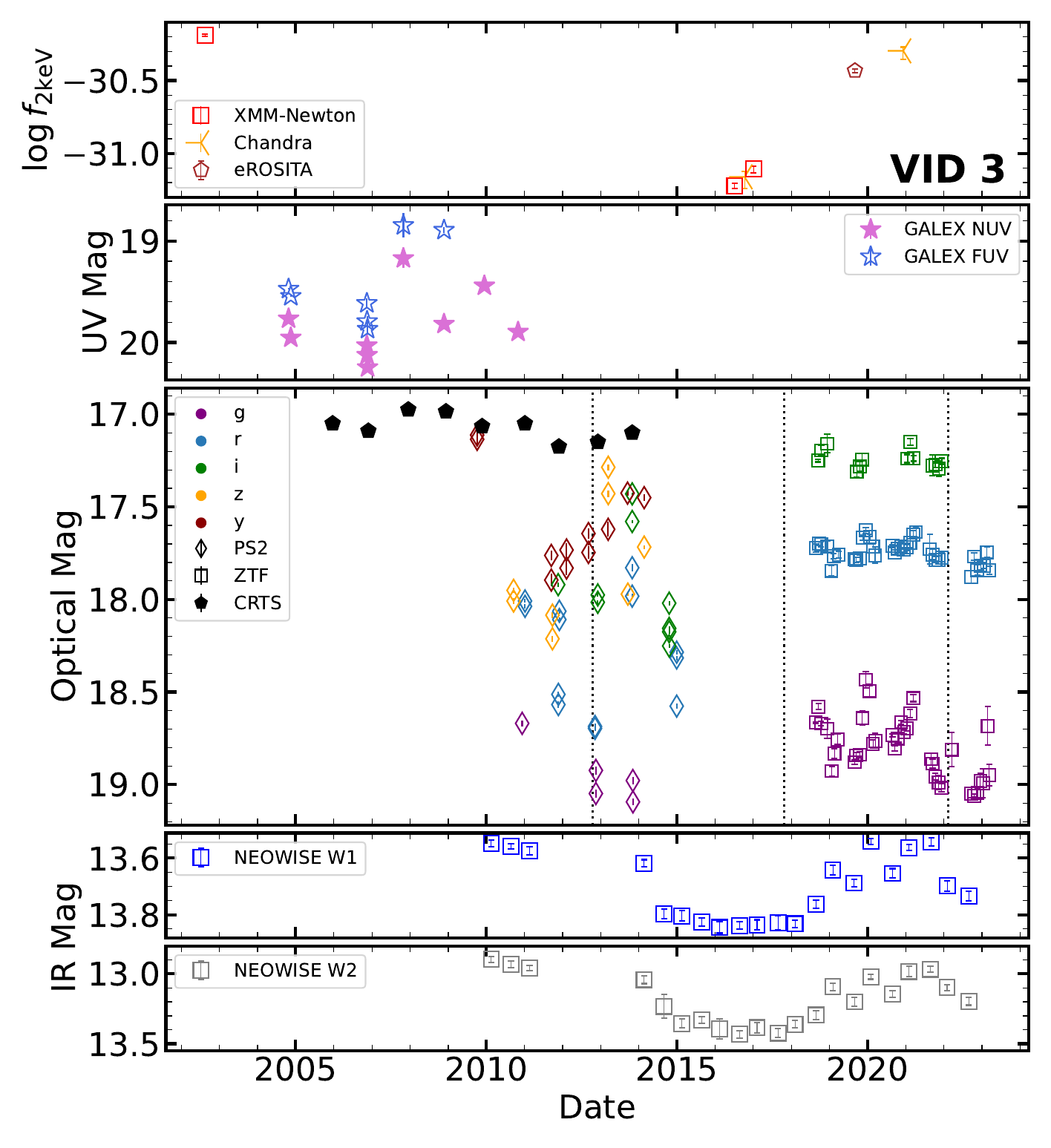}

\end{minipage}
\begin{minipage}{0.99\columnwidth}
      \includegraphics[clip,trim=0.5cm 0.cm 0 0cm, width=1.01\columnwidth]{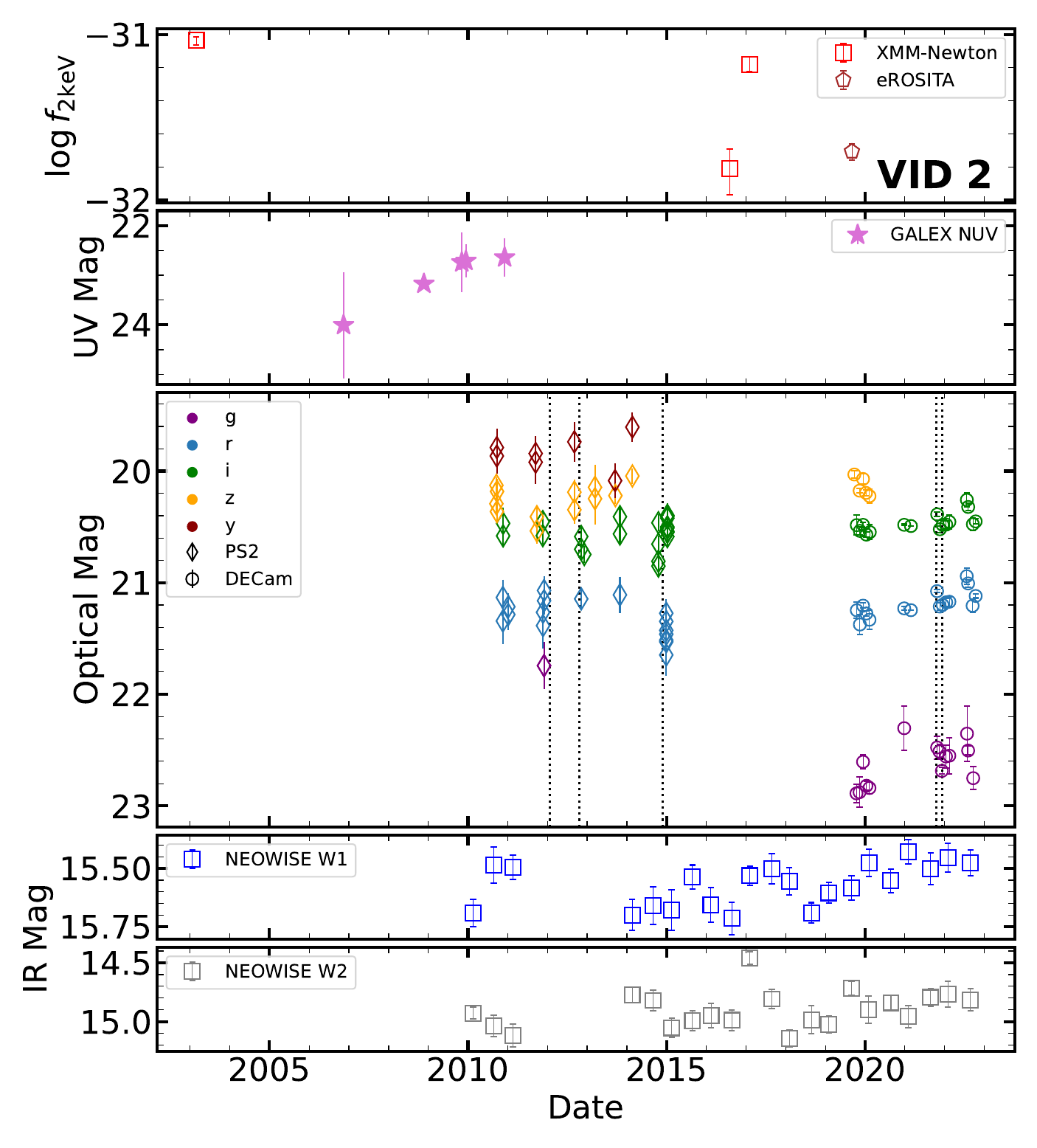}

       \includegraphics[clip,trim=0 0.cm 0 0cm, width=1.0\columnwidth]{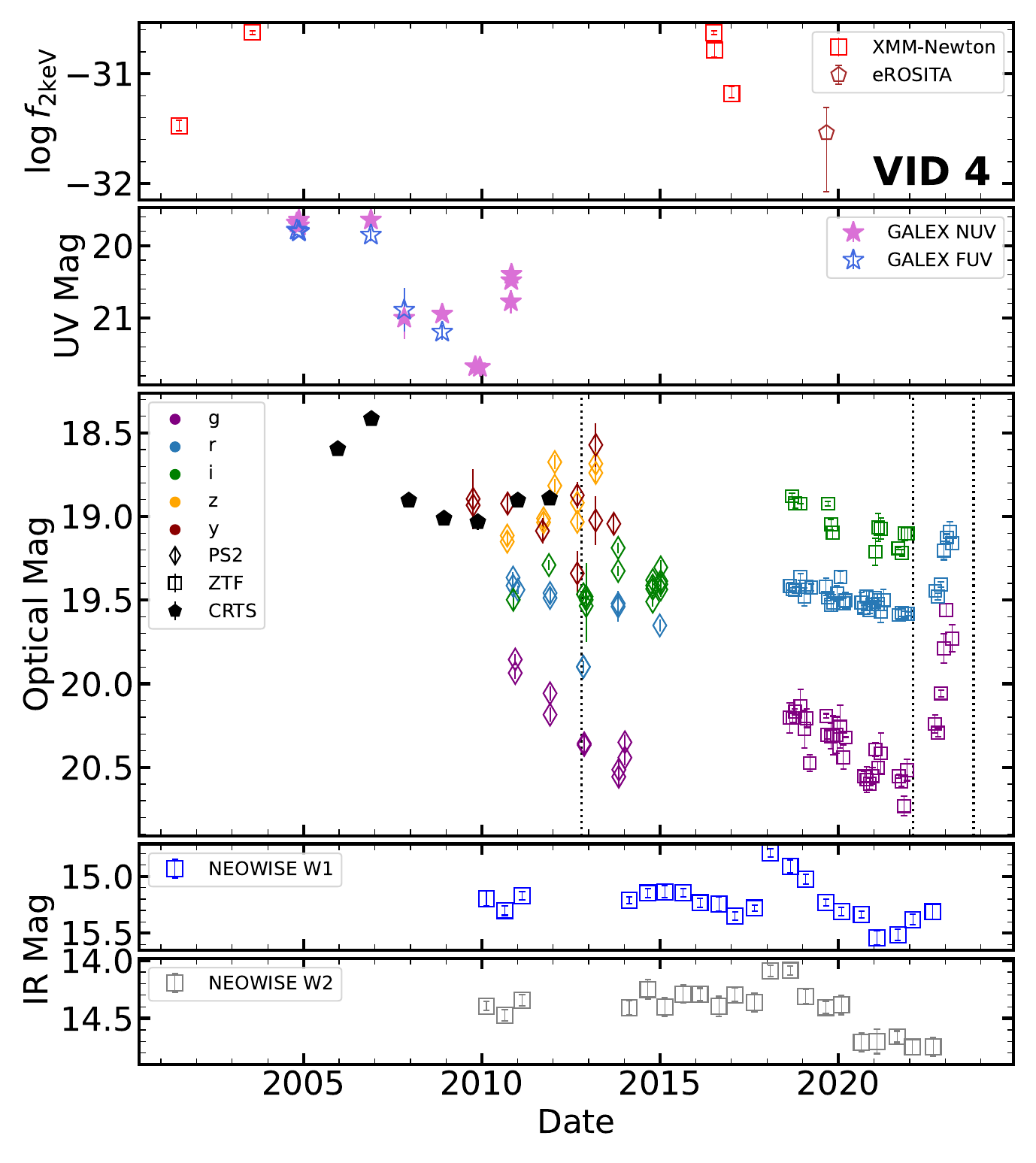}

\end{minipage}
\caption{The multiwavelength light curves of VIDs 1--4. For each source, the first panel shows the rest-frame 2 keV flux densities ($\rm erg~s^{-1}~cm^{-2}~Hz^{-1}$) in logarithm. The second panel shows the GALEX NUV- and FUV-band magnitudes. The third panel shows the optical magnitudes, including PS2 (open thin diamonds), ZTF (open squares), DECam (open circles), and CRTS (black pentagons). Purple, blue, green, yellow, and brown colors in the third panel correspond to $g$-, $r$-, $i$-, $z$-, and $y$-band magnitudes, respectively. The vertical dotted lines in the third panel indicate the epochs of optical spectroscopic observations. The last two panels show NEOWISE $W1$- and $W2$-band magnitudes. All the data points have been corrected for the Galactic extinction.}
\label{fig:alllc1}
\end{figure*}

\begin{figure*}
\centering
\begin{minipage}{1.01\columnwidth}
      \includegraphics[clip,trim=0 0.cm 0 0cm, width=1.0\columnwidth]{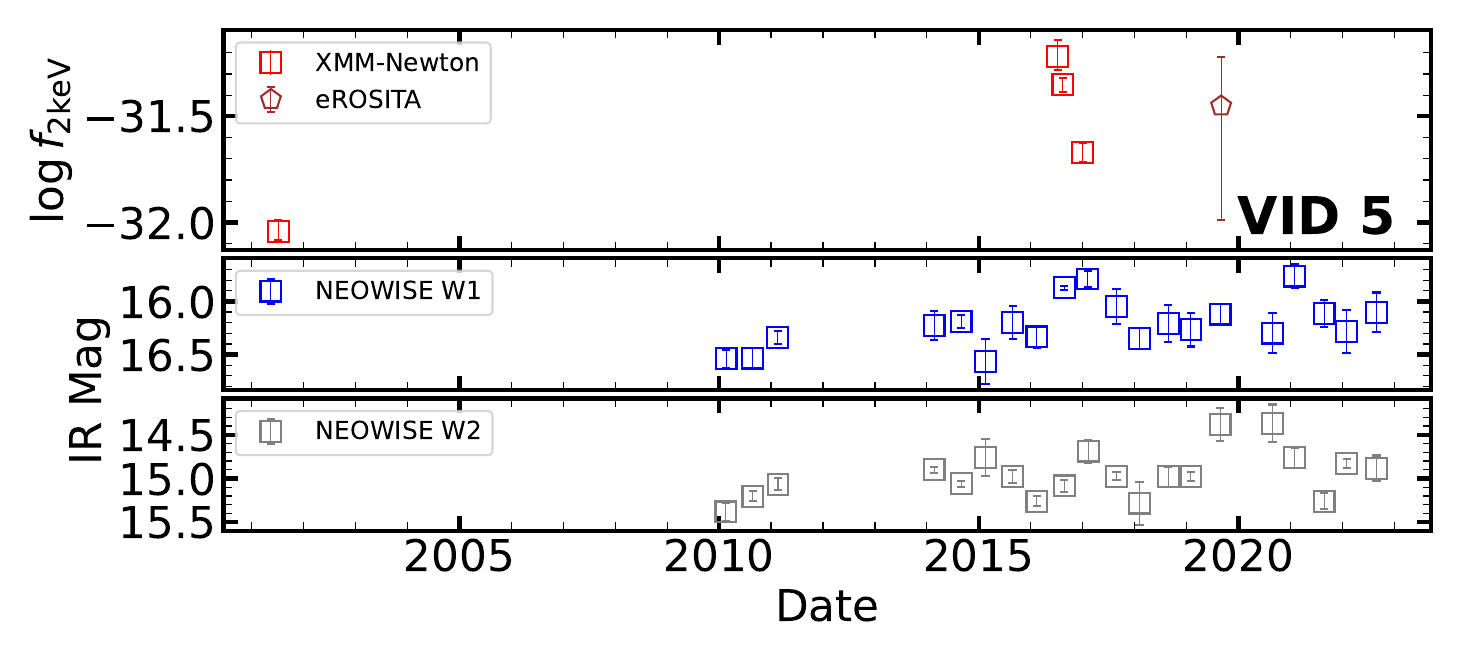}
      
     \vspace{0.2cm}
     \includegraphics[clip,trim=0 0.cm 0 0cm, width=1.0\columnwidth]{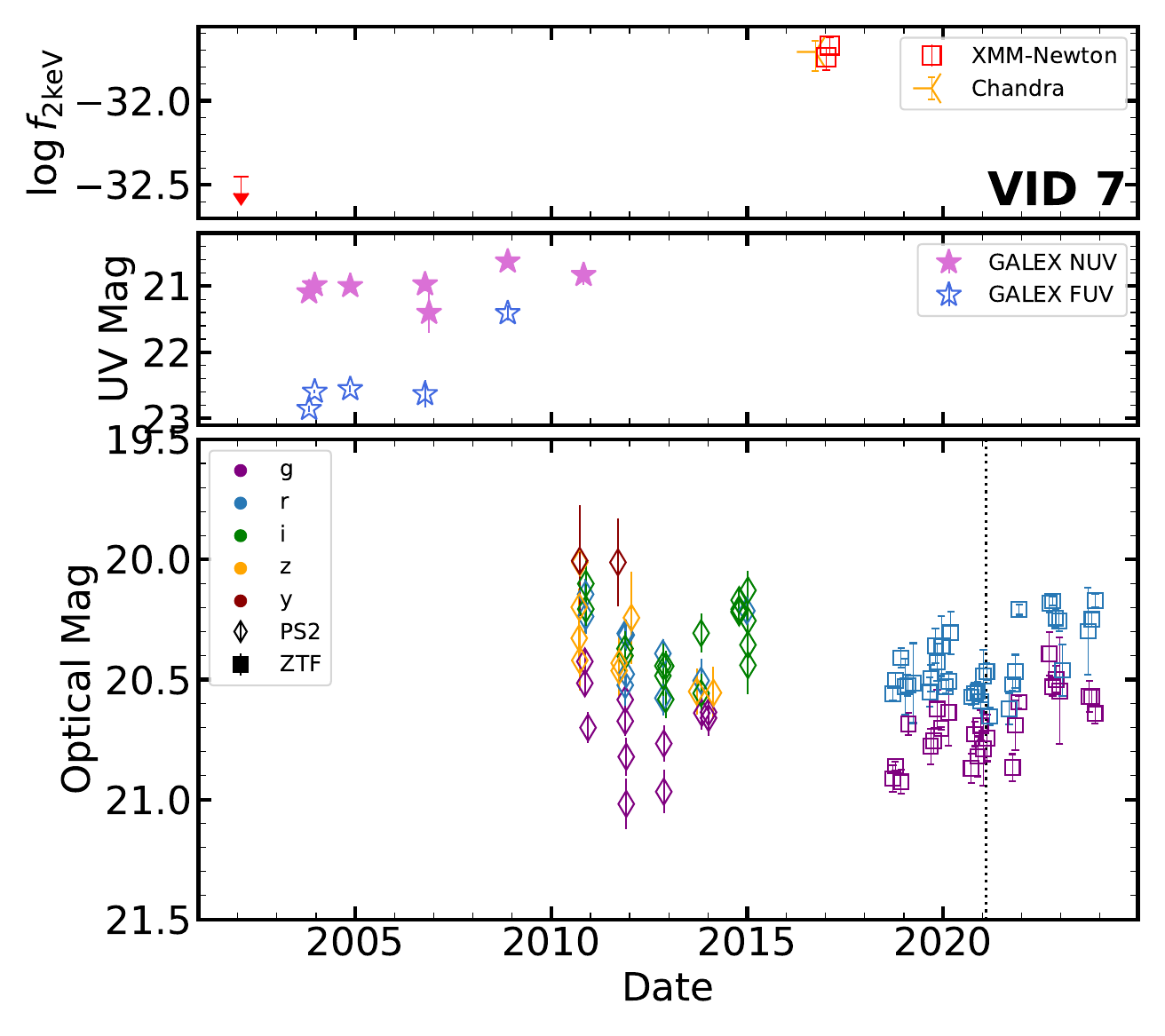}

\end{minipage}
\begin{minipage}{0.99\columnwidth}
      \includegraphics[clip,trim=0 0.cm 0 0cm, width=1.02\columnwidth]{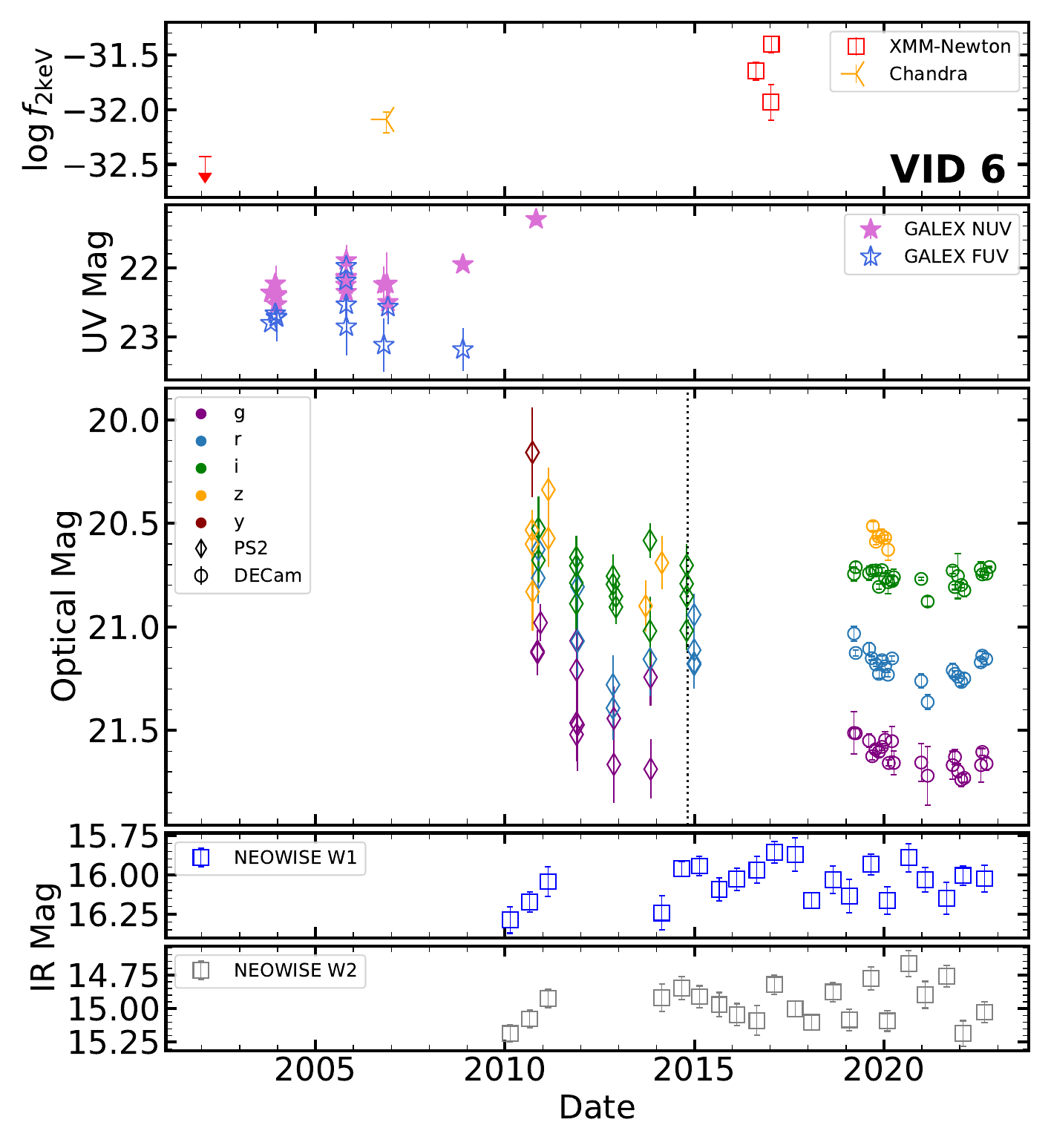}

       \includegraphics[clip,trim=0 0.cm 0 0cm, width=1.0\columnwidth]{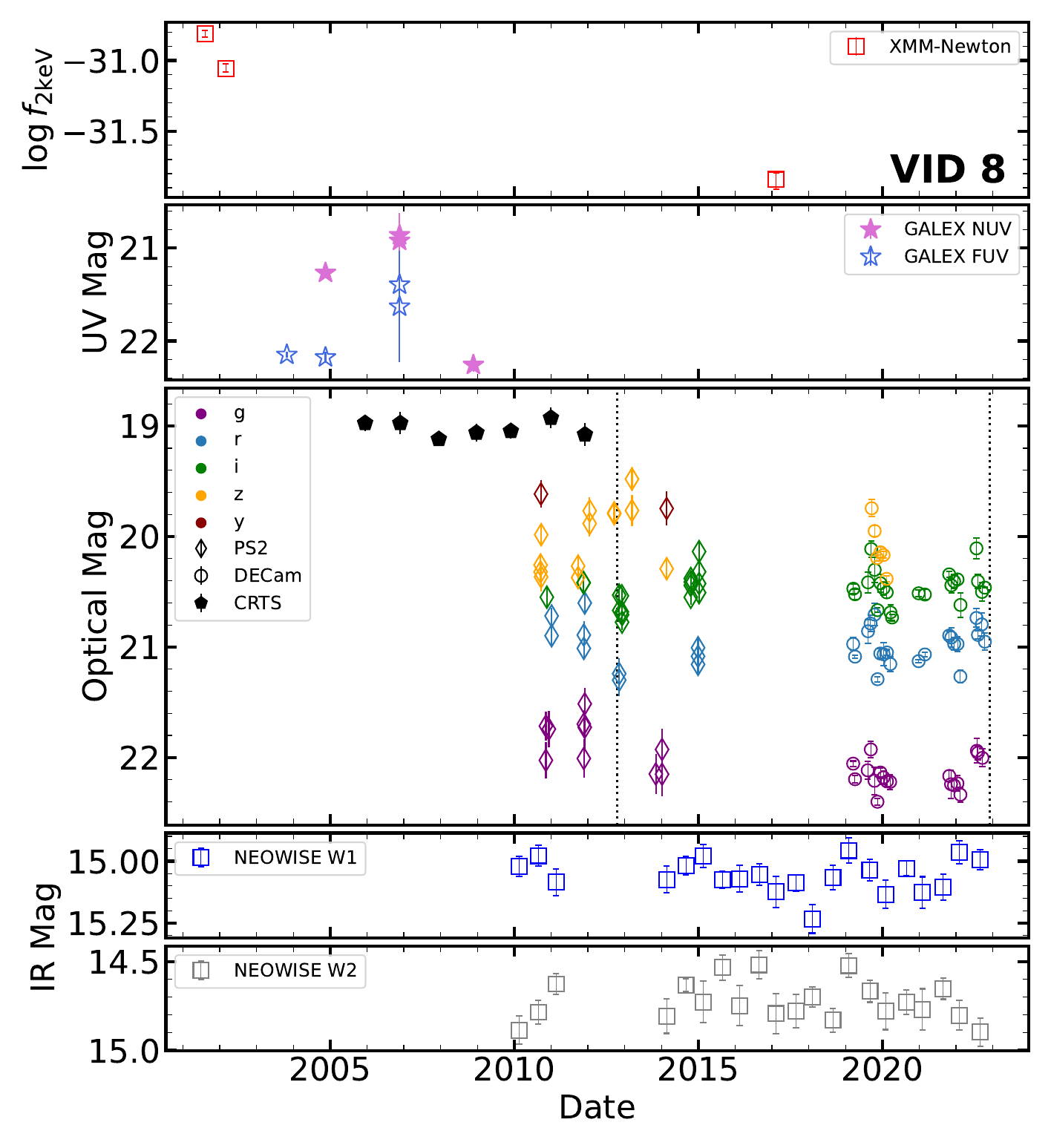}

\end{minipage}
\caption{The same as Figure \ref{fig:alllc1}, but for VIDs 5--8. For VID 5, it is too faint to be detected in the GALEX or optical monitoring surveys. VID 7 is not detected in the NEOWISE individual exposures and does not have an IR light curve.}
\label{fig:alllc2}
\end{figure*}

\subsection{\hbox{X-ray} and Multiwavelength Light Curves}
\label{subsec:lc}

Based on the measured $f_{\rm 2keV}$ values in Table \ref{tab:xray_variability_X_spec_info}, we plot the \hbox{X-ray} light curve of each source. The eROSITA 2 keV flux densities (Section \ref{subsec:X-ray data}) are also added. The eROSITA exposures for these sources were conducted from 2019-08-26 to 2019-09-08, and we use 2019-09-01 as the time-axis coordinate for the eROSITA measurement in the \hbox{X-ray} light curves. The $f_{\rm 2keV}$ light curves are presented in Figure \ref{fig:alllc1} and Figure \ref{fig:alllc2}. Table \ref{tab:properties_summary} summarizes the maximum amplitudes of the \hbox{X-ray} variability.

We construct multiwavelength light curves using the data collected in Section \ref{subsec:multi-wave_lc_construction}. We use the PSF magnitudes of the ZTF, Pan-STARRS, and HELM surveys, with problematic data filtered using the flag provided by the catalogs. The ZTF and HELM light curves are binned per month, while the CRTS and NEOWISE light curves are binned per year and per half a year, respectively. The errors of the binned light curves are estimated using a bootstrap approach. These light curves are also shown in Figure \ref{fig:alllc1} and Figure \ref{fig:alllc2}, with the maximum variability amplitudes of the binned \hbox{$g$-band} and $W1$-band light curves ($\Delta g$ and $\Delta W1$) listed in Table \ref{tab:properties_summary}. We use only the ZTF or HELM light curves to calculate the maximum $\Delta g$, as the uncertainties of the Pan-STARRS light curves are larger.

The \hbox{X-ray} light curves of these eight sources exhibit various patterns on timescales of the order of years. VID 2 and VIDs 4--6 have rapid variability within several months. The \hbox{X-ray} flux for VID 4 and VID 5 initially increase and then decrease, while for VID 1 and VID 3, it initially decreases and then increases. The \hbox{X-ray} light curves of VID 6 and VID 8 show a continuous upward and downward trend, respectively. VID 7 has \hbox{X-ray} measurements in only two epochs, which show an increasing trend over time. The \hbox{X-ray} flux of VID 2 exhibits multiple fluctuations. Such diverse characteristics of these sources are related to both the origin of the variability and the sampling of the X-ray observations.

\begin{figure}
 \includegraphics[width=0.45\textwidth]{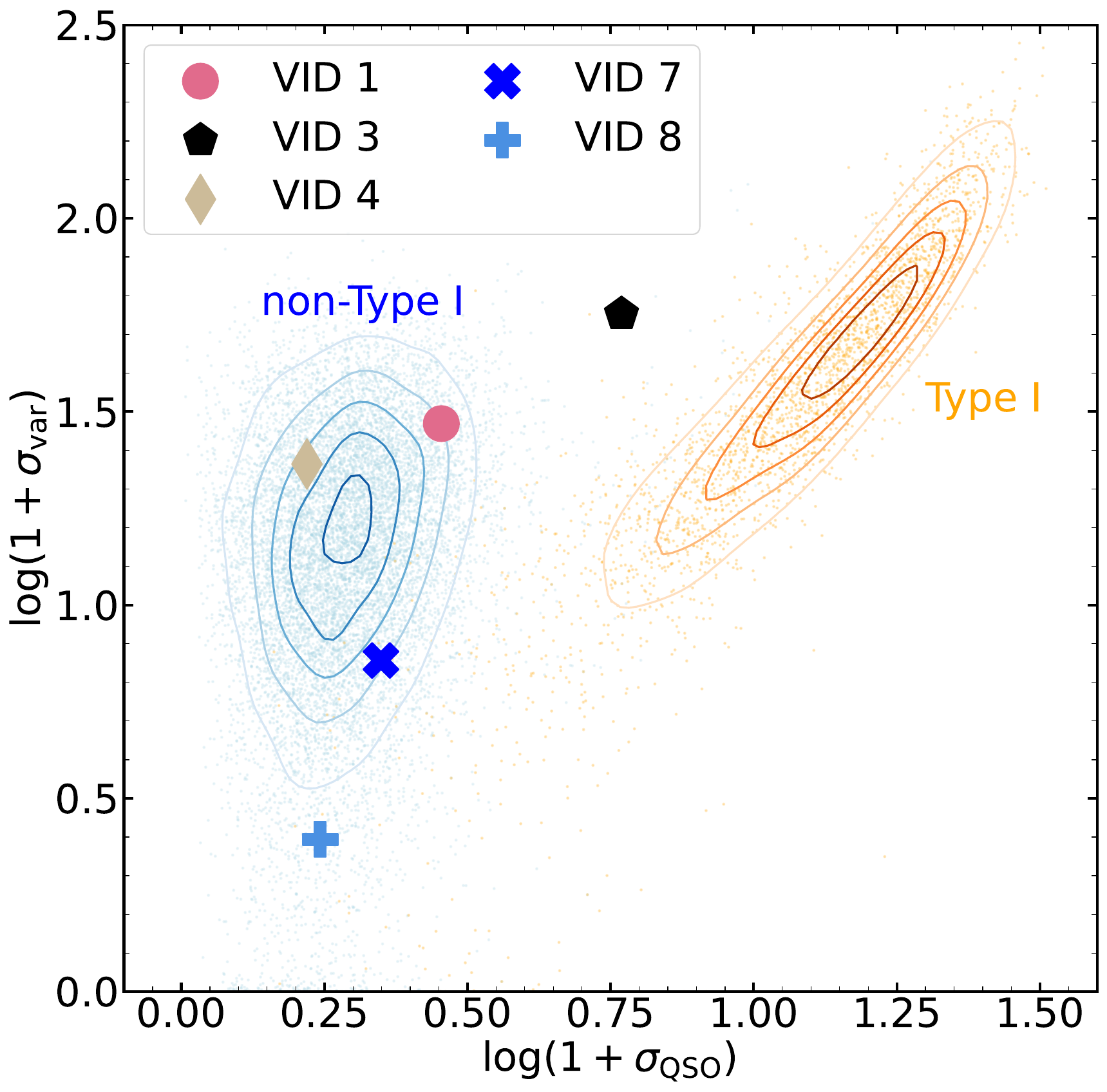}
 \centering
 \caption{Distribution of the five AGNs with ZTF light curves (VIDs 1, 3, 4, 7, 8) on the $\sigma_{\rm QSO}$--$\sigma_{\rm var}$ plane. The data points of typical Type I AGNs (orange dots and contours) and non-Type I AGNs (blue dots and contours) from \citet{2024ApJ...966..128W} are overlaid for comparison.}

 \label{fig:opt_vat_compare}
\end{figure}

All but two of these eight sources have multiwavelength light curves from IR to UV. VID 5 is too faint to be detected in the GALEX or optical monitoring surveys. VID 7 is not detected in the NEOWISE single exposures and thus it has no IR light curve. The \hbox{multiwavelength} light curves of these sources show two different characteristics as described below. 

VID 1, VID 3, and VID 4 all show significant and consistent multiwavelength variability. Their optical and IR variability amplitudes are larger than those of typical AGNs, with maximum $\Delta g \geq 0.5$ and $\Delta W1 \geq 0.5$ \citep[see Table \ref{tab:properties_summary};][]{2017ApJ...834..111C}. The variability timescales are of the order of years, similar to those of other changing-look AGNs \citep[e.g.,][]{2015ApJ...800..144L,2018ApJ...862..109Y,2022ApJ...933..180G}. The relative contribution of the AGN and the host galaxy strongly affects the observed variability amplitude since the galaxy's contribution can dilute the AGN's variability, making the overall variability amplitude smaller. These three sources have more intense variability in the bluer bands, which are less contaminated by the host galaxy. Their UV light curves show the largest variability amplitudes among the \hbox{IR--UV} light curves, and the $g$-band light curves show larger variability amplitude than other optical bands. Note that the images of the CRTS survey are taken unfiltered in the optical band to maximize throughput. As a result, the CRTS light curves suffer from large galaxy contamination and show relatively low variability amplitude. VID 5 has no optical light curves, but its IR light curves also show large variability ($\Delta W1 = 0.8$), with a trend that is consistent with its \hbox{X-ray} light curve.

The multiwavelength light curves of VID 2 and VIDs 6--8 show a different pattern. They show large \hbox{X-ray} variability. However, their optical and IR light curves show mild variability ($\Delta g \leq 0.5$ and $\Delta W1 \leq 0.5$). Their multiwavelength variability has no obvious coordination. The different variability characteristics of these four AGNs suggest a different nature of their variability compared with the former four AGNs.

We compare the optical variability amplitudes of our sources to typical Type I AGNs based on the variability statistics calculated by the \texttt{qso\_fit} software \citep{2011AJ....141...93B}. For a given light curve, \texttt{qso\_fit} provides two variability metrics: (i) $\sigma_{\rm var}$ assesses the significance of variability; (ii) $\sigma_{\rm QSO}$ quantifies the extent to which the sources variability is better described by a damped random walk (DRW) model, which can well describe the AGN UV-optical continuum variability \citep[e.g.,][]{2009ApJ...698..895K,2010ApJ...721.1014M,2011ApJ...735...80Z}, rather than a time-independent variable Gaussian signal. The computation was performed using the ZTF light curves for VIDs 1, 3, 4, 7, 8, which are binned per 3 days and cleaned following the same procedures described in Section 2 of \citet{2024ApJ...966..128W}. VIDs 2, 5, 6 do not have ZTF light curves; the HELM light curves of VID 2 and VID 6 have limited epochs and show little variability.

The distribution of VIDs 1, 3, 5, 7, 8 on the $\sigma_{\rm QSO}$--$\sigma_{\rm var}$ plane is shown in Figure \ref{fig:opt_vat_compare}, overlaid with the data points of Type I AGNs and non-Type I AGNs\footnote{These sources were termed as ``Type II AGNs'' in \citet{2024ApJ...966..128W}, which is not an accurate classification of sources without clear broad emission lines. To avoid confusion, we use the term ``non-Type I AGNs'' here.} from \citet{2024ApJ...966..128W} for comparison. The $\sigma_{\rm QSO}$ and $\sigma_{\rm var}$ of our sources and the \citet{2024ApJ...966..128W} sources are all calculated using the ZTF $g$-band light curves, except for VID 7 where the $r$-band light curve was used as its redshift is much higher than the \citet{2024ApJ...966..128W} sources ($z<0.35$). Figure \ref{fig:opt_vat_compare} demonstrates that typical Type I AGNs exhibit large $\sigma_{\rm var}$ and $\sigma_{\rm QSO}$, while non-Type I AGNs have smaller $\sigma_{\rm var}$ and $\sigma_{\rm QSO}$. VID 1 and VIDs 3--4 have similar $\sigma_{\rm var}$ with Type I AGNs, while their $\sigma_{\rm QSO}$ are relatively small, suggesting that their variability patterns are different from typical Type I AGNs (the DRW model). VIDs 7--8 have both smaller $\sigma_{\rm var}$ and $\sigma_{\rm QSO}$ than typical Type I AGNs, also suggestive of different variability patterns. \citet{2024ApJ...966..128W} considered an object to be a changing-look AGN candidate if its $\log (1+ \sigma_{\rm QSO}) < 0.3 ~(>0.8)$ and the variability classification (Figure \ref{fig:opt_vat_compare}) differs from the spectroscopic classification. Adopting these criteria, VID 4 and VID 8 are changing-look AGN
candidates, consistent with our classification below (Section \ref{subsec:accretion_rate_change}).

\begin{figure*}
\centering
\begin{minipage}{0.98\columnwidth}
      \includegraphics[clip,trim=0 1.2cm 0 0.2cm, width=1.0\columnwidth]{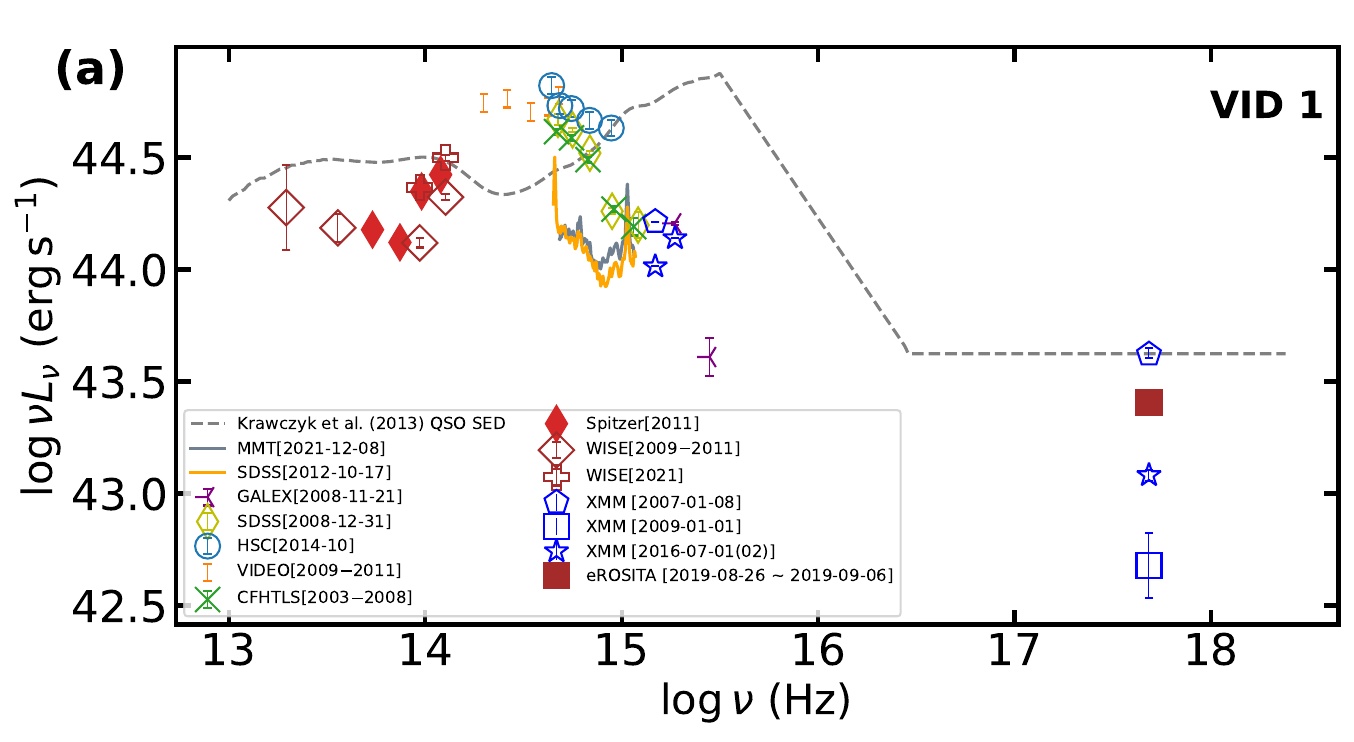}
      
     \vspace{0.2cm}
     \includegraphics[clip,trim=0 1.1cm 0 0.9cm, width=1.0\columnwidth]{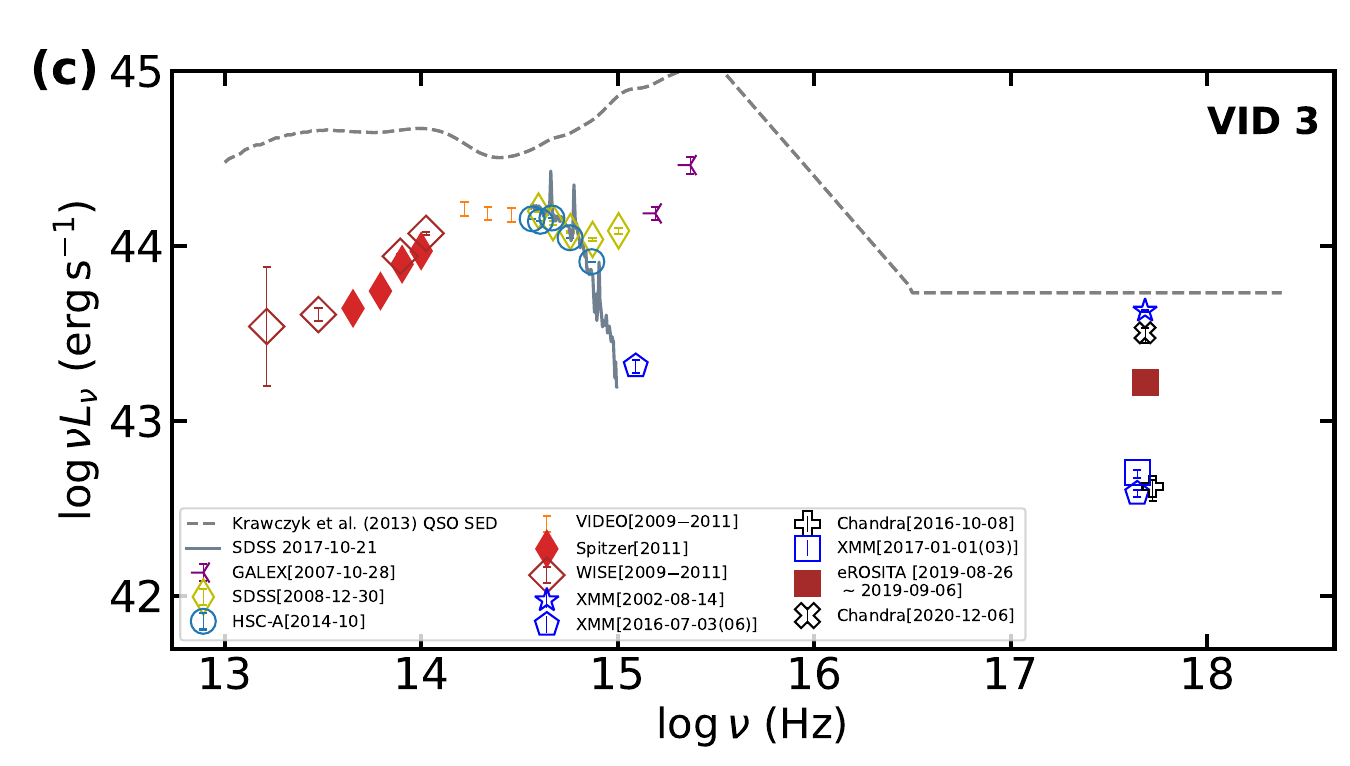}
     
    \includegraphics[clip,trim=0 1.1cm 0 0cm, width=1.0\columnwidth]{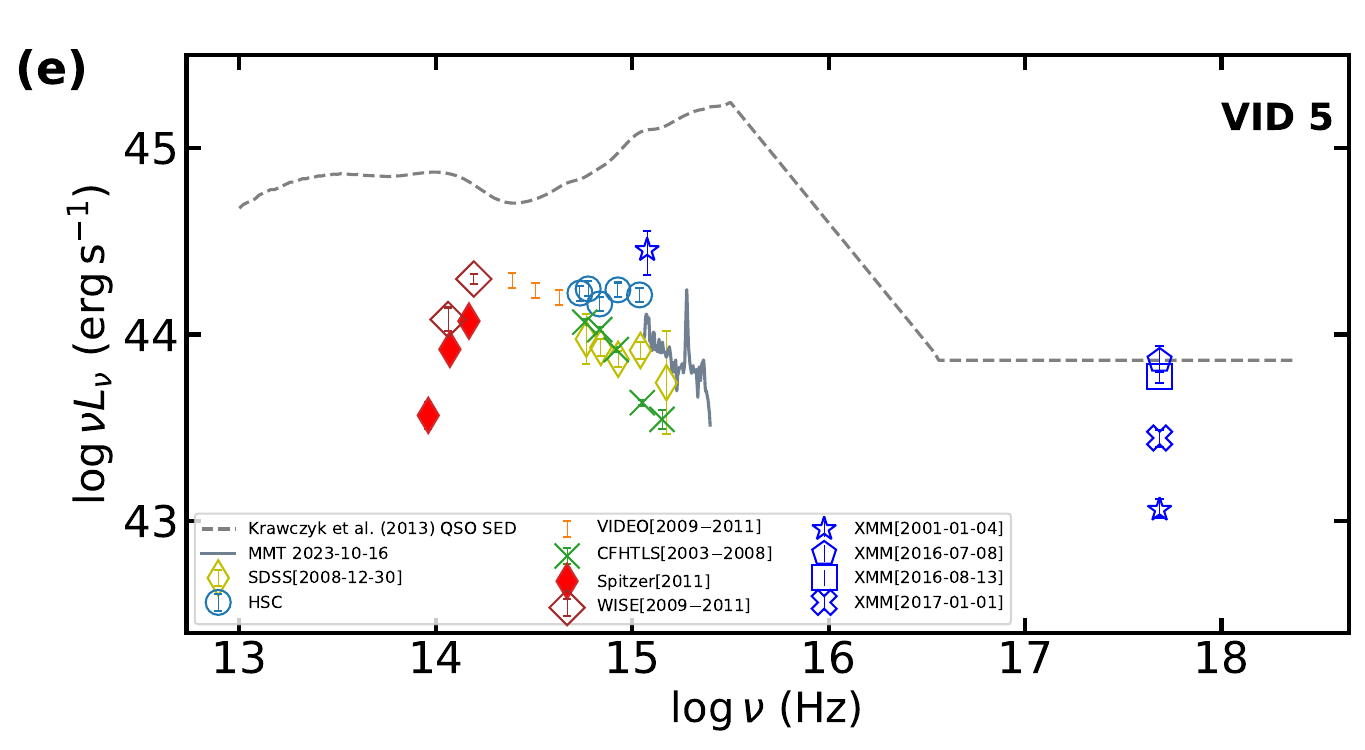}

     \includegraphics[clip,trim=0 0cm 0 0.4cm, width=1.0\columnwidth]{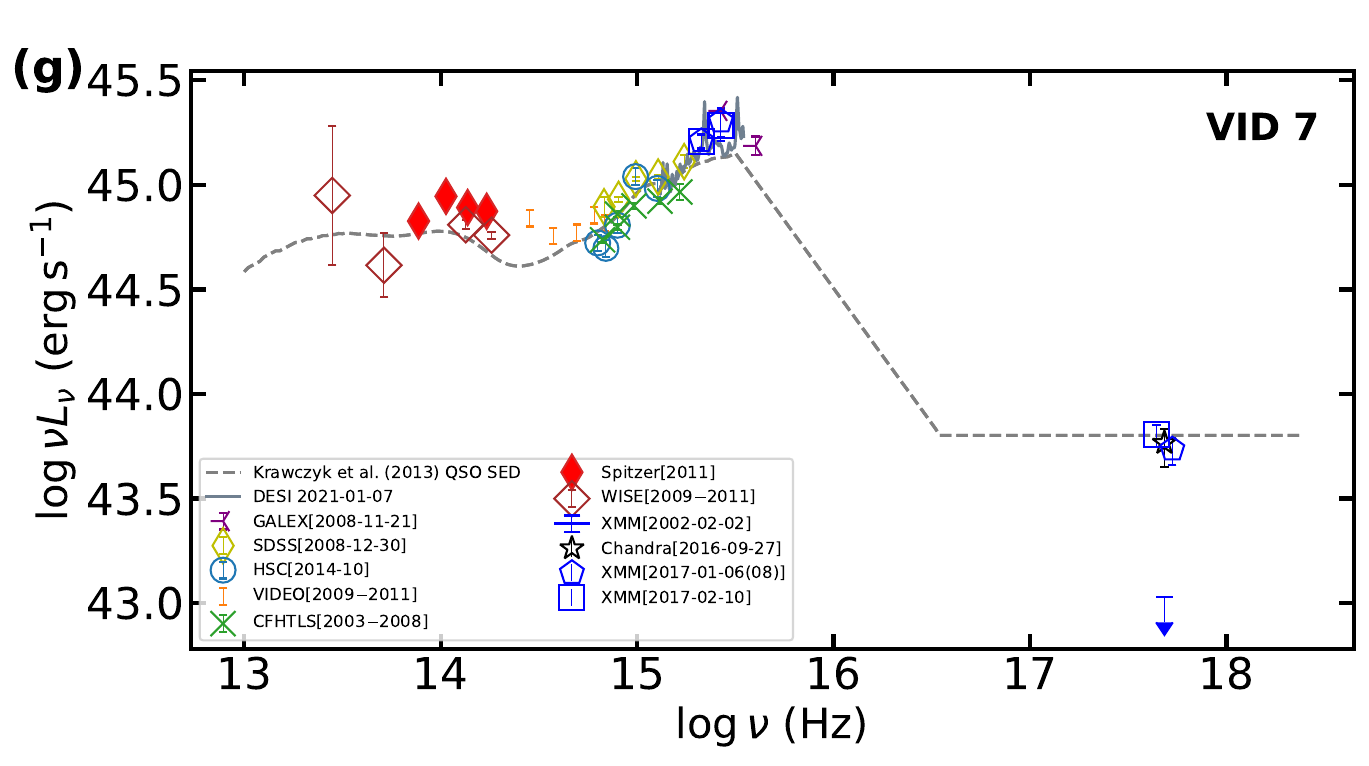}

\end{minipage}
\begin{minipage}{0.98\columnwidth}
      \includegraphics[clip,trim=0 1.1cm 0 0.7cm, width=1.0\columnwidth]{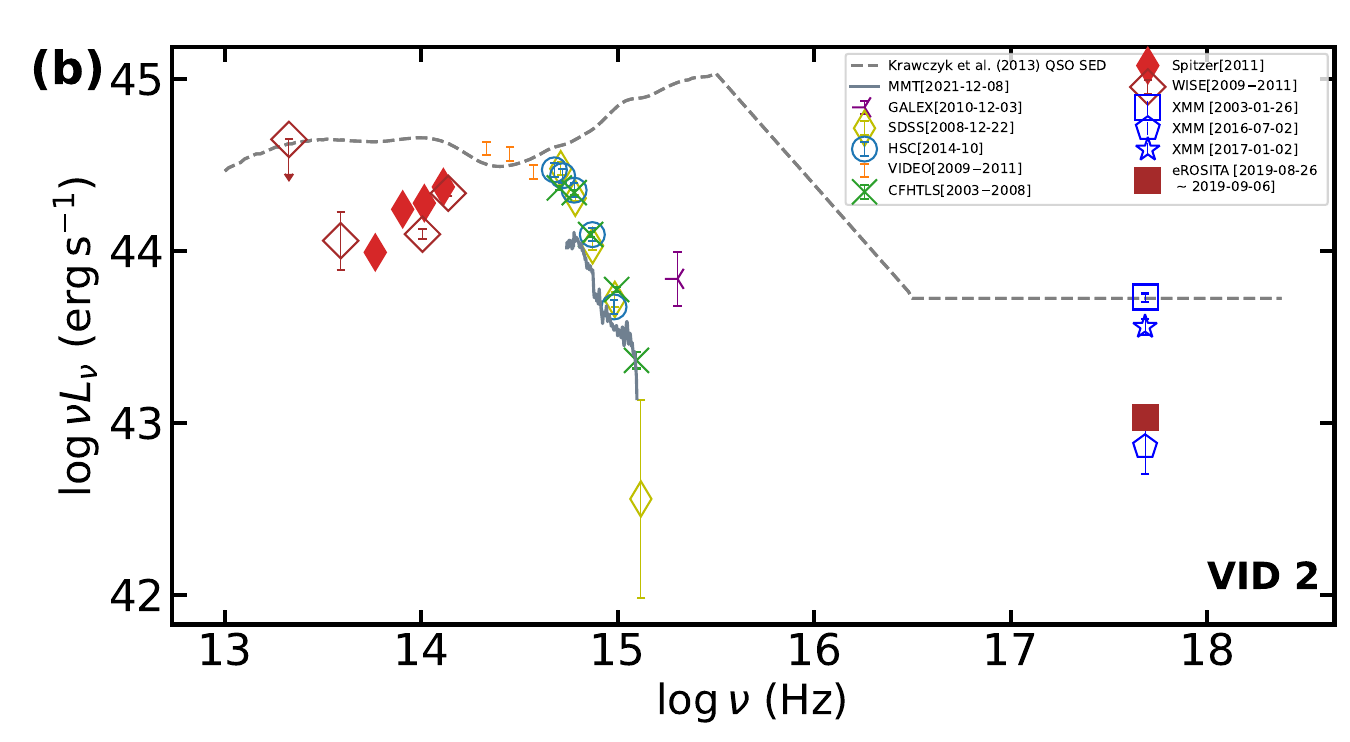}

       \includegraphics[clip,trim=0 1.1cm 0 0cm, width=1.0\columnwidth]{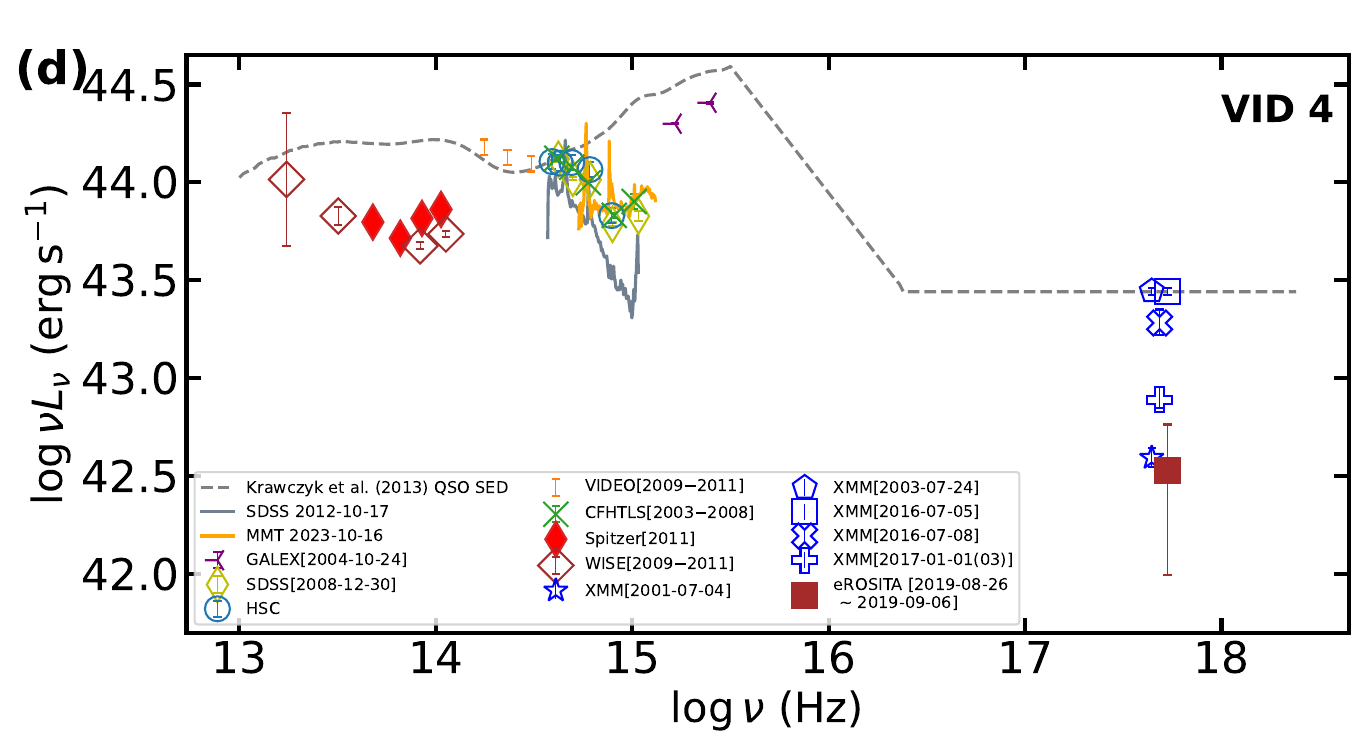}

      \includegraphics[clip,trim=0 1.1cm 0 0cm, width=1.00\columnwidth]{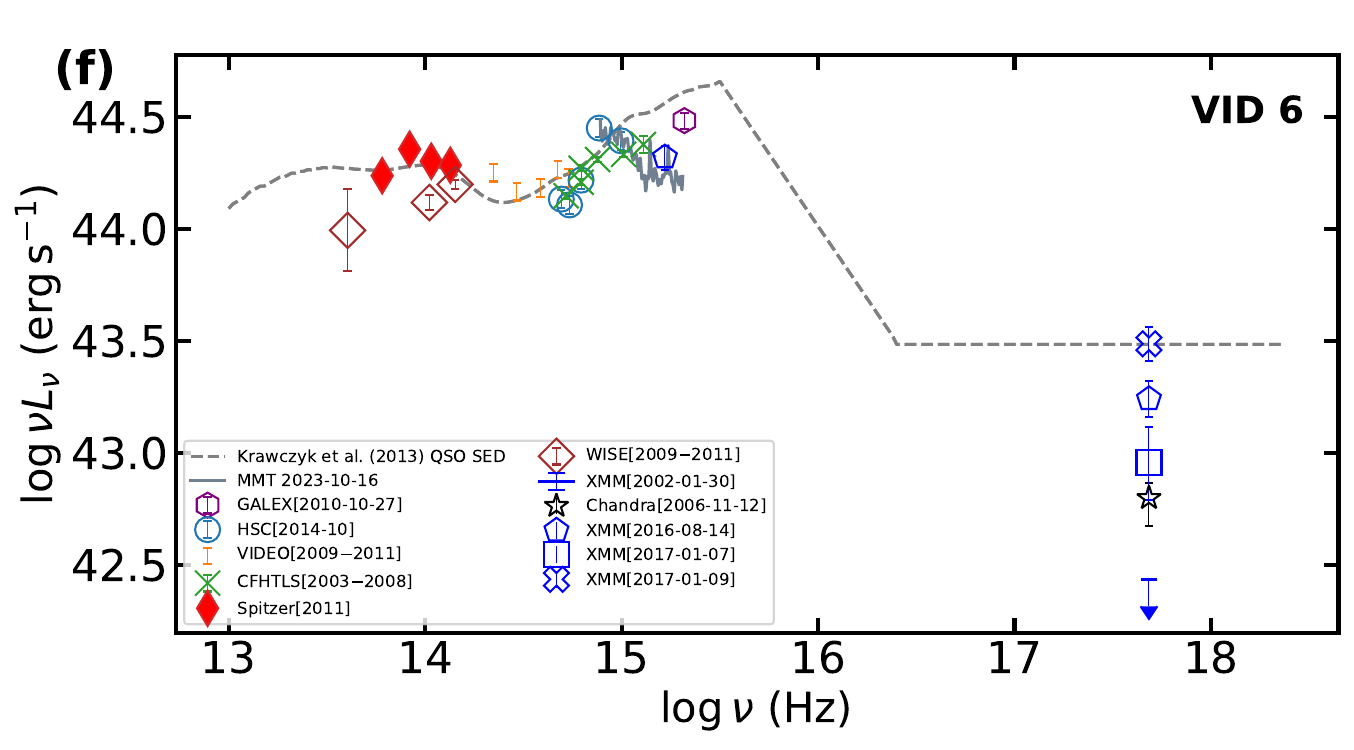}

     \includegraphics[clip,trim=0 0cm 0 0cm, width=1.0\columnwidth]{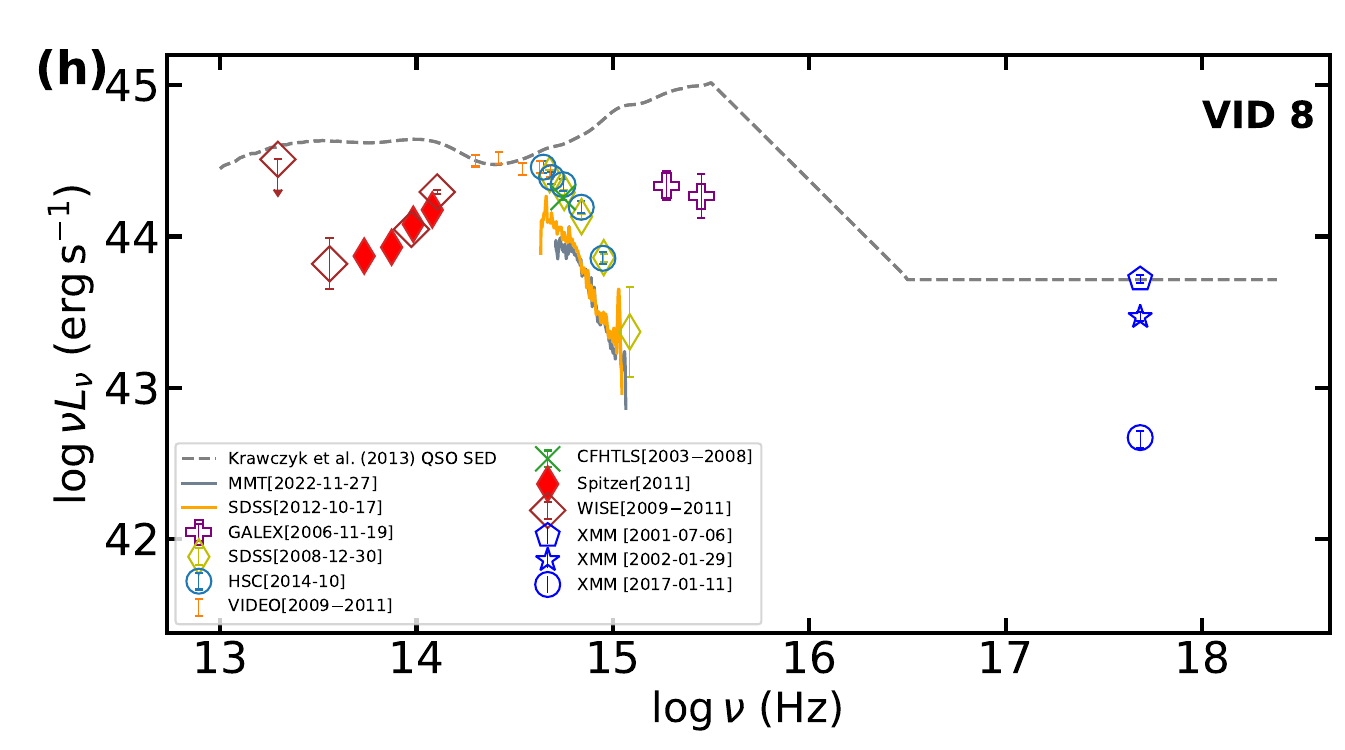}

\end{minipage}
\caption{IR-to-X-ray SED of VIDs 1--8. The available \xmm OM measurements are also shown in the plot. The \xmm and Chandra 2 keV luminosities are derived from the best-fit simple power-law models (Section \ref{3subsec:xray_spec}). The eROSITA 2 keV luminosities are estimated from the 0.6--2.3 keV flux (Section \ref{subsec:X-ray data}). Some 2 keV luminosity data points are displaced slightly along the X-axis for display purposes. For each source, the gray dashed curve shows the mean quasar SED from \cite{2013ApJS..206....4K} normalized to the highest-state 2 keV luminosities; the \hbox{X-ray} component is a $\Gamma$ = 2 \hbox{power-law} continuum with the 2 keV luminosity determined from the \cite{2006AJ....131.2826S} $\alpha_{\rm OX} \textrm{--} L_{\rm 2500 \AA}$ relation.}
\label{fig:all_SED}
\end{figure*}

\subsection{Spectral Energy Distributions}
\label{subsec:sed}
We adopt the IR-to-UV SEDs of the eight AGNs from \citet{2022ApJS..262...15Z}. The IR--UV photometric data in \citet{2022ApJS..262...15Z} are collected from the Spitzer Data Fusion catalog \citep[][]{2015fers.confE..27V}, the Wide-field Infrared Survey Explorer \citep[WISE;][]{2010AJ....140.1868W}, the VISTA Deep Extragalactic Observations survey \citep[VIDEO;][]{2013MNRAS.428.1281J}, SDSS \citep[][]{2000AJ....120.1579Y}, HSC \citep{2022PASJ...74..247A}, Canada-France-Hawaii Telescope Legacy Survey \citep[CFHTLS;][]{2012yCat.2317....0H}, and GALEX \citep{2005ApJ...619L...1M} catalogs. All the SED data have been corrected for the Galactic extinction using the $E_{B-V}$ value obtained from \cite{2011ApJ...737..103S} and the Milky Way extinction law ($R_V = 3.1$) from \citet{2019ApJ...886..108F}. The available \xmm OM measurements are also added to the SED plot. For each AGN's SED, we add the \xmm and Chandra 2 keV luminosities determined from the best-fit results. For the eROSITA observations, we add the 2 keV luminosity estimation. For comparison, we include the mean quasar SED in \cite{2013ApJS..206....4K}. Due to the lack of simultaneous optical--UV and X-ray observations, we simply normalize the mean SED to the \hbox{highest-state} 2 keV luminosity of each source. The \hbox{X-ray} component of the mean SED is modified to reflect the \cite{2006AJ....131.2826S} $\alpha_{\rm OX} \textrm{--} L_{\rm 2500 \AA}$ relation. The SEDs of all the sources are displayed in Figure \ref{fig:all_SED}. We also summarize the multi-state SED shape of each source in Table \ref{tab:properties_summary}.

The SEDs of VID 6 and VID 7 are broadly consistent with the mean SED template, and the deviation in the UV band of VID 6 indicates some degree of extinction. The other six sources' SEDs show some host-galaxy contamination in the IR--optical part. Thus, when considering the IR--UV variability of these AGNs, the influence of host contamination should be taken into account. For VID 1 and VIDs 3--5, their optical--UV SEDs have large scatter, indicating significant variability, which is consistent with their light curves (see Section \ref{subsec:lc}). For VID 3 and VID 4, their SEDs considering the SDSS \hbox{$u$- and} $g$-band and GALEX measurements appear to have an AGN shape, but the luminosities are lower than expected by the templates. Their SDSS and GALEX measurements correspond to relatively high states. The HSC and OM measurements of VID 3 and the SDSS spectrum of VID 4 correspond to their low states, in which the AGN emission strength is low and the SEDs are dominated by the host galaxies. For VID 2, VID 5, and VID 8, their optical--UV part SED is lower than that expected from the high-state \hbox{X-ray} luminosity.

The UV band SEDs of some sources (e.g., VID 2, VID 4, and VID 8) are brighter than the extrapolation of their optical SEDs, which may be due to the \hbox{non-simultaneous} observation of different bands. For example, the UV SED of VID 8 may capture its high state, while the optical SED corresponds to a low-state SED and is dominated by the host galaxy. Such a condition can be well illustrated by the SED of VID 3. Its SDSS $u$- and $g$-band and GALEX measurements lead to an SED that is consistent with an AGN shape, but if we did not have these SDSS measurements, its optical--UV SED would just show a similar behavior to VID 8.

\begin{table*}
\caption{Properties summary of each sources}
\label{tab:properties_summary}
\hspace*{-1.6cm}
\begin{tabular}{lccccccc}
\hline
\hline
Target&Max &\hbox{X-ray}& Optical  &multi-state &Max &Max  &Possible\\
&$A_{\rm var,2keV}$$^a$&absorption&Spectra$^b$&SED shape&$\Delta g$&$\Delta W1$&origin$^c$\\
\hline
VID 1&$8.8^{+3.5}_{-2.5}$ (1.4 yr) &No&1 1 0&significant host galaxy contribution&$0.9$&$0.7$&CS \\
&&&&broadly consistent with template in high state (HSC)&&\\
\hline
VID 2&$12.0^{+2.6}_{-2.1}$ (8.6 yr)&Yes&0 0 0&dominated by host galaxy&$0.3$&$0.3$&CO    \\
&&&&IR--UV SED lower than expected from \hbox{X-ray}&&\\
\hline
VID 3&$11.1^{+0.5}_{-0.5}$ (9.6 yr)&No&1 0 0&broadly consistent with template in high state&$0.6$&$0.5$&CS      \\
&&&&\hbox{X-ray} normal in the low state&&\\
\hline
VID 4&$7.1^{+0.9}_{-0.8}$ (1.6 yr)&No&1 1 1&broadly consistent with template in high state&$1.2$&$0.8$&CS      \\
\hline
VID 5&$6.3^{+1.4}_{-1.1}$ (8.4 yr)&No&1 1 --&significant host galaxy contribution&$-$&$0.8$&CS     \\
&&&&IR--UV SED lower than expected from \hbox{X-ray}&&\\
\hline
VID 6&$>11.3$ (9.3 yr)&Unclear&1 1 --&broadly consistent with template in high state&$0.2$&$0.4$&CO    \\

\hline
VID 7&$>6.0$ (7.3 yr)&Unclear&1 1 --&consistent with template in high state&$0.5$&$-$&CO       \\
\hline
VID 8&$11.2^{+2.0}_{-1.3}$ (10.7 yr)&No&1 0 0&dominated by host galaxy&$0.4$&$0.3$ &CS \\
&&&&IR--UV SED lower than expected from \hbox{X-ray}&&\\
\hline
\end{tabular}
\tablenotetext{a}{The rest-frame timescale over which the maximum X-ray variability amplitude is achieved is shown in parentheses.}
\tablenotetext{b}{{A three-digit flag of the characteristics of the spectra of the target. From left to right: showing AGN broad line, continuum dominated by AGN, showing significant variability. For each digit, the number ‘1’ means the source has the corresponding characteristic, and the number ‘0’ means the opposite. If the characteristic cannot be determined given the available data, the expression is ‘--’.}}
\tablenotetext{c}{``CS'' represents changing accretion state, ``CO'' represents changing obscuration.}
\end{table*}

\begin{table*}
\caption{\hbox{X-ray} Spectral Fitting Results of the Partial Covering Model}
\label{tab:pcf_fitting_results}
\hspace*{2.3cm}
\begin{tabular}{lcccc}
\hline
\hline
Target&Observation Start Date& $\log N_{\rm H}$ &$f_{\rm cover}$&W-stat/dof \\
(1)&(2) &(3)  &(4) &(5)  \\
\hline
VID 2&2016 Jul 02+2016 Jul 03&$24.5_{-0.3}$&$0.88_{-0.02}^{+0.02}$&437.9/401 \\
&2017 Jan 02&$22.0_{-0.2}^{+0.3}$&$0.54_{-0.11}^{+0.12}$&  339.0/314 \\
\hline
VID 6&2006 Nov 12&$ 24.3_{-0.5}$&$0.81_{-0.04}^{+0.04}$&22.4/29\\
&2016 Aug 14&$24.2_{-1.3}$&$0.34_{-0.07}^{+0.07}$&  101.2/151 \\
&2017 Jan 07&$23.9_{-0.7}$&$0.54_{-0.07}^{+0.08}$&  78.8/87 \\

\hline
\end{tabular}
\tablecomments{
Col. (1): object name.
Col. (2): observation start date.
Col. (3): log of column density in the units of $\rm cm^{-2}$.
Col. (4): covering factor of the \hbox{X-ray} absorber. 
Col. (5): W-stat value divided by the degrees of freedom.
}

\end{table*}

\section{Discussion}
\label{sec:Discussion}

In this section, we discuss the physical origin of the \hbox{X-ray} variability exhibited by our sources, taking into account their observed properties as described above. As introduced in Section \ref{sec:intro}, extreme \hbox{X-ray} variability of AGNs is commonly attributed to changing accretion state, changing obscuration, or TDEs.

The available data suggest that TDEs are not the primary explanation for our sources. Firstly, TDEs in pre-existing AGNs are rarely detected with current selection methods, though they may occur at a higher rate than TDEs around inactive SMBHs \citep[e.g.,][]{2024arXiv240518500K,2024arXiv240612096W}. Secondly, the multiwavelength features of our sources contradict to typical TDEs. For example, VIDs 1--3 show an \hbox{X-ray} re-brightening trend. VID 4 shows \hbox{X-ray} high states in two observations with an interval of $\sim 10$ years. VID 5 shows rapid \hbox{X-ray} flux dimming. VIDs 4--5 and VIDs 6--8 show significant Mg II emission lines, and they do not show apparent He II emission lines (Figure \ref{fig:all_opt_spec}). Most of our sources also have \hbox{X-ray} \hbox{power-law} indices typical of AGNs (see Table \ref{tab:xray_variability_X_spec_info}). However, we note that there is a strong diversity in the X-ray and optical spectroscopic behavior of TDEs. Thus we cannot completely rule out the possibility that some of these extreme X-ray variability events are due to unusual TDEs, such as those involving supermassive black hole binaries or repeated tidal stripping \citep[e.g.,][]{2020NatCo..11.5876S,2023ApJ...942L..33W}.

In the following, we focus on the changing accretion state and the changing obscuration scenarios for each source. We provide detailed information about each source in individual subsections in the Appendix to facilitate easy reference for individual object traits. Criteria for changes of accretion states include strong and coordinated multiwavelength variability (e.g., $\Delta g > 0.5$ and $\Delta W1 > 0.5$) and a lack of absorption in the X-ray spectra. In contrast, changing obscuration AGNs are expected to show small optical-IR variability along with signs of absorption in the X-ray spectra.


\subsection{Accretion Rate Change Induced Variability}
\label{subsec:accretion_rate_change}

\begin{figure*}
 \includegraphics[width=0.99\textwidth]{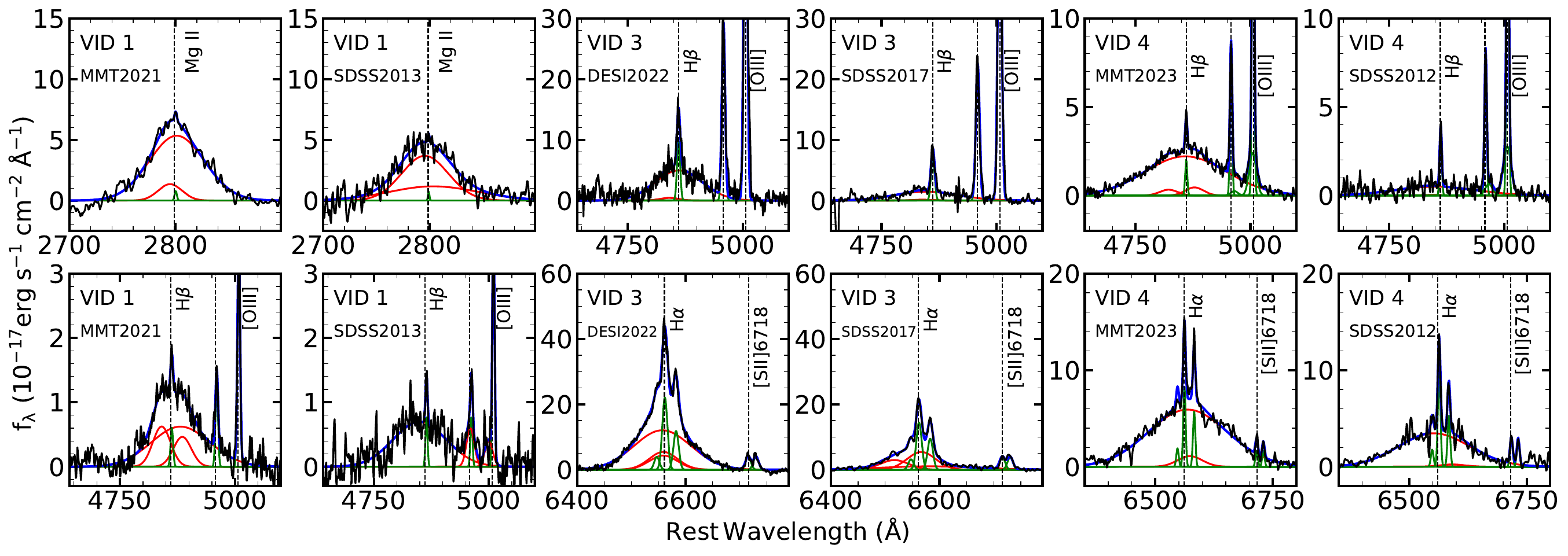}
 \centering
 \caption{Zoomed-in views of the multi-epoch continuum-subtracted optical spectra of VID 1, VID 3, and VID 4 in the broad emission line regions, fitted with broad (red curves) and/or narrow (green curves) Gaussian profiles. We show two spectra for each source (two columns), and each column displays two broad lines in one spectrum.}
 \label{fig:zoom_lines}
\end{figure*}

Considering the optical and \hbox{X-ray} spectra, multiwavelength light curves, and SED characteristics described in the above sections, we suggest that the extreme \hbox{X-ray} variability of VID 1, VIDs 3--5, and VID 8 is induced by changing accretion state. For VID 1 and VIDs 3–4, we fit their multi-epoch optical spectra using the method described in Section \ref{subsubsec:optical_fit}. The fitting results are plotted in Figure \ref{fig:zoom_lines} to illustrate the variability of their emission lines.

For VID 4, it exhibits a changing-look behavior in its multi-epoch optical spectra (see Section \ref{subsec:optical_spec}). The SDSS spectrum observed in 2012 shows no broad H$\beta$ component, whereas the MMT spectrum observed in 2023 reveals the emergence of a significant broad H$\beta$ line. Considering also its nominal \hbox{X-ray} photon index, its extreme \hbox{X-ray} variability can be attributed to the changing accretion state scenario. 


For VID 1, its significant and coordinated IR--\hbox{X-ray} variability suggests changing accretion state. Although its multi-epoch spectra do not show significant variability (see Figure \ref{fig:all_opt_spec} and Figure \ref{fig:zoom_lines}), the $g$- and $r$-band variability of the ZTF light curves of VID 1 are $\sim 0.9$ mag and $\sim 0.5$ mag, respectively. Such variability amplitudes roughly meet the selection criteria of changing-state AGN candidates \citep[$|\Delta g|>1$ mag and $|\Delta r|>0.5$ mag, as described in][]{2019ApJ...874....8M}, considering that it has some host-galaxy contamination as shown in its SED (Figure \ref{fig:all_SED}).

The IR--\hbox{X-ray} light curves of VID 3 also show coordinated variability but with relatively small amplitude in the optical and IR bands. VID 3 is significantly variable in its broad emission lines as shown in Figure \ref{fig:zoom_lines}. The equivalent widths of the broad components of its H$\alpha$ and H$\beta$ lines increase from $276.5 \pm 16.9 ~\AA$ and $53.0 \pm 5.4~\AA$ to $546.3 \pm 20.2~\AA$ and $118.1 \pm 14.4~\AA$, respectively. Since changing obscuration should not result in coordinated variability in the optical and IR bands, we suggest that its extreme \hbox{X-ray} variability is caused by the changing accretion state scenario. The spectra and SED of VID 3 show that it has significant host-galaxy contamination in the optical and IR bands, which may dilute the variability. Both VID 1 and VID 3 show typical \hbox{X-ray} photon indices in their high and low states (see Table \ref{tab:xray_variability_X_spec_info}), which also support the changing accretion state scenario. The limited variability of their multi-epoch spectra is likely due to the fact that we did not capture their high states in the three spectroscopic observations, given the relatively low optical fluxes in their light curves around the time when the spectra were taken (Figure \ref{fig:alllc1}).

VID 8 shows no significant variability in its \hbox{multi-epoch} optical spectra or optical--IR light curves. However, it shows no H$\beta$ line and a relatively weak H$\alpha$ line, while the Mg II line is strong and persists in the two spectroscopic observations (See Section \ref{subsec:optical_spec}). Such Mg II emitters can be interpreted as recently \hbox{turned-off} AGNs, where the Mg II line is not as responsive to the continuum variability as the H$\alpha$/H$\beta$ lines due to the different excitation mechanisms, radiative transfer effects, and the larger average distance of Mg II gas from the ionizing source \citep{2020ApJ...888...58G}. Thus, the Mg II line can sustain after the H$\alpha$/H$\beta$ lines have disappeared, as has been observed in some \hbox{changing-look} AGNs that show visible broad Mg II lines even in the dim states \citep[e.g.,][]{2019ApJ...874....8M,2020MNRAS.493.5773Y}. This explanation is consistent with the \hbox{X-ray} dimming trend of VID 8 with no signs of \hbox{X-ray} absorption, which indicates a decrease in the luminosity of the accretion-disk continuum induced by changing accretion state. Between the MMT and SDSS observations, the broad Mg II line likely sustained for $\sim 10$ years, indicating that the luminosity of the continuum did not change much during this period, which is also consistent with the fact that its optical light curves show mild variability in the last ten years.

For VID 5, due to the lack of multiwavelength light curves, the origin of its extreme \hbox{X-ray} variability cannot be well constrained. The scatter in its SED shows that it has significant variability in the optical--UV band. Its IR light curves also show a large variability amplitude ($\Delta W1=0.8$), which is comparable to VID 1 and VID 3. The low-state \hbox{X-ray} spectrum of VID 5 also shows no signs of absorption. Therefore, its extreme \hbox{X-ray} variability is probably induced by changing accretion state, but we still cannot rule out the changing obscuration scenario for this source.

\begin{figure}
 \includegraphics[width=0.45\textwidth]{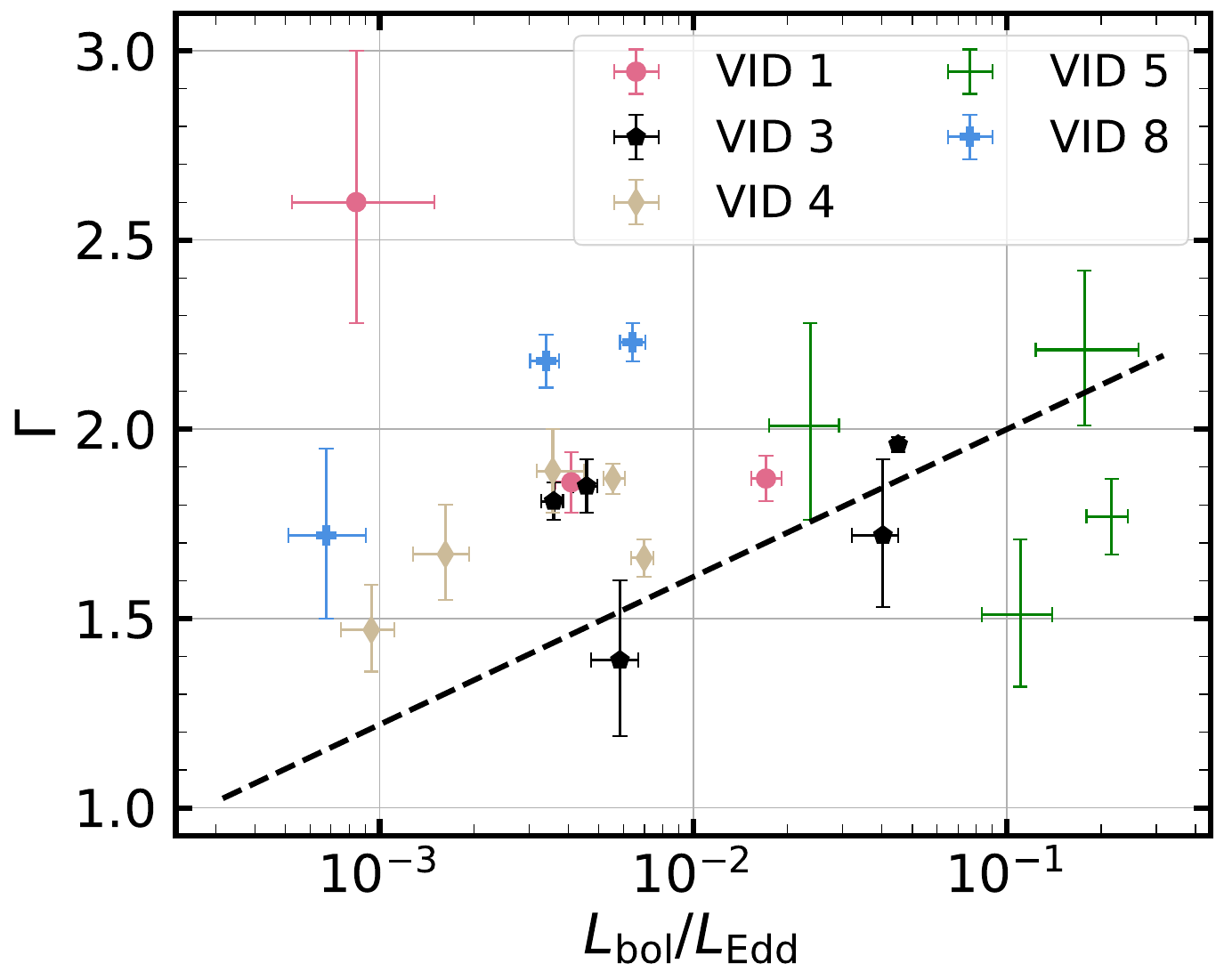}
 \centering
 \caption{Distribution of photon index and Eddington ratio for VIDs 1, 3, 4, 5, 8. For comparison, we also show a relation found by \citet{2013MNRAS.433.2485B} as black dashed line.}
 \label{fig:gamma_LX}
\end{figure}

The AGN X-ray power-law photon index is expected to evolve with the accretion rate, displaying a ``softer-when-brighter'' behavior \citep[e.g.,][]{2006ApJ...646L..29S,2009MNRAS.399.1597S}. The underlying mechanism remains poorly understood. One possibility is that the corona cools as the accretion rate increases \citep[e.g.,][]{2017MNRAS.468.3489K,2018MNRAS.480.1819R}, leading to a softer spectrum \citep[e.g.,][]{2007MNRAS.381.1235V}. We show the distribution of the photon indices\footnote{Hard X-ray ($>$2 keV) photon indices should be used in the study of the $\Gamma$--$L_{\rm bol}/L_{\rm Edd}$ relation. We verified that the hard X-ray photon indices do not differ significantly from the 0.3--10 keV photon indices listed in Table \ref{tab:xray_variability_X_spec_info}.} and $L_{\rm bol}/L_{\rm Edd}$ for VIDs 1, 3, 4, 5, 8 in Figure \ref{fig:gamma_LX}, along with a relation of $\Gamma = 0.39 \log (L_{\rm bol}/L_{\rm Edd}) + 2.39$ as found by \citet{2013MNRAS.433.2485B} for comparison. The $L_{\rm bol}$ value of each X-ray epoch is estimated from $L_{\rm X}$ using the same method as described in Section \ref{subsec:BHmass_estimation}. Most of these five sources do appear to follow a softer-when-brighter trend in general, considering the measurement uncertainties. The clear violation appears to be VID 1. It has low Eddington ratios ($\lesssim 0.01$) in general, and a negative correlation between $\Gamma$ and $L_{\rm bol}/L_{\rm Edd}$ for low luminosity AGNs has been reported before \citep[e.g.,][]{2009MNRAS.399..349G}, likely arising from accretion disks (e.g., advection-dominated accretion flows) different from those in luminous AGNs. Overall, the $\Gamma$ versus $L_{\rm bol}/L_{\rm Edd}$ distribution for these objects is consistent with the changing accretion state scenario.

\subsection{Changing Obscuration Scenario}
\label{subsec:change_of_obscuration}

Given the limited variability of the optical spectra and mild long-term optical/IR variability of VID 2, VID 6, and VID 7, their strong \hbox{X-ray} variability can be naturally explained by changing obscuration from a \hbox{dust-free} absorber. Although VID 8 also shows similar behavior, the absence of the H$\beta$ line indicates that changing accretion state is a more likely explanation for its \hbox{X-ray} variability as discussed in Section \ref{subsec:accretion_rate_change}. 

The \hbox{X-ray} photon index in the lowest state of VID 2 is $1.20^{+0.18}_{-0.17}$, indicating significant absorption. We added an intrinsic absorption component ({\sc zphabs}) to fit the spectrum. The intrinsic absorption model significantly improves the fitting with $\Delta W/{\rm dof} = 8.2/1$, and the resulting photon index and intrinsic absorption column density are $2.00^{+0.40}_{-0.30}$ and $N_{\rm H} = 6.26_{-3.03}^{+4.00} \times 10^{21} \rm ~cm^{-2}$, respectively. However, correcting for the intrinsic absorption results in an intrinsic \hbox{$f_{\rm 2keV} = 2_{-0.4}^{+0.4}\times10^{-32}$ $\rm erg~cm^{-2}~s^{-1}~Hz^{-1}$}, which is still significantly lower than its high-state 2~keV flux density ($8.8_{-0.4}^{+0.6}\times10^{-32}$ $\rm erg~cm^{-2}~s^{-1}~Hz^{-1}$) observed in 2003 Jan 26. Therefore, this simple intrinsic absorption model cannot explain the \hbox{X-ray} variability of VID 2 self-consistently, and more complex absorption might be present. For VID 6, it shows photon indices of $\sim 3$ in its \hbox{lowest-state} spectra and that cannot be explained by a simple intrinsic absorption model either. The extreme X-ray variability, steep X-ray spectral shapes, and extreme
X-ray weakness in the low states (Figure \ref{fig:all_SED}f) of VID 6 are reminiscent of
the remarkable X-ray properties of PHL 1811 \citep[e.g.,][]{2022ApJ...936...95W} that can be explained with \hbox{Compton-thick} \hbox{partial-covering} absorption \citep[the steep spectra are dominated by a variable leaked component; Figure
6 and Table 3 of ][]{2022ApJ...936...95W}.

We therefore tested a \hbox{partial-covering} absorption model ({\sc phabs*zpcf*zpowerlw}) to fit the spectra of VID 2 and VID 6, fixing the intrinsic continuum (i.e., $\Gamma$ and normalization of {\sc zpowerlw}) to the best-fit power-law model of their highest-state spectra, respectively. The results are shown in Table \ref{tab:pcf_fitting_results}. Four $N_{\rm H}$ values do not have upper errors, as the spectrum is dominated by the leaked component and is not sensitive to $N_{\rm H}$. Overall, the fitting results are acceptable considering the low $W$-stat/dof values and residuals. The \hbox{partial-covering} absorption scenario provides \hbox{self-consistent} explanations of the \hbox{X-ray} variability of VID 2 and VID 6, which is mainly induced by the varying column density and leaked fraction (\hbox{partial-covering} fraction) of the absorber. These absorbers are likely \hbox{dust-free} and do not affect the optical emission significantly. 

For VID 7, it was not detected in the \hbox{lowest-state}, and we cannot constrain its low-state \hbox{X-ray} spectral properties. But the lack of optical variability and its relatively high accretion rate ($L/L_{\rm Edd} = 0.132$) suggest the changing obscuration scenario, since changing-look AGNs tend to be found at low Eddington ratio systems \citep[e.g.,][]{2019ApJ...874....8M,2022ApJ...933..180G}. The change of obscuration may also be responsible for its significant FUV variability (Figure \ref{fig:alllc2}) if the X-ray absorber is dusty gas. For VID 7, we cannot completely rule out the changing accretion state scenario, since the state change could occur during 2003--2010, where we lack optical monitoring.

\subsection{Relative \hbox{X-ray} Emission Intensity}

\begin{figure}
 \includegraphics[width=0.47\textwidth]{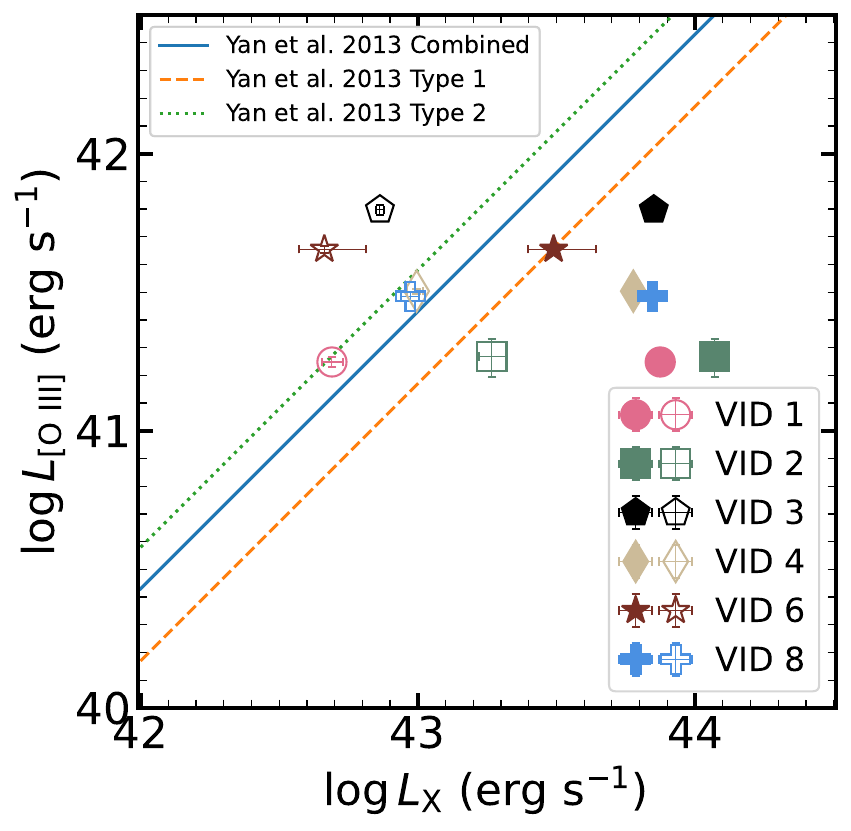}
 \centering
 \caption{[O \uppercase\expandafter{\romannumeral3}] line luminosity vs. 2--10 keV \hbox{X-ray} luminosity for the six extremely \hbox{X-ray} variable AGNs that have [O \uppercase\expandafter{\romannumeral3}] line coverage. VID 5 and VID 7 do not have [O III] spectral coverage since they are at higher redshifts. The filled and open symbols correspond to the high-state and low-state $L_{\rm X}$ measurements, respectively.} The solid line, dashed line, and dotted line indicate the median $\log L_{\rm X}/L_{\rm [O~\uppercase\expandafter{\romannumeral3}]}$ ratios found by \citet{2011ApJ...728...38Y} for their combined sample, Type I sample, and Type II sample, respectively.

 \label{fig:LOIII_LX}
\end{figure}

Although the IR--UV SED data of our sources are not simultaneous, by comparing the SEDs and the templates scaled to the high-state \hbox{X-ray} luminosity in Figure \ref{fig:all_SED}, we can roughly examine the relative \hbox{X-ray} emission intensity of each source. We note that the effect of the variability of some sources cannot be ignored (e.g., VID 1 and VID 4; see their light curves in Figure \ref{fig:alllc1} and Figure \ref{fig:alllc2}). The [O \uppercase\expandafter{\romannumeral3}] line can also be used as a rough indicator of the AGN activity strength on long timescales \citep[$\sim$ hundreds of years; e.g.,][]{2009ApJ...698..623D,2010ApJ...722..212T,2017MNRAS.468.1433P}. For the six sources with [O \uppercase\expandafter{\romannumeral3}] line coverage, we measure their [O \uppercase\expandafter{\romannumeral3}] luminosities ($L_{\rm [O~\uppercase\expandafter{\romannumeral3}]}$) using the { \sc GLEAM} package \citep{2021AJ....161..158S} and compare them with their high-state and low-state \hbox{X-ray} luminosities. For VIDs 1--2, VID 4, and VID 8, we measure their $L_{\rm [O~\uppercase\expandafter{\romannumeral3}]}$ using the MMT spectra. For VID 3 and VID 6, we measure their $L_{\rm [O~\uppercase\expandafter{\romannumeral3}]}$ using the SDSS spectra. For sources with multi-epoch spectra, measuring the $L_{\rm [O~\uppercase\expandafter{\romannumeral3}]}$ using a different spectrum does not significantly affect the results. Figure \ref{fig:LOIII_LX} shows the $L_{\rm [O~\uppercase\expandafter{\romannumeral3}]} \textrm{--} L_{\rm X}$ plot of these sources, along with the median $\log L_{\rm X}/L_{\rm [O~\uppercase\expandafter{\romannumeral3}]}$ ratios from \citet{2011ApJ...728...38Y} for comparison.

The IR--UV SEDs of VID 6 and VID 7 are broadly consistent with that expected from their high-state \hbox{X-ray} luminosity, and the high-state $\log L_{\rm X}/L_{\rm [O~\uppercase\expandafter{\romannumeral3}]}$ ratio of VID 6 is also consistent with that found by the \citet{2011ApJ...728...38Y} \hbox{Type I} sample. These results are consistent with their interpretation as changing-obscuration AGNs. Their intrinsic X-ray emission strength (revealed in high-state) is consistent with their IR--UV SEDs and $L_{\rm [O~\uppercase\expandafter{\romannumeral3}]}$, while their low-state X-ray weakness is caused by obscuration. Such obscuration is in small scale and does not significantly affect the IR--optical emission.

For VID 2, VID 5, and VID 8, their IR--UV SEDs are dominated by their host galaxies and significantly lower than that expected from the high-state \hbox{X-ray} luminosity. The high-state $L_{\rm X}$ values of VIDs 1--4 and VID 8 are also higher than those expected from their $L_{\rm [O~\uppercase\expandafter{\romannumeral3}]}$. Such X-ray bright behavior is similar to the \hbox{X-ray} selected but IR-optical faint AGNs reported by previous work \citep[e.g.,][]{2019ApJ...876...50L,2023arXiv231012330L}. Such sources may be explained by dust-deficient AGNs that show relatively weak hot/warm dust emission \citep[e.g.,][]{2010Natur.464..380J,2017ApJ...835..257L}, or their X-ray emission may be boosted by, e.g., an enhanced corona or jets \citep[e.g.,][]{2020MNRAS.496..245Z,2021MNRAS.505.1954Z}. According to the classification of \citet{2023MNRAS.522.3506Z}, which is based on the deep ATLAS and MIGHTEE data, none of our eight sources is a radio-loud AGN. Thus, their \hbox{X-ray} emission should not be boosted by an enhanced corona or jets. VID 2 is likely a dust-deficient Type II AGN, showing weak IR--optical continuum emission. However, the dust-deficient scenario still cannot explain the weak optical emission of VID 5 and VID 8. Their extreme X-ray variability are both attributed to the changing accretion state scenario. One possible explanation for their weak optical emission is that such changing-state AGNs do not follow the $\alpha_{\rm OX} \textrm{--} L_{\rm 2500 \AA}$ relation, and their IR--UV SEDs are weak compared to the X-ray emission in the high state. Another possible explanation is variability effects. The high-state duty cycle of these sources may be low, and they are in the low accretion rate state most of the time. So the high-state \hbox{X-ray} emission we caught shows relatively higher strength compared with their normal intrinsic IR--optical continuum or the long-term averaged $\rm [O~\uppercase\expandafter{\romannumeral3}]$ emission strength. The rapid \hbox{X-ray} flux decline of VID 5 provides additional support for this explanation. Some of the \hbox{X-ray} bright but IR--optical faint AGNs found by \citet{2023arXiv231012330L} may also be explained by this scenario. The low-state $L_{\rm X}$ values of VIDs 1, 4, and 8 are consistent with the expectations of their $L_{\rm [O~\uppercase\expandafter{\romannumeral3}]}$ (Figure \ref{fig:LOIII_LX}), suggesting a scenario where the recent SMBH growth is dominated by the low accretion rate state. For VID 3, the $L_{\rm [O~\uppercase\expandafter{\romannumeral3}]}$ comparison suggests a dominating state that is in between the X-ray high and low states.

\section{Conclusion}
\label{sec:Conclusion}
In this paper, we present a systematic investigation of the extremely \hbox{X-ray} variable AGNs in the \hbox{5.3-square-degree} XMM-SERVS XMM-LSS region, taking advantage of the superb multiwavelength data in this field. We summarize our study as follows:

\begin{enumerate}
\item We selected extreme variable AGN candidates with a criterion of variability amplitude $>$ 10 in terms of 0.2--12.0 keV count rate between any two observations using measurements provided by the 4XMM catalog and upper limits from the RapidXMM server. Eight sources spectroscopically confirmed as AGNs/galaxies are identified (VIDs 1--8). See Section \ref{subsec:sample}.

\item We reduced and analyzed the archival \hbox{XMM-Newton} and Chandra data of these eight AGNs. They show extreme \hbox{X-ray} variability with $f_{\rm 2keV}$ amplitudes ranging from 6 to 12. The \hbox{lowest-state} X-ray spectrum of VID 2 shows small photon index ($1.20^{+0.18}_{-0.17}$) and suggests absorption, while the X-ray spectra of the other sources do not show apparent absorption signatures. See Section \ref{subsec:X-ray data} and Section \ref{3subsec:xray_spec}.

\item We comprehensively analyzed the optical and \hbox{X-ray} spectra, multiwavelength light curves, and SED characteristics of these AGNs, including the newly obtained optical spectra observed by MMT, to assess the origin of their extreme \hbox{X-ray} variability. Some of the sources show coordinated variability in their IR--UV light curves, while some show mild variability in these bands. See Section \ref{subsec:optical_spec}--\ref{subsec:sed}.

\item Considering the significant and coordinated multiwavelength variability and absence of \hbox{X-ray} absorption in low-state spectra, we suggest the extreme \hbox{X-ray} variability of VID 1, VIDs 3--5, and VID 8 to be induced by changing accretion state. Among them, VID 4 has been confirmed by a new MMT spectrum to be a \hbox{changing-look} AGN. However, for VID 5, we still cannot rule out the changing obscuration scenario due to the lack of multiwavelength light curves. See Section \ref{subsec:accretion_rate_change}.

\item The extreme \hbox{X-ray} variability of VID 2, VID 6, and VID 7 is interpreted with the changing obscuration scenario, given the little variability of their optical spectra and mild long-term optical/IR variability. The low-state \hbox{X-ray} photon index of VID 2 indicates significant absorption, while VID 6 and VID 7 were not detected in their low state and lack constraints. The absorbers of these sources are likely the \hbox{dust-free} gas associated with clumpy accretion-disk winds, having variable column density and covering factor along the line of sight. See Section \ref{subsec:change_of_obscuration}.
\end{enumerate}

Our results demonstrate that large area \hbox{X-ray} surveys with multi-epoch observations are essential for finding a representative sample of extreme \hbox{X-ray} variable AGNs. Future investigation of a larger sample of such sources should be able to constrain the fraction of different origins and improve our understanding of AGN \hbox{X-ray} variability. Long-term \hbox{X-ray} and multiwavelength monitoring of such sources with short cadence will provide constraints on the duty cycle of the changing accretion state and changing obscuration scenarios. W-CDF-S (4.9 $\rm deg^2$) and ELAIS-S1 (3.4 $\rm deg^2$) are the other two XMM-SERVS fields with similar area, X-ray depth, and multiwavelength coverage \citep{2021ApJS..256...21N}. We expect to find a comparable number of extreme \hbox{X-ray} variable AGNs in these two fields. Besides, combing the all-sky eROSITA survey \citep[e.g.,][]{2012arXiv1209.3114M,2021A&A...647A...1P} with archival XMM-Newton data will enable us to find more extreme \hbox{X-ray} variable AGNs in a larger area \citep[e.g.,][]{2024MNRAS.528.1264K}. Incorporating the \hbox{X-ray} data with other multiwavelength data provided by large sky surveys such as CSST, Euclid, and LSST will enable us to better clarify their nature \citep{2011arXiv1110.3193L,2019ApJ...873..111I,zhan2021wide}.

\begin{acknowledgments}
We thank Zhiwei Pan, Shengxiu Sun, and Hengxiao Guo for helpful discussions. The MMT spectra reported here were obtained at the MMT Observatory, a joint facility of the Smithsonian Institution and the University of Arizona. We acknowledge support from the National Key R\&D Program of China (2021YFA1600404), the China Manned Space Project (CMS-CSST-2021-A05, CMS-CSST-2021-A06), the National Natural Science Foundation of China (11991053, 11991050, 12225301), and CNSA program D050102. W.N.B. acknowledges support from XMM-Newton grant 80NSSC22K1806, NSF grants AST-2106990 and AST-2407089, and the Penn State Eberly Endowment. This paper employs a list of Chandra datasets, obtained by the Chandra X-ray Observatory, contained in~\dataset[DOI: 10.25574/cdc.314]{https://doi.org/10.25574/cdc.314}.

\end{acknowledgments}

\facilities{XMM-Newton (EPIC and OM), Chandra, MMT:6.5m}

\software{\texttt{astropy} \citep{2013A&A...558A..33A,2018AJ....156..123A,2022ApJ...935..167A}, 
          \texttt{CIAO} \citep{2006SPIE.6270E..1VF},
          \texttt{dustmap} \citep{2018JOSS....3..695G},
          \texttt{gleam} \citep{2021AJ....161..158S},
          \texttt{matplotlib} \citep{Hunter2007}, 
          \texttt{numpy} \citep{Harris2020}, 
          \texttt{QSOFITMORE} \citep{2018ascl.soft09008G,2021zndo...5810042F}, 
          \texttt{SAS} \citep{2004ASPC..314..759G},
          \texttt{specpro}
          \citep{2011PASP..123..638M},
          \texttt{TOPCAT} \citep{2005ASPC..347...29T},
          \texttt{Veusz} \citep{2023ascl.soft07017S},
          \texttt{xspec} \citep{arnaud1996astronomical}
          }

\appendix
\restartappendixnumbering

\section{Details of each individual source}
\subsection{VID 1}
\label{subsec:VID1}

The SDSS survey reported VID 1 as a broad-line AGN at $z = 0.439$. This object has seven \xmm observations, and after combining the spectra observed at similar dates as described in Section \ref{3subsec:xray_spec}, it has three XMM-Newton spectra observed on 2007 Jan 07, 2009 Jan 01, and 2016 Jul 01, respectively. The 2 keV flux light curve, as well as other multi-wavelength light curves of VID 1, are shown in Figure \ref{fig:alllc1}(a). VID 1 dimmed by an amplitude of $8.8^{+3.5}_{-2.5}$ ($15.3^{+8.8}_{-6.6}$) in terms of $f_{\rm 2keV}$ ($L_{\rm X}$) between 2007 Jan 07 and 2009 Jan 01. Then it brightened again in the following XMM-Newton and eROSITA observations, with the $f_{\rm 2keV}$ increasing by a factor of $5.3^{+2.2}_{-1.6}$ between 2009 Jan 01 and 2019 Sept. The three XMM-Newton spectra can all be described well by the simple power-law model. The best-fit \hbox{power-law} photon indices of the 2007 Jan 07 and 2016 Jul 01 spectra are $1.87^{+0.06}_{-0.06}$ and $1.86^{+0.08}_{-0.08}$, respectively, typical of Type I quasars. The effective photon index estimated from the eROSITA observation is $\sim 2.1$. The low state spectrum in 2009 Jan 01 shows a relatively higher \hbox{power-law} photon index ($2.60^{+0.40}_{-0.32}$). Although it is not well constrained due to the small photon count number, such a photon index indicates no absorption.

The MMT, SDSS, and DESI spectra of VID 1 are shown in Figure \ref{fig:all_opt_spec}(a). The MMT and DESI spectra are \hbox{$<$ 30\%} brighter than the SDSS spectrum in the blueward continuum, Mg II line, and H$\beta$ line (Figure \ref{fig:zoom_lines}), but overall the spectra show little variability. The SED of VID 1 shows significant galaxy contamination (see Figure \ref{fig:all_SED}a). The spectra of VID 1 are also shown in Figure \ref{fig:all_SED}(a), and it is significantly lower than the photometric SED, as the fiber only samples the spectrum from the central region dominated by the AGN. VID 1 agreed with the $\alpha_{\rm OX} \textrm{--} L_{\rm 2500 \AA}$ relation well in 2016 Jul 01. The GALEX data points shown in Figure \ref{fig:all_SED}(a) are quasi-simultaneous with the \xmm observation on 2009 Jan 01, and the SED they constructed shows that the X-ray luminosity deviates downward from the $\alpha_{\rm OX} \textrm{--} L_{\rm 2500 \AA}$ relationship. 

The optical and IR light curves of VID 1 (Figure \ref{fig:alllc1}a) show consistent and significant variability, and the bluer optical bands have larger variability amplitude, as they are less contaminated by galaxies. The maximum variability amplitudes of ZTF $g$-band and NEOWISE W1 and W2 band are $\sim 0.7$ mag, $\sim 0.5$ mag, and $\sim 0.7$ mag, respectively. The CRTS images are taken unfiltered to maximize throughput, thus the CRTS light curve suffers large contamination from the galaxy and shows relatively small variability. The non-significant variability of their multi-epoch spectra is likely due to the fact that we did not capture their high state in the three spectroscopic observations, given the relatively low optical fluxes in its light curves when the spectra were observed (Figure \ref{fig:alllc1}).

Considering its large optical--IR variability and lack of X-ray absorption, we conclude that the extreme X-ray variability of VID 1 is likely caused by changing accretion state (see Section \ref{subsec:accretion_rate_change}).


\subsection{VID 2}
\label{subsec:VID2}
VID 2 is a $z = 0.556$ object with typical galaxy spectrum (see Figure \ref{fig:all_opt_spec}b). It has four \xmm observations, and the spectra observed on 2016 Jul 02 and 2016 Jul 03 are combined in the spectral fitting as described in \ref{3subsec:xray_spec}. Figure \ref{fig:alllc1}(b) shows the 2 keV flux light curve, as well as other multi-wavelength light curves of VID 1. Compared with the high state in 2003 Jan 26, VID 2 dimmed by an amplitude of $12.0^{+2.6}_{-2.1}$ ($6.2^{+2.9}_{-1.2}$) in terms of $f_{\rm 2keV}$ ($L_{\rm X}$) when observed on 2016 Jul 02. Then it re-brightened with the $f_{\rm 2keV}$ ($L_{\rm X}$) increasing by a factor of $8.9^{+2.0}_{-1.8}$ ($5.1^{+2.4}_{-1.2}$) on 2017 Jan 02 and dimmed again when observed by eROSITA during 2019 Aug -- 2019 Sept. The best-fit \hbox{power-law} photon indices are $1.64^{+0.08}_{-0.08}$, $1.20^{+0.18}_{-0.17}$, and $1.44^{+0.11}_{-0.11}$ for the spectra observed on 2003 Jan 26, 2016 Jul 02, and 2017 Jan 02, respectively. The effective photon index estimated from the eROSITA band ratio is $\sim 0.8$. 

The MMT, SDSS, and DESI spectra of VID 2 (Figure \ref{fig:all_opt_spec}b) show no AGN signature and little variability. The IR-optical part SED of VID 2 (Figure \ref{fig:all_SED}b) also shows typical galaxy features. However, it shows multiple features that satisfied the X-ray AGN criteria (see Section \ref{subsec:optical_spec}) and can be reliably classified as an X-ray AGN. VID 2 has no ZTF light curve due to it low brightness, and it do not show substantial long-term variability in the Pan-STARRS, DECam, and NEOWISE light curves.

The low photon indices in its low states suggest absorption. The X-ray spectra of VID 2 can be described by a partial-covering absorption model self-consistently. We conclude that the extreme X-ray variability of VID 2 is likely caused by changing obscuration (see Section \ref{subsec:change_of_obscuration}).

\subsection{VID 3}
\label{subsec:VID3}

\begin{figure}
\hspace{-0.2cm}
 \includegraphics[width=0.47\textwidth]{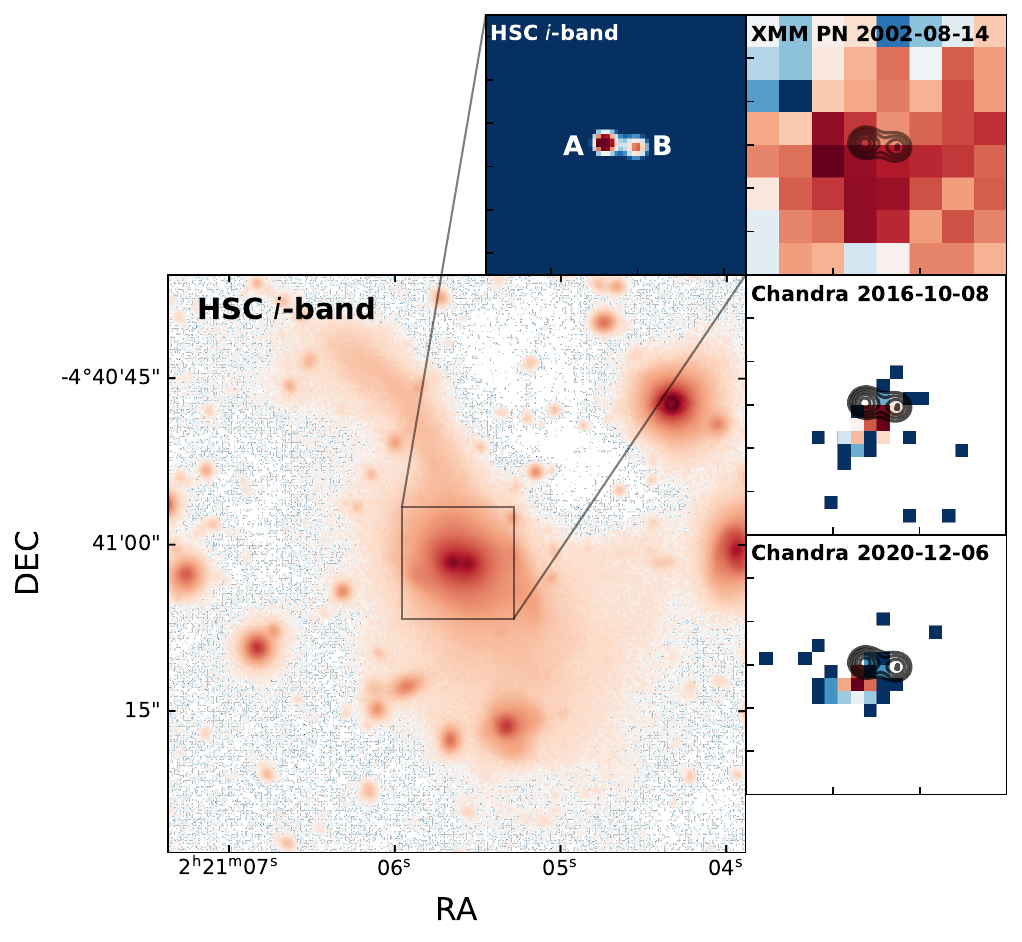}
 \centering
 \caption{The HSC $i$-band image of VID 3. The zoomed panels show the HSC image of the central nuclear pairs and the \hbox{X-ray} images of the same region. For comparision, the contour of the two optical point sources is shown in the \hbox{X-ray} images.}
 \label{fig:img_J0221}
\end{figure}

VID 3 is a merging galaxy at $z = 0.199$ with two nuclei in the central region, which have a separation of $\sim 1''$. Figure \ref{fig:img_J0221} shows the HSC $i$-band image of VID 3 and the corresponding \hbox{zoom-in} \xmm and Chandra image of its central region. We denote the left point source as A and the right point source as B. \cite{2021ApJ...922...83T} has performed Gemini/GMOS-N spectroscopic observation of VID 3 and confirms that source A is a broad-line AGN, while source B does not require broad component in the spectral fitting \citep[see Figure 20 of][]{2021ApJ...922...83T}. They suggest source B to be a SMBH not strongly activated yet or in a post-AGN phase. The high-resolution Chandra X-ray image also shows that the X-ray emission is more likely from source A.

VID 3 has seven \xmm observations and two Chandra observations, and after combining the spectra observed at similar dates as described in Section \ref{3subsec:xray_spec}, it has five X-ray spectra observed between 2002 Aug 14 and 2020 Dec 06. Figure \ref{fig:alllc1}(c) shows the 2 keV flux light curve, as well as other multi-wavelength light curves of VID 3. It dimmed by an amplitude of $11.1^{+0.5}_{-0.5}$ ($9.8^{+1.0}_{-0.8}$) in terms of $f_{\rm 2keV}$ ($L_{\rm X}$) between 2007 Aug 14 and 2016 Jul 03. After that it showed continuing trend of brightening in the following observations and almost returned to its peak brightness on 2020 Dec 06, with the $f_{\rm 2keV}$ ($L_{\rm X}$) increasing by a factor of $8.8^{+0.7}_{-1.1}$ ($8.3^{+1.5}_{-1.3}$) compared with 2016 Jul 03. The X-ray spectra of VID 3 can all be described well by the simple power-law model. The best-fit \hbox{power-law} photon indices of most of the spectra are $1.8 \sim 1.9$, except that the value of the spectrum observed in 2016 Oct 08 is $1.51^{+0.18}_{-0.18}$. The estimated effective photon index from the eROSITA band ratio is $\sim 1.9$. These photon indices indicate no significant absorption for both the high state and low state of VID 3.

VID 3 has two public SDSS spectra (see Figure \ref{fig:all_opt_spec}c), which are both observed centered on source A, showing AGN Balmer broad lines and galaxy shape continuum. In Figure \ref{fig:all_opt_spec}(c), we show that the SDSS spectrum observed on 2012 Oct 17 scaled by a factor of 0.6 (gray dotted line) agrees well with the SDSS spectrum observed on 2017 Oct 21. Thus, the difference between them is more likely caused by inaccurate flux calibration. But the DESI spectrum show an enhancement in both the continuum and the emission lines compared with the 2017 SDSS spectrum (see Figure \ref{fig:zoom_lines} below). The equivalent widths of the broad components of the H$\alpha$ and H$\beta$ lines increased from $276.5 \pm 16.9~\AA$ and $53.0 \pm 5.4~\AA$ in 2017 to $546.3 \pm 20.2~\AA$ and $118.1 \pm 14.4~\AA$ in 2022, respectively (Figure \ref{fig:zoom_lines}).

The IR-optical part SED of VID 3 is dominated by the host galaxy (Figure \ref{fig:all_SED}c). The simultaneous OM and X-ray measurement on 2016 Jul 03 shows that VID 3 follows the $\alpha_{\rm OX} \textrm{--} L_{\rm 2500 \AA}$ relation well in its low state, but at high state it is uncertain due to the lack of simultaneous observation. The GALEX light curves in Figure \ref{fig:alllc1}(c) show large variability in UV. The optical and IR light curves of VID 3 show consistent variability compared with the X-ray light curve, and the bluer optical bands less contaminated by host galaxy have larger variability amplitude. The maximum variability amplitudes of ZTF $g$-band and NEOWISE $W1$ and $W2$ band are $\sim 0.4$ mag, $\sim 0.3$ mag, and $\sim 0.5$ mag, respectively. 

Considerig the coordinated variability in the optical--IR and the lack of X-ray absorption, we suggest that its extreme \hbox{X-ray} variability is caused by the changing accretion state (see Section \ref{subsec:accretion_rate_change}).

\subsection{VID 4}
\label{subsec:VID4}

The SDSS survey reported VID 4 as a broad-line AGN at $z = 0.276$. This object has six \xmm observations between 2001 Jul 04 and 2017 Jan 03, and the observations on 2017 Jan 01 and 2017 Jan 03 are combined in the spectral fitting as described in Section \ref{3subsec:xray_spec}. Between 2001 Jul 04 and 2003 Jul 24, VID 4 brightened by an amplitude of $7.1^{+0.9}_{-0.8}$ ($6.0^{+1.4}_{-1.0}$) in terms of $f_{\rm 2keV}$ ($L_{\rm X}$). The spectrum observed on 2016 Jul 05 shows a similar flux to 2003 Jul 24. After that it showed a rapid dimming trend in the following observations, and the $f_{\rm 2keV}$ measured by the latest eROSITA observation from August to September 2019 decreased by a factor of $8.2^{+20.0}_{-3.4}$ compared with the high state in 2003 Jul 24. The five spectra can all be described well by the simple power-law model. The best-fit \hbox{power-law} photon indices of the high state spectra (see Table \ref{tab:xray_variability_X_spec_info}) are typical of Type I quasars. The spectrum in 2001 Jul 04 shows a relatively lower \hbox{power-law} photon index of $1.47^{+0.12}_{-0.11}$, and the effective photon index of the eROSIRA observation is estimated to be $\sim 1.9$. The photon indices in the low states still indicate no significant absorption.

An new optical spectrum of VID 4 was taken by MMT on 2023 Oct 16 and it is presented in Figure \ref{fig:all_opt_spec}(d), alongside the SDSS spectrum obtained on October 17, 2012 and the DESI spectrum obtained on February 07, 2022. The MMT observation reveals a significant enhancement in the quasar continuum emission, as well as the broad Balmer emission, in comparison to the SDSS and DESI spectra. The variability of the spectra is consistent with the brightening trend of the ZTF $g$- and $r$-band light curves in Figure \ref{fig:alllc1}(d). Specifically, the broad H$\beta$ emission, which is weak (or absent) in the SDSS spectrum, exhibits a notable increase in the MMT spectrum (Figure \ref{fig:zoom_lines}).

The SED of VID 4 shows galaxy contamination in the IR-optical part and large scatter in the UV part (see Figure \ref{fig:all_SED}d). Although the UV and X-ray data in the SED are not observed simultaneously, their fluxes in the high state and low state respectively roughly conform to the $\alpha_{\rm OX} \textrm{--} L_{\rm 2500 \AA}$ relation, indicating that their variability are caused by the same physical mechanism. The GALEX light curve of VID 4 shows significant variability, with maximum variability amplitude of $\sim 2.0$ mag and $\sim 1.4$ mag (corresponding to difference of $\sim 6$ times and $\sim 4$ times) in the NUV and FUV bands, respectively. The NUV and FUV variability is consistent with the overlapped CRTS light curve, but the optical variability amplitude is smaller due to the galaxy contamination. The overlapped ZTF and IR light curves show significant variability, which is consistent with the trend of the X-ray light curve. The maximum variability amplitudes of ZTF $g$-band and NEOWISE W1 and W2 band are $\sim 1.0$ mag, $\sim 0.8$ mag, and $\sim 0.7$ mag, respectively. 

Considering its nominal \hbox{X-ray} photon index and significant variability of the optical spectra, the extreme \hbox{X-ray} variability of VID 4 is attributed to the changing accretion state scenario (see Section \ref{subsec:accretion_rate_change}). 

\subsection{VID 5}
\label{subsec:VID5}
VID 5 was first reported as a galaxy at $z = 0.787$ by the PRIsm MUlti-object Survey \citep[PRIMUS;][]{2011ApJ...741....8C}, which uses low-dispersion prism and slitmasks to obtain spectra. An optical spectrum of VID 5 was taken by the MMT on 2023 Oct 16. The MMT spectrum shows significant broad Mg II line, and we measured its redshift more precisely to be 0.7703 using the \texttt{specpro} tool \citep{2011PASP..123..638M}.

This object has four \xmm observations. Between 2001 Jan 04 and 2016 Jul 08, brightened by an amplitude of $6.3^{+1.4}_{-1.1}$ ($6.5^{+2.2}_{-1.4}$) in terms of $f_{\rm 2keV}$ ($L_{\rm X}$). Then it dimmed rapidly in the following two XMM-Newton observations, with a factor of $2.6^{+0.6}_{-0.4}$ ($1.7^{+0.5}_{-0.5}$) in terms of $f_{\rm 2keV}$ ($L_{\rm X}$) from 2016 Jul 08 to 2017 Jan 01. The eROSITA measurement of VID 5 has large uncertainty and it is not used in our study. The four XMM-Newton spectra can all be described well by the simple power-law model. The best-fit \hbox{power-law} photon indices of the high state spectra (see Table \ref{tab:xray_variability_X_spec_info}) are typical of Type I quasars. The spectrum in 2017 Jan 01 shows a relatively lower \hbox{power-law} photon index of $1.51^{+0.20}_{-0.19}$, but it still indicates no significant absorption.

As VID 5 is too faint, it is not detected on the GALEX survey and optical light curve monitoring surveys. But the scatter of the optical part SED of VID 5 indicates significant optical variability (Figure \ref{fig:alllc1}e), and it's heavily contaminated by the host galaxy. The IR light curves also show large variability (Figure \ref{fig:alllc2}a), with the maximum variability amplitudes of $\sim 0.8$ mag and $\sim 1.0$ mag for NEOWISE $W1$ and $W2$ band, respectively. The mean SED of all quasars in \cite{2013ApJS..206....4K} for comparison is scaled to the 2001 Jan 04 \xmm OM B band measurement. The simultaneous OM and X-ray measurements of VID 5 on 2001 Jan 04 indicate X-ray weakness at that time. The expected IR-UV SED from the high state X-ray luminosity is higher than the available data point, yet simultaneous data is lacking. 

Given its large IR variability amplitude and the lack of significant absorption in the low-state X-ray spectra, the extreme X-ray variability is likely due to a changing accretion state (see Section \ref{subsec:accretion_rate_change}). However, we cannot entirely rule out a changing obscuration scenario for this source.

\subsection{VID 6}
\label{subsec:VID6}
The SDSS survey reported VID 6 as an AGN with broad-line features at $z = 0.661$, and its spectrum is shown in Figure \ref{fig:all_opt_spec}(f). This object has four \xmm observations and one Chandra observation. It is not detected in the observation on 2002 Jan 30 and yielded a 2 keV flux upper limit of $3.8 \times 10^{-33}$\fluxd as described in Section \ref{subsec:X-ray data}. VID 6 shows a brightening trend in the following two observations till 2016 Aug 14, with a variability amplitude of larger than 9.6 in terms of $f_{\rm 2keV}$. Then it dimmed a little on 2017 Jan 07 and re-brightened rapidly with a factor of $3.4^{+1.7}_{-1.2}$ ($5.2^{+4.4}_{-2.7}$) in terms of $f_{\rm 2keV}$ ($L_{\rm X}$) within two days. The maximum variability amplitude between 2002 Jan 30 and 2017 Jan 09 is larger than 17. The four spectra of VID 6 during 2006 Nov 12 and 2017 Jan 09 can all be described well by the simple power-law model, but the best-fit \hbox{power-law} photon indices (see Table \ref{tab:xray_variability_X_spec_info}) are higher than typical Type I quasars.

The optical-IR light curves of VID 6 in Figure \ref{fig:alllc2}(b) show little variability. The upward trend of the GALEX light curve is consistent with the X-ray light curve. The SED of VID 6 is generally consistent with the template (Figure \ref{fig:all_SED}f). The simultaneous OM and X-ray measurement on 2016 Aug 14 follows the $\alpha_{\rm OX} \textrm{--} L_{\rm 2500 \AA}$ relation well. The Chandra X-ray flux shows a little downward deviation compared with the quasi-simultaneous GALEX NUV measurement, but it may be due to the non-simultaneous observations. The X-ray, optical, and IR light curves show that VID 6 have rapid variability after 2016. The maximum variability amplitudes of DECam $g$-band and NEOWISE $W1$ and $W2$ band are $\sim 0.1$ mag, $\sim 0.4$ mag, and $\sim 0.5$ mag, respectively. 

Given the little optical-IR variability, we suggest that the extreme X-ray variability of VID 6 is caused by changing obscuration. We show that the X-ray spectra of VID 6 can be described by a partial-covering absorption model self-consistently (see Section \ref{subsec:change_of_obscuration}).

\subsection{VID 7}
\label{subsec:VID7}

The DESI EDR spectrum \citep{2023arXiv230606308D} of VID 7 is shown in Figure \ref{fig:all_opt_spec}(g), confirming it as an typical broad-line quasar at $z = 1.059$. It has four \xmm observations and one Chandra observation. The \xmm X-ray spectra observed on 2017 Jan 06 and 2017 Jan 08 show little variability and are combined in the spectral fitting. VID 7 is not detected in the first observation on 2002 Feb 02, and the 2 keV flux upper limit is estimated to be $4.8 \times 10^{-33}$ \fluxd as described in Section \ref{subsec:X-ray data}. In the following three observations during 2016 Sept 27 to 2017 Feb 10, it is significantly brightened, with a maximum factor of $>9.2$ in terms of $f_{\rm 2keV}$. The three X-ray spectra of VID 7 can all be described well by the simple power-law model with best-fit \hbox{power-law} photon indices (see Table \ref{tab:xray_variability_X_spec_info}) of typical Type I quasars.

The SED of VID 7 is shown in Figure \ref{fig:all_SED}(g). Its IR–UV SED is broadly consistent with those of typical quasars. The little scatter in the optical part of the SED indicates no significant variability, which is consistent with the light curves in Figure \ref{fig:alllc2}(c). VID 7 is not detected in the NEOWISE survey. The simultaneous OM and X-ray measurements on 2017 Jan 06 and 2017 Feb 10 follow the $\alpha_{\rm OX} \textrm{--} L_{\rm 2500 \AA}$ relation well.  

Given the little optical variability amplitude and relatively high accretion rate of VID 7, we suggest that its extreme X-ray variability is caused by the changing obscuration (see Section \ref{subsec:change_of_obscuration}). However, we still can not rule out the changing accretion state scenario, since it is likely the changing accretion state occur during 2003--2010, and we lack optical monitoring during this period.

\subsection{VID 8}
\label{subsec:VID 8}

SDSS classified VID 8 as a broad-line AGN at $z = 0.446$. This object has four \xmm observations, and the two X-ray spectra observed on 2017 Jan 11 are combined in the spectral fitting as described in Section \ref{3subsec:xray_spec}. The 2 keV flux light curve of VID 8 in Figure \ref{fig:alllc2}(d) shows a  continuous downward trend. It shows a maximum variability amplitude of $11.2^{+2.0}_{-1.3}$ ($7.4^{+2.4}_{-1.9}$) in terms of $f_{\rm 2keV}$ ($L_{\rm X}$) between the two observations of 2001 Jul 06 and 2017 Jan 11. The simple power-law model can fit all the three spectra well with photon indices of $2.23^{+0.05}_{-0.05}$, $2.18^{+0.07}_{-0.07}$, and $1.72^{+0.23}_{-0.22}$, respectively. Adding an intrinsic absorption component ({\sc zphabs}) for the fitting of the low state spectra observed on 2017 Jan 11 does not improve the fits, and we set upper limits on the intrinsic $N_{\rm H}$ of $1.8 \times 10^{21} \rm ~cm^{-2}$, assuming the intrinsic photon index to be 2.2.

The MMT and SDSS spectra of VID 8 are shown in Figure \ref{fig:all_opt_spec}(h). The two spectra are consistent with each other with little variability. The spectra have galaxy dominated continuum and significant Mg II broad line. For comparison, we also include the composite spectrum of typical SDSS quasar from \citet{2001AJ....122..549V}, scaled to have a similar Mg II line intensity with VID 8. It shows no H$\beta$ broad line in both the MMT and SDSS spectra, and the AGN continuum level and H$\alpha$ line intensity is also less than that expected from the Mg II line. The IR-to-UV SED (Figure \ref{fig:all_SED}h) is lower than the expectation from the high state X-ray luminosity from the $\alpha_{\rm OX} \textrm{--} L_{\rm 2500 \AA}$ relation, but it may be due to the lack of simultaneous observations.

VID 8 shows no significant variability in its \hbox{multi-epoch} optical spectra or optical--IR light curves, but it also do not show absorption signature in its low-state X-ray spectrum. Its Mg II line is strong and persists in the two spectroscopic observations. Such Mg II emitters can be interpreted as recently \hbox{turned-off} AGNs. Therefore, we suggest that its extreme X-ray variability is caused by changing accretion state (see Section \ref{subsec:accretion_rate_change}).


\begin{thebibliography}{}
\expandafter\ifx\csname natexlab\endcsname\relax\def\natexlab#1{#1}\fi
\providecommand{\url}[1]{\href{#1}{#1}}
\providecommand{\dodoi}[1]{doi:~\href{http://doi.org/#1}{\nolinkurl{#1}}}
\providecommand{\doeprint}[1]{\href{http://ascl.net/#1}{\nolinkurl{http://ascl.net/#1}}}
\providecommand{\doarXiv}[1]{\href{https://arxiv.org/abs/#1}{\nolinkurl{https://arxiv.org/abs/#1}}}

\bibitem[{{Aihara} {et~al.}(2022){Aihara}, {AlSayyad}, {Ando}, {Armstrong}, {Bosch}, {Egami}, {Furusawa}, {Furusawa}, {Harasawa}, {Harikane}, {Hsieh}, {Ikeda}, {Ito}, {Iwata}, {Kodama}, {Koike}, {Kokubo}, {Komiyama}, {Li}, {Liang}, {Lin}, {Lupton}, {Lust}, {MacArthur}, {Mawatari}, {Mineo}, {Miyatake}, {Miyazaki}, {More}, {Morishima}, {Murayama}, {Nakajima}, {Nakata}, {Nishizawa}, {Oguri}, {Okabe}, {Okura}, {Ono}, {Osato}, {Ouchi}, {Pan}, {Plazas Malag{\'o}n}, {Price}, {Reed}, {Rykoff}, {Shibuya}, {Simunovic}, {Strauss}, {Sugimori}, {Suto}, {Suzuki}, {Takada}, {Takagi}, {Takata}, {Takita}, {Tanaka}, {Tang}, {Taranu}, {Terai}, {Toba}, {Turner}, {Uchiyama}, {Vijarnwannaluk}, {Waters}, {Yamada}, {Yamamoto}, \& {Yamashita}}]{2022PASJ...74..247A}
{Aihara}, H., {AlSayyad}, Y., {Ando}, M., {et~al.} 2022, \pasj, 74, 247, \dodoi{10.1093/pasj/psab122}

\bibitem[{Arnaud(1996)}]{arnaud1996astronomical}
Arnaud, K. 1996, in ASP Conf., Vol.~17

\bibitem[{{Astropy Collaboration} {et~al.}(2013){Astropy Collaboration}, {Robitaille}, {Tollerud}, {Greenfield}, {Droettboom}, {Bray}, {Aldcroft}, {Davis}, {Ginsburg}, {Price-Whelan}, {Kerzendorf}, {Conley}, {Crighton}, {Barbary}, {Muna}, {Ferguson}, {Grollier}, {Parikh}, {Nair}, {Unther}, {Deil}, {Woillez}, {Conseil}, {Kramer}, {Turner}, {Singer}, {Fox}, {Weaver}, {Zabalza}, {Edwards}, {Azalee Bostroem}, {Burke}, {Casey}, {Crawford}, {Dencheva}, {Ely}, {Jenness}, {Labrie}, {Lim}, {Pierfederici}, {Pontzen}, {Ptak}, {Refsdal}, {Servillat}, \& {Streicher}}]{2013A&A...558A..33A}
{Astropy Collaboration}, {Robitaille}, T.~P., {Tollerud}, E.~J., {et~al.} 2013, \aap, 558, A33, \dodoi{10.1051/0004-6361/201322068}

\bibitem[{{Astropy Collaboration} {et~al.}(2018){Astropy Collaboration}, {Price-Whelan}, {Sip{\H{o}}cz}, {G{\"u}nther}, {Lim}, {Crawford}, {Conseil}, {Shupe}, {Craig}, {Dencheva}, {Ginsburg}, {VanderPlas}, {Bradley}, {P{\'e}rez-Su{\'a}rez}, {de Val-Borro}, {Aldcroft}, {Cruz}, {Robitaille}, {Tollerud}, {Ardelean}, {Babej}, {Bach}, {Bachetti}, {Bakanov}, {Bamford}, {Barentsen}, {Barmby}, {Baumbach}, {Berry}, {Biscani}, {Boquien}, {Bostroem}, {Bouma}, {Brammer}, {Bray}, {Breytenbach}, {Buddelmeijer}, {Burke}, {Calderone}, {Cano Rodr{\'\i}guez}, {Cara}, {Cardoso}, {Cheedella}, {Copin}, {Corrales}, {Crichton}, {D'Avella}, {Deil}, {Depagne}, {Dietrich}, {Donath}, {Droettboom}, {Earl}, {Erben}, {Fabbro}, {Ferreira}, {Finethy}, {Fox}, {Garrison}, {Gibbons}, {Goldstein}, {Gommers}, {Greco}, {Greenfield}, {Groener}, {Grollier}, {Hagen}, {Hirst}, {Homeier}, {Horton}, {Hosseinzadeh}, {Hu}, {Hunkeler}, {Ivezi{\'c}}, {Jain}, {Jenness}, {Kanarek}, {Kendrew}, {Kern}, {Kerzendorf}, {Khvalko}, {King}, {Kirkby}, {Kulkarni},
  {Kumar}, {Lee}, {Lenz}, {Littlefair}, {Ma}, {Macleod}, {Mastropietro}, {McCully}, {Montagnac}, {Morris}, {Mueller}, {Mumford}, {Muna}, {Murphy}, {Nelson}, {Nguyen}, {Ninan}, {N{\"o}the}, {Ogaz}, {Oh}, {Parejko}, {Parley}, {Pascual}, {Patil}, {Patil}, {Plunkett}, {Prochaska}, {Rastogi}, {Reddy Janga}, {Sabater}, {Sakurikar}, {Seifert}, {Sherbert}, {Sherwood-Taylor}, {Shih}, {Sick}, {Silbiger}, {Singanamalla}, {Singer}, {Sladen}, {Sooley}, {Sornarajah}, {Streicher}, {Teuben}, {Thomas}, {Tremblay}, {Turner}, {Terr{\'o}n}, {van Kerkwijk}, {de la Vega}, {Watkins}, {Weaver}, {Whitmore}, {Woillez}, {Zabalza}, \& {Astropy Contributors}}]{2018AJ....156..123A}
{Astropy Collaboration}, {Price-Whelan}, A.~M., {Sip{\H{o}}cz}, B.~M., {et~al.} 2018, \aj, 156, 123, \dodoi{10.3847/1538-3881/aabc4f}

\bibitem[{{Astropy Collaboration} {et~al.}(2022){Astropy Collaboration}, {Price-Whelan}, {Lim}, {Earl}, {Starkman}, {Bradley}, {Shupe}, {Patil}, {Corrales}, {Brasseur}, {N{\"o}the}, {Donath}, {Tollerud}, {Morris}, {Ginsburg}, {Vaher}, {Weaver}, {Tocknell}, {Jamieson}, {van Kerkwijk}, {Robitaille}, {Merry}, {Bachetti}, {G{\"u}nther}, {Aldcroft}, {Alvarado-Montes}, {Archibald}, {B{\'o}di}, {Bapat}, {Barentsen}, {Baz{\'a}n}, {Biswas}, {Boquien}, {Burke}, {Cara}, {Cara}, {Conroy}, {Conseil}, {Craig}, {Cross}, {Cruz}, {D'Eugenio}, {Dencheva}, {Devillepoix}, {Dietrich}, {Eigenbrot}, {Erben}, {Ferreira}, {Foreman-Mackey}, {Fox}, {Freij}, {Garg}, {Geda}, {Glattly}, {Gondhalekar}, {Gordon}, {Grant}, {Greenfield}, {Groener}, {Guest}, {Gurovich}, {Handberg}, {Hart}, {Hatfield-Dodds}, {Homeier}, {Hosseinzadeh}, {Jenness}, {Jones}, {Joseph}, {Kalmbach}, {Karamehmetoglu}, {Ka{\l}uszy{\'n}ski}, {Kelley}, {Kern}, {Kerzendorf}, {Koch}, {Kulumani}, {Lee}, {Ly}, {Ma}, {MacBride}, {Maljaars}, {Muna}, {Murphy}, {Norman},
  {O'Steen}, {Oman}, {Pacifici}, {Pascual}, {Pascual-Granado}, {Patil}, {Perren}, {Pickering}, {Rastogi}, {Roulston}, {Ryan}, {Rykoff}, {Sabater}, {Sakurikar}, {Salgado}, {Sanghi}, {Saunders}, {Savchenko}, {Schwardt}, {Seifert-Eckert}, {Shih}, {Jain}, {Shukla}, {Sick}, {Simpson}, {Singanamalla}, {Singer}, {Singhal}, {Sinha}, {Sip{\H{o}}cz}, {Spitler}, {Stansby}, {Streicher}, {{\v{S}}umak}, {Swinbank}, {Taranu}, {Tewary}, {Tremblay}, {de Val-Borro}, {Van Kooten}, {Vasovi{\'c}}, {Verma}, {de Miranda Cardoso}, {Williams}, {Wilson}, {Winkel}, {Wood-Vasey}, {Xue}, {Yoachim}, {Zhang}, {Zonca}, \& {Astropy Project Contributors}}]{2022ApJ...935..167A}
{Astropy Collaboration}, {Price-Whelan}, A.~M., {Lim}, P.~L., {et~al.} 2022, \apj, 935, 167, \dodoi{10.3847/1538-4357/ac7c74}

\bibitem[{{Bachev} {et~al.}(2009){Bachev}, {Grupe}, {Boeva}, {Ovcharov}, {Valcheva}, {Semkov}, {Georgiev}, \& {Gallo}}]{2009MNRAS.399..750B}
{Bachev}, R., {Grupe}, D., {Boeva}, S., {et~al.} 2009, \mnras, 399, 750, \dodoi{10.1111/j.1365-2966.2009.15301.x}

\bibitem[{{Bianchi} {et~al.}(2017){Bianchi}, {Shiao}, \& {Thilker}}]{2017ApJS..230...24B}
{Bianchi}, L., {Shiao}, B., \& {Thilker}, D. 2017, \apjs, 230, 24, \dodoi{10.3847/1538-4365/aa7053}

\bibitem[{{Blanchard} {et~al.}(2017){Blanchard}, {Nicholl}, {Berger}, {Guillochon}, {Margutti}, {Chornock}, {Alexander}, {Leja}, \& {Drout}}]{2017ApJ...843..106B}
{Blanchard}, P.~K., {Nicholl}, M., {Berger}, E., {et~al.} 2017, \apj, 843, 106, \dodoi{10.3847/1538-4357/aa77f7}

\bibitem[{{Boller} {et~al.}(1996){Boller}, {Brandt}, \& {Fink}}]{1996A&A...305...53B}
{Boller}, T., {Brandt}, W.~N., \& {Fink}, H. 1996, \aap, 305, 53, \dodoi{10.48550/arXiv.astro-ph/9504093}

\bibitem[{{Brandt} \& {Alexander}(2015)}]{2015A&ARv..23....1B}
{Brandt}, W.~N., \& {Alexander}, D.~M. 2015, \aapr, 23, 1, \dodoi{10.1007/s00159-014-0081-z}

\bibitem[{{Brandt} {et~al.}(1995){Brandt}, {Pounds}, \& {Fink}}]{1995MNRAS.273L..47B}
{Brandt}, W.~N., {Pounds}, K.~A., \& {Fink}, H. 1995, \mnras, 273, L47, \dodoi{10.1093/mnras/273.1.L47}

\bibitem[{{Brightman} {et~al.}(2013){Brightman}, {Silverman}, {Mainieri}, {Ueda}, {Schramm}, {Matsuoka}, {Nagao}, {Steinhardt}, {Kartaltepe}, {Sanders}, {Treister}, {Shemmer}, {Brandt}, {Brusa}, {Comastri}, {Ho}, {Lanzuisi}, {Lusso}, {Nandra}, {Salvato}, {Zamorani}, {Akiyama}, {Alexander}, {Bongiorno}, {Capak}, {Civano}, {Del Moro}, {Doi}, {Elvis}, {Hasinger}, {Laird}, {Masters}, {Mignoli}, {Ohta}, {Schawinski}, \& {Taniguchi}}]{2013MNRAS.433.2485B}
{Brightman}, M., {Silverman}, J.~D., {Mainieri}, V., {et~al.} 2013, \mnras, 433, 2485, \dodoi{10.1093/mnras/stt920}

\bibitem[{{Butler} \& {Bloom}(2011)}]{2011AJ....141...93B}
{Butler}, N.~R., \& {Bloom}, J.~S. 2011, \aj, 141, 93, \dodoi{10.1088/0004-6256/141/3/93}

\bibitem[{{Caplar} {et~al.}(2017){Caplar}, {Lilly}, \& {Trakhtenbrot}}]{2017ApJ...834..111C}
{Caplar}, N., {Lilly}, S.~J., \& {Trakhtenbrot}, B. 2017, \apj, 834, 111, \dodoi{10.3847/1538-4357/834/2/111}

\bibitem[{{Chartas} {et~al.}(2009){Chartas}, {Kochanek}, {Dai}, {Poindexter}, \& {Garmire}}]{2009ApJ...693..174C}
{Chartas}, G., {Kochanek}, C.~S., {Dai}, X., {Poindexter}, S., \& {Garmire}, G. 2009, ApJ, 693, 174, \dodoi{10.1088/0004-637X/693/1/174}

\bibitem[{{Chen} {et~al.}(2018){Chen}, {Brandt}, {Luo}, {Ranalli}, {Yang}, {Alexander}, {Bauer}, {Kelson}, {Lacy}, {Nyland}, {Tozzi}, {Vito}, {Cirasuolo}, {Gilli}, {Jarvis}, {Lehmer}, {Paolillo}, {Schneider}, {Shemmer}, {Smail}, {Sun}, {Tanaka}, {Vaccari}, {Vignali}, {Xue}, {Banerji}, {Chow}, {H{\"a}u{\ss}ler}, {Norris}, {Silverman}, \& {Trump}}]{2018MNRAS.478.2132C}
{Chen}, C. T.~J., {Brandt}, W.~N., {Luo}, B., {et~al.} 2018, \mnras, 478, 2132, \dodoi{10.1093/mnras/sty1036}

\bibitem[{{Coil} {et~al.}(2011){Coil}, {Blanton}, {Burles}, {Cool}, {Eisenstein}, {Moustakas}, {Wong}, {Zhu}, {Aird}, {Bernstein}, {Bolton}, \& {Hogg}}]{2011ApJ...741....8C}
{Coil}, A.~L., {Blanton}, M.~R., {Burles}, S.~M., {et~al.} 2011, \apj, 741, 8, \dodoi{10.1088/0004-637X/741/1/8}

\bibitem[{{Dai} {et~al.}(2010){Dai}, {Kochanek}, {Chartas}, {Koz{\l}owski}, {Morgan}, {Garmire}, \& {Agol}}]{2010ApJ...709..278D}
{Dai}, X., {Kochanek}, C.~S., {Chartas}, G., {et~al.} 2010, \apj, 709, 278, \dodoi{10.1088/0004-637X/709/1/278}

\bibitem[{{Denney} {et~al.}(2009){Denney}, {Peterson}, {Dietrich}, {Vestergaard}, \& {Bentz}}]{2009ApJ...692..246D}
{Denney}, K.~D., {Peterson}, B.~M., {Dietrich}, M., {Vestergaard}, M., \& {Bentz}, M.~C. 2009, \apj, 692, 246, \dodoi{10.1088/0004-637X/692/1/246}

\bibitem[{{DESI Collaboration} {et~al.}(2022){DESI Collaboration}, {Abareshi}, {Aguilar}, {Ahlen}, {Alam}, {Alexander}, {Alfarsy}, {Allen}, {Allende Prieto}, {Alves}, {Ameel}, {Armengaud}, {Asorey}, {Aviles}, {Bailey}, {Balaguera-Antol{\'\i}nez}, {Ballester}, {Baltay}, {Bault}, {Beltran}, {Benavides}, {BenZvi}, {Berti}, {Besuner}, {Beutler}, {Bianchi}, {Blake}, {Blanc}, {Blum}, {Bolton}, {Bose}, {Bramall}, {Brieden}, {Brodzeller}, {Brooks}, {Brownewell}, {Buckley-Geer}, {Cahn}, {Cai}, {Canning}, {Capasso}, {Carnero Rosell}, {Carton}, {Casas}, {Castander}, {Cervantes-Cota}, {Chabanier}, {Chaussidon}, {Chuang}, {Circosta}, {Cole}, {Cooper}, {da Costa}, {Cousinou}, {Cuceu}, {Davis}, {Dawson}, {de la Cruz-Noriega}, {de la Macorra}, {de Mattia}, {Della Costa}, {Demmer}, {Derwent}, {Dey}, {Dey}, {Dhungana}, {Ding}, {Dobson}, {Doel}, {Donald-McCann}, {Donaldson}, {Douglass}, {Duan}, {Dunlop}, {Edelstein}, {Eftekharzadeh}, {Eisenstein}, {Enriquez-Vargas}, {Escoffier}, {Evatt}, {Fagrelius}, {Fan}, {Fanning},
  {Fawcett}, {Ferraro}, {Ereza}, {Flaugher}, {Font-Ribera}, {Forero-Romero}, {Frenk}, {Fromenteau}, {G{\"a}nsicke}, {Garcia-Quintero}, {Garrison}, {Gazta{\~n}aga}, {Gerardi}, {Gil-Mar{\'\i}n}, {Gontcho a Gontcho}, {Gonzalez-Morales}, {Gonzalez-de-Rivera}, {Gonzalez-Perez}, {Gordon}, {Graur}, {Green}, {Grove}, {Gruen}, {Gutierrez}, {Guy}, {Hahn}, {Harris}, {Herrera}, {Herrera-Alcantar}, {Honscheid}, {Howlett}, {Huterer}, {Ir{\v{s}}i{\v{c}}}, {Ishak}, {Jelinsky}, {Jiang}, {Jimenez}, {Jing}, {Joyce}, {Jullo}, {Juneau}, {Kara{\c{c}}ayl{\i}}, {Karamanis}, {Karcher}, {Karim}, {Kehoe}, {Kent}, {Kirkby}, {Kisner}, {Kitaura}, {Koposov}, {Kov{\'a}cs}, {Kremin}, {Krolewski}, {L'Huillier}, {Lahav}, {Lambert}, {Lamman}, {Lan}, {Landriau}, {Lane}, {Lang}, {Lange}, {Lasker}, {Le Guillou}, {Leauthaud}, {Le Van Suu}, {Levi}, {Li}, {Magneville}, {Manera}, {Manser}, {Marshall}, {Martini}, {McCollam}, {McDonald}, {Meisner}, {Mena-Fern{\'a}ndez}, {Meneses-Rizo}, {Mezcua}, {Miller}, {Miquel}, {Montero-Camacho}, {Moon},
  {Moustakas}, {Mueller}, {Mu{\~n}oz-Guti{\'e}rrez}, {Myers}, {Nadathur}, {Najita}, {Napolitano}, {Neilsen}, {Newman}, {Nie}, {Ning}, {Niz}, {Norberg}, {Noriega}, {O'Brien}, {Obuljen}, {Palanque-Delabrouille}, {Palmese}, {Zhiwei}, {Pappalardo}, {PENG}, {Percival}, {Perruchot}, {Pogge}, {Poppett}, {Porredon}, {Prada}, {Prochaska}, {Pucha}, {P{\'e}rez-Fern{\'a}ndez}, {P{\'e}rez-R{\`a}fols}, {Rabinowitz}, {Raichoor}, {Ramirez-Solano}, {Ram{\'\i}rez-P{\'e}rez}, {Ravoux}, {Reil}, {Rezaie}, {Rocher}, {Rockosi}, {Roe}, {Roodman}, {Ross}, {Rossi}, {Ruggeri}, {Ruhlmann-Kleider}, {Sabiu}, {Safonova}, {Said}, {Saintonge}, {Salas Catonga}, {Samushia}, {Sanchez}, {Saulder}, {Schaan}, {Schlafly}, {Schlegel}, {Schmoll}, {Scholte}, {Schubnell}, {Secroun}, {Seo}, {Serrano}, {Sharples}, {Sholl}, {Silber}, {Silva}, {Sirk}, {Siudek}, {Smith}, {Sprayberry}, {Staten}, {Stupak}, {Tan}, {Tarl{\'e}}, {Tie}, {Tojeiro}, {Ure{\~n}a-L{\'o}pez}, {Valdes}, {Valenzuela}, {Valluri}, {Vargas-Maga{\~n}a}, {Verde}, {Walther}, {Wang}, {Wang},
  {Weaver}, {Weaverdyck}, {Wechsler}, {Wilson}, {Yang}, {Yu}, {Yuan}, {Y{\`e}che}, {Zhang}, {Zhang}, {Zhao}, {Zhou}, {Zhou}, {Zou}, {Zou}, {Zou}, {Zu}, \& {DESI Collaboration}}]{2022AJ....164..207D}
{DESI Collaboration}, {Abareshi}, B., {Aguilar}, J., {et~al.} 2022, \aj, 164, 207, \dodoi{10.3847/1538-3881/ac882b}

\bibitem[{{DESI Collaboration} {et~al.}(2023){DESI Collaboration}, {Adame}, {Aguilar}, {Ahlen}, {Alam}, {Aldering}, {Alexander}, {Alfarsy}, {Allende Prieto}, {Alvarez}, {Alves}, {Anand}, {Andrade-Oliveira}, {Armengaud}, {Asorey}, {Avila}, {Aviles}, {Bailey}, {Balaguera-Antol{\'\i}nez}, {Ballester}, {Baltay}, {Bault}, {Bautista}, {Behera}, {Beltran}, {BenZvi}, {Beraldo e Silva}, {Bermejo-Climent}, {Berti}, {Besuner}, {Beutler}, {Bianchi}, {Blake}, {Blum}, {Bolton}, {Brieden}, {Brodzeller}, {Brooks}, {Brown}, {Buckley-Geer}, {Burtin}, {Cabayol-Garcia}, {Cai}, {Canning}, {Cardiel-Sas}, {Carnero Rosell}, {Castander}, {Cervantes-Cota}, {Chabanier}, {Chaussidon}, {Chaves-Montero}, {Chen}, {Chuang}, {Claybaugh}, {Cole}, {Cooper}, {Cuceu}, {Davis}, {Dawson}, {de Belsunce}, {de la Cruz}, {de la Macorra}, {de Mattia}, {Demina}, {Demirbozan}, {DeRose}, {Dey}, {Dey}, {Dhungana}, {Ding}, {Ding}, {Doel}, {Doshi}, {Douglass}, {Edge}, {Eftekharzadeh}, {Eisenstein}, {Elliott}, {Escoffier}, {Fagrelius}, {Fan}, {Fanning},
  {Fawcett}, {Ferraro}, {Ereza}, {Flaugher}, {Font-Ribera}, {Forero-S{\'a}nchez}, {Forero-Romero}, {Frenk}, {G{\"a}nsicke}, {Garc{\'\i}a}, {Garc{\'\i}a-Bellido}, {Garcia-Quintero}, {Garrison}, {Gil-Mar{\'\i}n}, {Golden-Marx}, {Gontcho}, {Gonzalez-Morales}, {Gonzalez-Perez}, {Gordon}, {Graur}, {Green}, {Gruen}, {Guy}, {Hadzhiyska}, {Hahn}, {Han}, {Hanif}, {Herrera-Alcantar}, {Honscheid}, {Hou}, {Howlett}, {Huterer}, {Ir{\v{s}}i{\v{c}}}, {Ishak}, {Jacques}, {Jana}, {Jiang}, {Jimenez}, {Jing}, {Joudaki}, {Jullo}, {Juneau}, {Kizhuprakkat}, {Kara{\c{c}}ayl{\i}}, {Karim}, {Kehoe}, {Kent}, {Khederlarian}, {Kim}, {Kirkby}, {Kisner}, {Kitaura}, {Kneib}, {Koposov}, {Kov{\'a}cs}, {Kremin}, {Krolewski}, {L'Huillier}, {Lambert}, {Lamman}, {Lan}, {Landriau}, {Lang}, {Lange}, {Lasker}, {Le Guillou}, {Leauthaud}, {Levi}, {Li}, {Linder}, {Lyons}, {Magneville}, {Manera}, {Manser}, {Margala}, {Martini}, {McDonald}, {Medina}, {Medina-Varela}, {Meisner}, {Mena-Fern{\'a}ndez}, {Meneses-Rizo}, {Mezcua}, {Miquel}, {Montero-Camacho},
  {Moon}, {Moore}, {Moustakas}, {Mueller}, {Mundet}, {Mu{\~n}oz-Guti{\'e}rrez}, {Myers}, {Nadathur}, {Napolitano}, {Neveux}, {Newman}, {Nie}, {Nikutta}, {Niz}, {Norberg}, {Noriega}, {Paillas}, {Palanque-Delabrouille}, {Palmese}, {Zhiwei}, {Parkinson}, {Penmetsa}, {Percival}, {P{\'e}rez-Fern{\'a}ndez}, {P{\'e}rez-R{\`a}fols}, {Pieri}, {Poppett}, {Porredon}, {Pothier}, {Prada}, {Pucha}, {Raichoor}, {Ram{\'\i}rez-P{\'e}rez}, {Ramirez-Solano}, {Rashkovetskyi}, {Ravoux}, {Rocher}, {Rockosi}, {Ross}, {Rossi}, {Ruggeri}, {Ruhlmann-Kleider}, {Sabiu}, {Said}, {Saintonge}, {Samushia}, {Sanchez}, {Saulder}, {Schaan}, {Schlafly}, {Schlegel}, {Scholte}, {Schubnell}, {Seo}, {Shafieloo}, {Sharples}, {Sheu}, {Silber}, {Sinigaglia}, {Siudek}, {Slepian}, {Smith}, {Sprayberry}, {Stephey}, {Su{\'a}rez-P{\'e}rez}, {Sun}, {Tan}, {Tarl{\'e}}, {Tojeiro}, {Ure{\~n}a-L{\'o}pez}, {Vaisakh}, {Valcin}, {Valdes}, {Valluri}, {Vargas-Maga{\~n}a}, {Variu}, {Verde}, {Walther}, {Wang}, {Wang}, {Weaver}, {Weaverdyck}, {Wechsler}, {White},
  {Xie}, {Yang}, {Y{\`e}che}, {Yu}, {Yuan}, {Zhang}, {Zhang}, {Zhao}, {Zheng}, {Zhou}, {Zhou}, {Zou}, {Zou}, \& {Zu}}]{2023arXiv230606308D}
{DESI Collaboration}, {Adame}, A.~G., {Aguilar}, J., {et~al.} 2023, arXiv e-prints, arXiv:2306.06308, \dodoi{10.48550/arXiv.2306.06308}

\bibitem[{{DESI Collaboration} {et~al.}(2024){DESI Collaboration}, {Adame}, {Aguilar}, {Ahlen}, {Alam}, {Aldering}, {Alexander}, {Alfarsy}, {Prieto}, {Alvarez}, {Alves}, {Anand}, {Andrade-Oliveira}, {Armengaud}, {Asorey}, {Avila}, {Aviles}, {Bailey}, {Balaguera-Antol{\'\i}nez}, {Ballester}, {Baltay}, {Bault}, {Bautista}, {Behera}, {Beltran}, {BenZvi}, {Beraldo e Silva}, {Bermejo-Climent}, {Berti}, {Besuner}, {Beutler}, {Bianchi}, {Blake}, {Blum}, {Bolton}, {Brieden}, {Brodzeller}, {Brooks}, {Brown}, {Buckley-Geer}, {Burtin}, {Cabayol-Garcia}, {Cai}, {Canning}, {Cardiel-Sas}, {Carnero Rosell}, {Castander}, {Cervantes-Cota}, {Chabanier}, {Chaussidon}, {Chaves-Montero}, {Chen}, {Chen}, {Chuang}, {Claybaugh}, {Cole}, {Cooper}, {Cuceu}, {Davis}, {Dawson}, {de Belsunce}, {de la Cruz}, {de la Macorra}, {Della Costa}, {de Mattia}, {Demina}, {Demirbozan}, {DeRose}, {Dey}, {Dey}, {Dhungana}, {Ding}, {Ding}, {Doel}, {Doshi}, {Douglass}, {Edge}, {Eftekharzadeh}, {Eisenstein}, {Elliott}, {Ereza}, {Escoffier}, {Fagrelius},
  {Fan}, {Fanning}, {Fawcett}, {Ferraro}, {Flaugher}, {Font-Ribera}, {Forero-Romero}, {Forero-S{\'a}nchez}, {Frenk}, {G{\"a}nsicke}, {Garc{\'\i}a}, {Garc{\'\i}a-Bellido}, {Garcia-Quintero}, {Garrison}, {Gil-Mar{\'\i}n}, {Golden-Marx}, {Gontcho}, {Gonzalez-Morales}, {Gonzalez-Perez}, {Gordon}, {Graur}, {Green}, {Gruen}, {Guy}, {Hadzhiyska}, {Hahn}, {Han}, {Hanif}, {Herrera-Alcantar}, {Honscheid}, {Hou}, {Howlett}, {Huterer}, {Ir{\v{s}}i{\v{c}}}, {Ishak}, {Jacques}, {Jana}, {Jiang}, {Jimenez}, {Jing}, {Joudaki}, {Joyce}, {Jullo}, {Juneau}, {Kara{\c{c}}ayl{\i}}, {Karim}, {Kehoe}, {Kent}, {Khederlarian}, {Kim}, {Kirkby}, {Kisner}, {Kitaura}, {Kizhuprakkat}, {Kneib}, {Koposov}, {Kov{\'a}cs}, {Kremin}, {Krolewski}, {L'Huillier}, {Lahav}, {Lambert}, {Lamman}, {Lan}, {Landriau}, {Lang}, {Lange}, {Lasker}, {Leauthaud}, {Le Guillou}, {Levi}, {Li}, {Linder}, {Lyons}, {Magneville}, {Manera}, {Manser}, {Margala}, {Martini}, {McDonald}, {Medina}, {Medina-Varela}, {Meisner}, {Mena-Fern{\'a}ndez}, {Meneses-Rizo}, {Mezcua},
  {Miquel}, {Montero-Camacho}, {Moon}, {Moore}, {Moustakas}, {Mueller}, {Mundet}, {Mu{\~n}oz-Guti{\'e}rrez}, {Myers}, {Nadathur}, {Napolitano}, {Neveux}, {Newman}, {Nie}, {Nikutta}, {Niz}, {Norberg}, {Noriega}, {Paillas}, {Palanque-Delabrouille}, {Palmese}, {Pan}, {Parkinson}, {Penmetsa}, {Percival}, {P{\'e}rez-Fern{\'a}ndez}, {P{\'e}rez-R{\`a}fols}, {Pieri}, {Poppett}, {Porredon}, {Pothier}, {Prada}, {Pucha}, {Raichoor}, {Ram{\'\i}rez-P{\'e}rez}, {Ramirez-Solano}, {Rashkovetskyi}, {Ravoux}, {Rocher}, {Rockosi}, {Ross}, {Rossi}, {Ruggeri}, {Ruhlmann-Kleider}, {Sabiu}, {Said}, {Saintonge}, {Samushia}, {Sanchez}, {Saulder}, {Schaan}, {Schlafly}, {Schlegel}, {Scholte}, {Schubnell}, {Seo}, {Shafieloo}, {Sharples}, {Sheu}, {Silber}, {Sinigaglia}, {Siudek}, {Slepian}, {Smith}, {Soumagnac}, {Sprayberry}, {Stephey}, {Su{\'a}rez-P{\'e}rez}, {Sun}, {Tan}, {Tarl{\'e}}, {Tojeiro}, {Ure{\~n}a-L{\'o}pez}, {Vaisakh}, {Valcin}, {Valdes}, {Valluri}, {Vargas-Maga{\~n}a}, {Variu}, {Verde}, {Walther}, {Wang}, {Wang}, {Weaver},
  {Weaverdyck}, {Wechsler}, {White}, {Xie}, {Yang}, {Y{\`e}che}, {Yu}, {Yuan}, {Zhang}, {Zhang}, {Zhao}, {Zheng}, {Zhou}, {Zhou}, {Zou}, {Zou}, \& {Zu}}]{2024AJ....168...58D}
---. 2024, \aj, 168, 58, \dodoi{10.3847/1538-3881/ad3217}

\bibitem[{{Diamond-Stanic} {et~al.}(2009){Diamond-Stanic}, {Rieke}, \& {Rigby}}]{2009ApJ...698..623D}
{Diamond-Stanic}, A.~M., {Rieke}, G.~H., \& {Rigby}, J.~R. 2009, \apj, 698, 623, \dodoi{10.1088/0004-637X/698/1/623}

\bibitem[{{Done}(2010)}]{2010arXiv1008.2287D}
{Done}, C. 2010, arXiv e-prints, arXiv:1008.2287, \dodoi{10.48550/arXiv.1008.2287}

\bibitem[{{Drake} {et~al.}(2009){Drake}, {Djorgovski}, {Mahabal}, {Beshore}, {Larson}, {Graham}, {Williams}, {Christensen}, {Catelan}, {Boattini}, {Gibbs}, {Hill}, \& {Kowalski}}]{2009ApJ...696..870D}
{Drake}, A.~J., {Djorgovski}, S.~G., {Mahabal}, A., {et~al.} 2009, \apj, 696, 870, \dodoi{10.1088/0004-637X/696/1/870}

\bibitem[{{Duras} {et~al.}(2020){Duras}, {Bongiorno}, {Ricci}, {Piconcelli}, {Shankar}, {Lusso}, {Bianchi}, {Fiore}, {Maiolino}, {Marconi}, {Onori}, {Sani}, {Schneider}, {Vignali}, \& {La Franca}}]{2020A&A...636A..73D}
{Duras}, F., {Bongiorno}, A., {Ricci}, F., {et~al.} 2020, \aap, 636, A73, \dodoi{10.1051/0004-6361/201936817}

\bibitem[{{Fabian} {et~al.}(2017){Fabian}, {Alston}, {Cackett}, {Kara}, {Uttley}, \& {Wilkins}}]{2017AN....338..269F}
{Fabian}, A.~C., {Alston}, W.~N., {Cackett}, E.~M., {et~al.} 2017, Astronomische Nachrichten, 338, 269, \dodoi{10.1002/asna.201713341}

\bibitem[{{Fabricant} {et~al.}(2005){Fabricant}, {Fata}, {Roll}, {Hertz}, {Caldwell}, {Gauron}, {Geary}, {McLeod}, {Szentgyorgyi}, {Zajac}, {Kurtz}, {Barberis}, {Bergner}, {Brown}, {Conroy}, {Eng}, {Geller}, {Goddard}, {Honsa}, {Mueller}, {Mink}, {Ordway}, {Tokarz}, {Woods}, {Wyatt}, {Epps}, \& {Dell'Antonio}}]{2005PASP..117.1411F}
{Fabricant}, D., {Fata}, R., {Roll}, J., {et~al.} 2005, \pasp, 117, 1411, \dodoi{10.1086/497385}

\bibitem[{{Fitzpatrick} {et~al.}(2019){Fitzpatrick}, {Massa}, {Gordon}, {Bohlin}, \& {Clayton}}]{2019ApJ...886..108F}
{Fitzpatrick}, E.~L., {Massa}, D., {Gordon}, K.~D., {Bohlin}, R., \& {Clayton}, G.~C. 2019, \apj, 886, 108, \dodoi{10.3847/1538-4357/ab4c3a}

\bibitem[{{Flewelling} {et~al.}(2020){Flewelling}, {Magnier}, {Chambers}, {Heasley}, {Holmberg}, {Huber}, {Sweeney}, {Waters}, {Calamida}, {Casertano}, {Chen}, {Farrow}, {Hasinger}, {Henderson}, {Long}, {Metcalfe}, {Narayan}, {Nieto-Santisteban}, {Norberg}, {Rest}, {Saglia}, {Szalay}, {Thakar}, {Tonry}, {Valenti}, {Werner}, {White}, {Denneau}, {Draper}, {Hodapp}, {Jedicke}, {Kaiser}, {Kudritzki}, {Price}, {Wainscoat}, {Chastel}, {McLean}, {Postman}, \& {Shiao}}]{2020ApJS..251....7F}
{Flewelling}, H.~A., {Magnier}, E.~A., {Chambers}, K.~C., {et~al.} 2020, \apjs, 251, 7, \dodoi{10.3847/1538-4365/abb82d}

\bibitem[{{Fruscione} {et~al.}(2006){Fruscione}, {McDowell}, {Allen}, {Brickhouse}, {Burke}, {Davis}, {Durham}, {Elvis}, {Galle}, {Harris}, {Huenemoerder}, {Houck}, {Ishibashi}, {Karovska}, {Nicastro}, {Noble}, {Nowak}, {Primini}, {Siemiginowska}, {Smith}, \& {Wise}}]{2006SPIE.6270E..1VF}
{Fruscione}, A., {McDowell}, J.~C., {Allen}, G.~E., {et~al.} 2006, in Society of Photo-Optical Instrumentation Engineers (SPIE) Conference Series, Vol. 6270, Observatory Operations: Strategies, Processes, and Systems, ed. D.~R. {Silva} \& R.~E. {Doxsey}, 62701V, \dodoi{10.1117/12.671760}

\bibitem[{{Fu}(2021)}]{2021zndo...5810042F}
{Fu}, Y. 2021, {QSOFITMORE: a python package for fitting UV-optical spectra of quasars}, v1.1.0,  Zenodo, \dodoi{10.5281/zenodo.5810042}

\bibitem[{{Fu} {et~al.}(2022){Fu}, {Wu}, {Jiang}, {Zhang}, {Huo}, {Ai}, {Yang}, {Ma}, {Feng}, {Joshi}, {Hon}, {Wolf}, {Li}, {Jin}, {Yao}, {Pang}, {Wang}, {Lu}, {Wang}, {Zheng}, {Xu}, {Yu}, {Lun}, \& {Zuo}}]{2022ApJS..261...32F}
{Fu}, Y., {Wu}, X.-B., {Jiang}, L., {et~al.} 2022, \apjs, 261, 32, \dodoi{10.3847/1538-4365/ac7f3e}

\bibitem[{{Gabriel} {et~al.}(2004){Gabriel}, {Denby}, {Fyfe}, {Hoar}, {Ibarra}, {Ojero}, {Osborne}, {Saxton}, {Lammers}, \& {Vacanti}}]{2004ASPC..314..759G}
{Gabriel}, C., {Denby}, M., {Fyfe}, D.~J., {et~al.} 2004, in Astronomical Society of the Pacific Conference Series, Vol. 314, Astronomical Data Analysis Software and Systems (ADASS) XIII, ed. F.~{Ochsenbein}, M.~G. {Allen}, \& D.~{Egret}, 759

\bibitem[{{Gallo} {et~al.}(2023){Gallo}, {Miller}, \& {Costantini}}]{2023arXiv230210930G}
{Gallo}, L.~C., {Miller}, J.~M., \& {Costantini}, E. 2023, arXiv e-prints, arXiv:2302.10930, \dodoi{10.48550/arXiv.2302.10930}

\bibitem[{{Gezari}(2021)}]{2021ARA&A..59...21G}
{Gezari}, S. 2021, \araa, 59, 21, \dodoi{10.1146/annurev-astro-111720-030029}

\bibitem[{{Gibson} \& {Brandt}(2012)}]{2012ApJ...746...54G}
{Gibson}, R.~R., \& {Brandt}, W.~N. 2012, \apj, 746, 54, \dodoi{10.1088/0004-637X/746/1/54}

\bibitem[{{Gilfanov} \& {Merloni}(2014)}]{2014SSRv..183..121G}
{Gilfanov}, M., \& {Merloni}, A. 2014, \ssr, 183, 121, \dodoi{10.1007/s11214-014-0071-5}

\bibitem[{{G{\'o}rski} \& {Hivon}(2011)}]{2011ascl.soft07018G}
{G{\'o}rski}, K.~M., \& {Hivon}, E. 2011, {HEALPix: Hierarchical Equal Area isoLatitude Pixelization of a sphere}, Astrophysics Source Code Library, record ascl:1107.018

\bibitem[{{Green}(2018)}]{2018JOSS....3..695G}
{Green}, G.~M. 2018, The Journal of Open Source Software, 3, 695, \dodoi{10.21105/joss.00695}

\bibitem[{{Green} {et~al.}(2022){Green}, {Pulgarin-Duque}, {Anderson}, {MacLeod}, {Eracleous}, {Ruan}, {Runnoe}, {Graham}, {Roulston}, {Schneider}, {Ahlf}, {Bizyaev}, {Brownstein}, {del Casal}, {Dodd}, {Hoover}, {Matt}, {Merloni}, {Pan}, {Ramirez}, {Ridder}, \& {Moseley}}]{2022ApJ...933..180G}
{Green}, P.~J., {Pulgarin-Duque}, L., {Anderson}, S.~F., {et~al.} 2022, \apj, 933, 180, \dodoi{10.3847/1538-4357/ac743f}

\bibitem[{{Grupe} {et~al.}(2007){Grupe}, {Komossa}, \& {Gallo}}]{2007ApJ...668L.111G}
{Grupe}, D., {Komossa}, S., \& {Gallo}, L.~C. 2007, \apjl, 668, L111, \dodoi{10.1086/523042}

\bibitem[{{Gu} \& {Cao}(2009)}]{2009MNRAS.399..349G}
{Gu}, M., \& {Cao}, X. 2009, \mnras, 399, 349, \dodoi{10.1111/j.1365-2966.2009.15277.x}

\bibitem[{{Guainazzi}(2002)}]{2002MNRAS.329L..13G}
{Guainazzi}, M. 2002, \mnras, 329, L13, \dodoi{10.1046/j.1365-8711.2002.05132.x}

\bibitem[{{Guo} {et~al.}(2018){Guo}, {Shen}, \& {Wang}}]{2018ascl.soft09008G}
{Guo}, H., {Shen}, Y., \& {Wang}, S. 2018, {PyQSOFit: Python code to fit the spectrum of quasars}, Astrophysics Source Code Library, record ascl:1809.008.
\newblock \doeprint{1809.008}

\bibitem[{{Guo} {et~al.}(2020){Guo}, {Shen}, {He}, {Wang}, {Liu}, {Wang}, {Sun}, {Yang}, {Kong}, \& {Sheng}}]{2020ApJ...888...58G}
{Guo}, H., {Shen}, Y., {He}, Z., {et~al.} 2020, \apj, 888, 58, \dodoi{10.3847/1538-4357/ab5db0}

\bibitem[{{Guo} {et~al.}(2024){Guo}, {Zou}, {Fawcett}, {Canning}, {Juneau}, {Davis}, {Alexander}, {Jiang}, {Aguilar}, {Ahlen}, {Brooks}, {Claybaugh}, {de la Macorra}, {Doel}, {Fanning}, {Forero-Romero}, {Gontcho A Gontcho}, {Honscheid}, {Kisner}, {Kremin}, {Landriau}, {Meisner}, {Miquel}, {Moustakas}, {Nie}, {Pan}, {Poppett}, {Prada}, {Rezaie}, {Rossi}, {Siudek}, {Sanchez}, {Schubnell}, {Seo}, {Sui}, {Tarl{\'e}}, \& {Zhou}}]{2024ApJS..270...26G}
{Guo}, W.-J., {Zou}, H., {Fawcett}, V.~A., {et~al.} 2024, \apjs, 270, 26, \dodoi{10.3847/1538-4365/ad118a}

\bibitem[{Harris {et~al.}(2020)Harris, Millman, van~der Walt, Gommers, Virtanen, Cournapeau, Wieser, Taylor, Berg, Smith, Kern, Picus, Hoyer, van Kerkwijk, Brett, Haldane, del R{\'{i}}o, Wiebe, Peterson, G{\'{e}}rard-Marchant, Sheppard, Reddy, Weckesser, Abbasi, Gohlke, \& Oliphant}]{Harris2020}
Harris, C.~R., Millman, K.~J., van~der Walt, S.~J., {et~al.} 2020, Nature, 585, 357, \dodoi{10.1038/s41586-020-2649-2}

\bibitem[{{Huang} {et~al.}(2024){Huang}, {Luo}, {Brandt}, {Chen}, {Ni}, {Xue}, \& {Zhang}}]{2024arXiv241206923H}
{Huang}, J., {Luo}, B., {Brandt}, W.~N., {et~al.} 2024, arXiv e-prints, arXiv:2412.06923, \dodoi{10.48550/arXiv.2412.06923}

\bibitem[{{Huang} {et~al.}(2023){Huang}, {Luo}, {Brandt}, {Du}, {Garmire}, {Hu}, {Liu}, {Ni}, \& {Wang}}]{2023ApJ...950...18H}
---. 2023, \apj, 950, 18, \dodoi{10.3847/1538-4357/accd64}

\bibitem[{{Hudelot} {et~al.}(2012){Hudelot}, {Cuillandre}, {Withington}, {Goranova}, {McCracken}, {Magnard}, {Mellier}, {Regnault}, {Betoule}, {Aussel}, {Kavelaars}, {Fernique}, {Bonnarel}, {Ochsenbein}, \& {Ilbert}}]{2012yCat.2317....0H}
{Hudelot}, P., {Cuillandre}, J.~C., {Withington}, K., {et~al.} 2012, VizieR Online Data Catalog, II/317

\bibitem[{Hunter(2007)}]{Hunter2007}
Hunter, J.~D. 2007, Computing in Science \& Engineering, 9, 90, \dodoi{10.1109/MCSE.2007.55}

\bibitem[{{Ivezi{\'c}} {et~al.}(2019){Ivezi{\'c}}, {Kahn}, {Tyson}, {Abel}, {Acosta}, {Allsman}, {Alonso}, {AlSayyad}, {Anderson}, {Andrew}, {Angel}, {Angeli}, {Ansari}, {Antilogus}, {Araujo}, {Armstrong}, {Arndt}, {Astier}, {Aubourg}, {Auza}, {Axelrod}, {Bard}, {Barr}, {Barrau}, {Bartlett}, {Bauer}, {Bauman}, {Baumont}, {Bechtol}, {Bechtol}, {Becker}, {Becla}, {Beldica}, {Bellavia}, {Bianco}, {Biswas}, {Blanc}, {Blazek}, {Blandford}, {Bloom}, {Bogart}, {Bond}, {Booth}, {Borgland}, {Borne}, {Bosch}, {Boutigny}, {Brackett}, {Bradshaw}, {Brandt}, {Brown}, {Bullock}, {Burchat}, {Burke}, {Cagnoli}, {Calabrese}, {Callahan}, {Callen}, {Carlin}, {Carlson}, {Chandrasekharan}, {Charles-Emerson}, {Chesley}, {Cheu}, {Chiang}, {Chiang}, {Chirino}, {Chow}, {Ciardi}, {Claver}, {Cohen-Tanugi}, {Cockrum}, {Coles}, {Connolly}, {Cook}, {Cooray}, {Covey}, {Cribbs}, {Cui}, {Cutri}, {Daly}, {Daniel}, {Daruich}, {Daubard}, {Daues}, {Dawson}, {Delgado}, {Dellapenna}, {de Peyster}, {de Val-Borro}, {Digel}, {Doherty}, {Dubois},
  {Dubois-Felsmann}, {Durech}, {Economou}, {Eifler}, {Eracleous}, {Emmons}, {Fausti Neto}, {Ferguson}, {Figueroa}, {Fisher-Levine}, {Focke}, {Foss}, {Frank}, {Freemon}, {Gangler}, {Gawiser}, {Geary}, {Gee}, {Geha}, {Gessner}, {Gibson}, {Gilmore}, {Glanzman}, {Glick}, {Goldina}, {Goldstein}, {Goodenow}, {Graham}, {Gressler}, {Gris}, {Guy}, {Guyonnet}, {Haller}, {Harris}, {Hascall}, {Haupt}, {Hernandez}, {Herrmann}, {Hileman}, {Hoblitt}, {Hodgson}, {Hogan}, {Howard}, {Huang}, {Huffer}, {Ingraham}, {Innes}, {Jacoby}, {Jain}, {Jammes}, {Jee}, {Jenness}, {Jernigan}, {Jevremovi{\'c}}, {Johns}, {Johnson}, {Johnson}, {Jones}, {Juramy-Gilles}, {Juri{\'c}}, {Kalirai}, {Kallivayalil}, {Kalmbach}, {Kantor}, {Karst}, {Kasliwal}, {Kelly}, {Kessler}, {Kinnison}, {Kirkby}, {Knox}, {Kotov}, {Krabbendam}, {Krughoff}, {Kub{\'a}nek}, {Kuczewski}, {Kulkarni}, {Ku}, {Kurita}, {Lage}, {Lambert}, {Lange}, {Langton}, {Le Guillou}, {Levine}, {Liang}, {Lim}, {Lintott}, {Long}, {Lopez}, {Lotz}, {Lupton}, {Lust}, {MacArthur}, {Mahabal},
  {Mandelbaum}, {Markiewicz}, {Marsh}, {Marshall}, {Marshall}, {May}, {McKercher}, {McQueen}, {Meyers}, {Migliore}, {Miller}, {Mills}, {Miraval}, {Moeyens}, {Moolekamp}, {Monet}, {Moniez}, {Monkewitz}, {Montgomery}, {Morrison}, {Mueller}, {Muller}, {Mu{\~n}oz Arancibia}, {Neill}, {Newbry}, {Nief}, {Nomerotski}, {Nordby}, {O'Connor}, {Oliver}, {Olivier}, {Olsen}, {O'Mullane}, {Ortiz}, {Osier}, {Owen}, {Pain}, {Palecek}, {Parejko}, {Parsons}, {Pease}, {Peterson}, {Peterson}, {Petravick}, {Libby Petrick}, {Petry}, {Pierfederici}, {Pietrowicz}, {Pike}, {Pinto}, {Plante}, {Plate}, {Plutchak}, {Price}, {Prouza}, {Radeka}, {Rajagopal}, {Rasmussen}, {Regnault}, {Reil}, {Reiss}, {Reuter}, {Ridgway}, {Riot}, {Ritz}, {Robinson}, {Roby}, {Roodman}, {Rosing}, {Roucelle}, {Rumore}, {Russo}, {Saha}, {Sassolas}, {Schalk}, {Schellart}, {Schindler}, {Schmidt}, {Schneider}, {Schneider}, {Schoening}, {Schumacher}, {Schwamb}, {Sebag}, {Selvy}, {Sembroski}, {Seppala}, {Serio}, {Serrano}, {Shaw}, {Shipsey}, {Sick}, {Silvestri},
  {Slater}, {Smith}, {Smith}, {Sobhani}, {Soldahl}, {Storrie-Lombardi}, {Stover}, {Strauss}, {Street}, {Stubbs}, {Sullivan}, {Sweeney}, {Swinbank}, {Szalay}, {Takacs}, {Tether}, {Thaler}, {Thayer}, {Thomas}, {Thornton}, {Thukral}, {Tice}, {Trilling}, {Turri}, {Van Berg}, {Vanden Berk}, {Vetter}, {Virieux}, {Vucina}, {Wahl}, {Walkowicz}, {Walsh}, {Walter}, {Wang}, {Wang}, {Warner}, {Wiecha}, {Willman}, {Winters}, {Wittman}, {Wolff}, {Wood-Vasey}, {Wu}, {Xin}, {Yoachim}, \& {Zhan}}]{2019ApJ...873..111I}
{Ivezi{\'c}}, {\v{Z}}., {Kahn}, S.~M., {Tyson}, J.~A., {et~al.} 2019, \apj, 873, 111, \dodoi{10.3847/1538-4357/ab042c}

\bibitem[{{Jarvis} {et~al.}(2013){Jarvis}, {Bonfield}, {Bruce}, {Geach}, {McAlpine}, {McLure}, {Gonz{\'a}lez-Solares}, {Irwin}, {Lewis}, {Yoldas}, {Andreon}, {Cross}, {Emerson}, {Dalton}, {Dunlop}, {Hodgkin}, {Le}, {Karouzos}, {Meisenheimer}, {Oliver}, {Rawlings}, {Simpson}, {Smail}, {Smith}, {Sullivan}, {Sutherland}, {White}, \& {Zwart}}]{2013MNRAS.428.1281J}
{Jarvis}, M.~J., {Bonfield}, D.~G., {Bruce}, V.~A., {et~al.} 2013, \mnras, 428, 1281, \dodoi{10.1093/mnras/sts118}

\bibitem[{{Jiang} {et~al.}(2010){Jiang}, {Fan}, {Brandt}, {Carilli}, {Egami}, {Hines}, {Kurk}, {Richards}, {Shen}, {Strauss}, {Vestergaard}, \& {Walter}}]{2010Natur.464..380J}
{Jiang}, L., {Fan}, X., {Brandt}, W.~N., {et~al.} 2010, \nat, 464, 380, \dodoi{10.1038/nature08877}

\bibitem[{{Jiang} {et~al.}(2014){Jiang}, {Stone}, \& {Davis}}]{2014ApJ...796..106J}
{Jiang}, Y.-F., {Stone}, J.~M., \& {Davis}, S.~W. 2014, \apj, 796, 106, \dodoi{10.1088/0004-637X/796/2/106}

\bibitem[{{Jiang} {et~al.}(2019){Jiang}, {Stone}, \& {Davis}}]{2019ApJ...880...67J}
---. 2019, \apj, 880, 67, \dodoi{10.3847/1538-4357/ab29ff}

\bibitem[{{Just} {et~al.}(2007){Just}, {Brandt}, {Shemmer}, {Steffen}, {Schneider}, {Chartas}, \& {Garmire}}]{2007ApJ...665.1004J}
{Just}, D.~W., {Brandt}, W.~N., {Shemmer}, O., {et~al.} 2007, \apj, 665, 1004, \dodoi{10.1086/519990}

\bibitem[{{Kara} {et~al.}(2017){Kara}, {Garc{\'\i}a}, {Lohfink}, {Fabian}, {Reynolds}, {Tombesi}, \& {Wilkins}}]{2017MNRAS.468.3489K}
{Kara}, E., {Garc{\'\i}a}, J.~A., {Lohfink}, A., {et~al.} 2017, \mnras, 468, 3489, \dodoi{10.1093/mnras/stx792}

\bibitem[{{Kaur} \& {Stone}(2024)}]{2024arXiv240518500K}
{Kaur}, K., \& {Stone}, N.~C. 2024, arXiv e-prints, arXiv:2405.18500, \dodoi{10.48550/arXiv.2405.18500}

\bibitem[{{Kelly} {et~al.}(2009){Kelly}, {Bechtold}, \& {Siemiginowska}}]{2009ApJ...698..895K}
{Kelly}, B.~C., {Bechtold}, J., \& {Siemiginowska}, A. 2009, \apj, 698, 895, \dodoi{10.1088/0004-637X/698/1/895}

\bibitem[{{Krawczyk} {et~al.}(2013){Krawczyk}, {Richards}, {Mehta}, {Vogeley}, {Gallagher}, {Leighly}, {Ross}, \& {Schneider}}]{2013ApJS..206....4K}
{Krawczyk}, C.~M., {Richards}, G.~T., {Mehta}, S.~S., {et~al.} 2013, \apjs, 206, 4, \dodoi{10.1088/0067-0049/206/1/4}

\bibitem[{{Krivonos} {et~al.}(2024){Krivonos}, {Gilfanov}, {Medvedev}, {Sazonov}, \& {Sunyaev}}]{2024MNRAS.528.1264K}
{Krivonos}, R., {Gilfanov}, M., {Medvedev}, P., {Sazonov}, S., \& {Sunyaev}, R. 2024, \mnras, 528, 1264, \dodoi{10.1093/mnras/stae105}

\bibitem[{{LaMassa} {et~al.}(2019){LaMassa}, {Georgakakis}, {Vivek}, {Salvato}, {Ananna}, {Urry}, {MacLeod}, \& {Ross}}]{2019ApJ...876...50L}
{LaMassa}, S.~M., {Georgakakis}, A., {Vivek}, M., {et~al.} 2019, \apj, 876, 50, \dodoi{10.3847/1538-4357/ab108b}

\bibitem[{{LaMassa} {et~al.}(2015){LaMassa}, {Cales}, {Moran}, {Myers}, {Richards}, {Eracleous}, {Heckman}, {Gallo}, \& {Urry}}]{2015ApJ...800..144L}
{LaMassa}, S.~M., {Cales}, S., {Moran}, E.~C., {et~al.} 2015, \apj, 800, 144, \dodoi{10.1088/0004-637X/800/2/144}

\bibitem[{{Laureijs} {et~al.}(2011){Laureijs}, {Amiaux}, {Arduini}, {Augu{\`e}res}, {Brinchmann}, {Cole}, {Cropper}, {Dabin}, {Duvet}, {Ealet}, {Garilli}, {Gondoin}, {Guzzo}, {Hoar}, {Hoekstra}, {Holmes}, {Kitching}, {Maciaszek}, {Mellier}, {Pasian}, {Percival}, {Rhodes}, {Saavedra Criado}, {Sauvage}, {Scaramella}, {Valenziano}, {Warren}, {Bender}, {Castander}, {Cimatti}, {Le F{\`e}vre}, {Kurki-Suonio}, {Levi}, {Lilje}, {Meylan}, {Nichol}, {Pedersen}, {Popa}, {Rebolo Lopez}, {Rix}, {Rottgering}, {Zeilinger}, {Grupp}, {Hudelot}, {Massey}, {Meneghetti}, {Miller}, {Paltani}, {Paulin-Henriksson}, {Pires}, {Saxton}, {Schrabback}, {Seidel}, {Walsh}, {Aghanim}, {Amendola}, {Bartlett}, {Baccigalupi}, {Beaulieu}, {Benabed}, {Cuby}, {Elbaz}, {Fosalba}, {Gavazzi}, {Helmi}, {Hook}, {Irwin}, {Kneib}, {Kunz}, {Mannucci}, {Moscardini}, {Tao}, {Teyssier}, {Weller}, {Zamorani}, {Zapatero Osorio}, {Boulade}, {Foumond}, {Di Giorgio}, {Guttridge}, {James}, {Kemp}, {Martignac}, {Spencer}, {Walton}, {Bl{\"u}mchen}, {Bonoli},
  {Bortoletto}, {Cerna}, {Corcione}, {Fabron}, {Jahnke}, {Ligori}, {Madrid}, {Martin}, {Morgante}, {Pamplona}, {Prieto}, {Riva}, {Toledo}, {Trifoglio}, {Zerbi}, {Abdalla}, {Douspis}, {Grenet}, {Borgani}, {Bouwens}, {Courbin}, {Delouis}, {Dubath}, {Fontana}, {Frailis}, {Grazian}, {Koppenh{\"o}fer}, {Mansutti}, {Melchior}, {Mignoli}, {Mohr}, {Neissner}, {Noddle}, {Poncet}, {Scodeggio}, {Serrano}, {Shane}, {Starck}, {Surace}, {Taylor}, {Verdoes-Kleijn}, {Vuerli}, {Williams}, {Zacchei}, {Altieri}, {Escudero Sanz}, {Kohley}, {Oosterbroek}, {Astier}, {Bacon}, {Bardelli}, {Baugh}, {Bellagamba}, {Benoist}, {Bianchi}, {Biviano}, {Branchini}, {Carbone}, {Cardone}, {Clements}, {Colombi}, {Conselice}, {Cresci}, {Deacon}, {Dunlop}, {Fedeli}, {Fontanot}, {Franzetti}, {Giocoli}, {Garcia-Bellido}, {Gow}, {Heavens}, {Hewett}, {Heymans}, {Holland}, {Huang}, {Ilbert}, {Joachimi}, {Jennins}, {Kerins}, {Kiessling}, {Kirk}, {Kotak}, {Krause}, {Lahav}, {van Leeuwen}, {Lesgourgues}, {Lombardi}, {Magliocchetti}, {Maguire},
  {Majerotto}, {Maoli}, {Marulli}, {Maurogordato}, {McCracken}, {McLure}, {Melchiorri}, {Merson}, {Moresco}, {Nonino}, {Norberg}, {Peacock}, {Pello}, {Penny}, {Pettorino}, {Di Porto}, {Pozzetti}, {Quercellini}, {Radovich}, {Rassat}, {Roche}, {Ronayette}, {Rossetti}, {Sartoris}, {Schneider}, {Semboloni}, {Serjeant}, {Simpson}, {Skordis}, {Smadja}, {Smartt}, {Spano}, {Spiro}, {Sullivan}, {Tilquin}, {Trotta}, {Verde}, {Wang}, {Williger}, {Zhao}, {Zoubian}, \& {Zucca}}]{2011arXiv1110.3193L}
{Laureijs}, R., {Amiaux}, J., {Arduini}, S., {et~al.} 2011, arXiv e-prints, arXiv:1110.3193, \dodoi{10.48550/arXiv.1110.3193}

\bibitem[{{Li} {et~al.}(2023){Li}, {Shen}, {Ho}, {Brandt}, {Grier}, {Hall}, {Homayouni}, {Koekemoer}, {Schneider}, \& {Trump}}]{2023ApJ...954..173L}
{Li}, J. I.~H., {Shen}, Y., {Ho}, L.~C., {et~al.} 2023, \apj, 954, 173, \dodoi{10.3847/1538-4357/acddda}

\bibitem[{{Liu} {et~al.}(2022){Liu}, {Luo}, {Brandt}, {Huang}, {Pu}, {Yi}, \& {Yu}}]{2022ApJ...930...53L}
{Liu}, H., {Luo}, B., {Brandt}, W.~N., {et~al.} 2022, \apj, 930, 53, \dodoi{10.3847/1538-4357/ac6265}

\bibitem[{{Liu} {et~al.}(2020){Liu}, {Li}, {Liu}, {Lu}, {Yuan}, {Dou}, \& {Shen}}]{2020ApJ...894...93L}
{Liu}, Z., {Li}, D., {Liu}, H.-Y., {et~al.} 2020, \apj, 894, 93, \dodoi{10.3847/1538-4357/ab880f}

\bibitem[{{Luo} {et~al.}(2015){Luo}, {Brandt}, {Hall}, {Wu}, {Anderson}, {Garmire}, {Gibson}, {Plotkin}, {Richards}, {Schneider}, {Shemmer}, \& {Shen}}]{2015ApJ...805..122L}
{Luo}, B., {Brandt}, W.~N., {Hall}, P.~B., {et~al.} 2015, \apj, 805, 122, \dodoi{10.1088/0004-637X/805/2/122}

\bibitem[{{Luo} {et~al.}(2017){Luo}, {Brandt}, {Xue}, {Lehmer}, {Alexander}, {Bauer}, {Vito}, {Yang}, {Basu-Zych}, {Comastri}, {Gilli}, {Gu}, {Hornschemeier}, {Koekemoer}, {Liu}, {Mainieri}, {Paolillo}, {Ranalli}, {Rosati}, {Schneider}, {Shemmer}, {Smail}, {Sun}, {Tozzi}, {Vignali}, \& {Wang}}]{2017ApJS..228....2L}
{Luo}, B., {Brandt}, W.~N., {Xue}, Y.~Q., {et~al.} 2017, \apjs, 228, 2, \dodoi{10.3847/1538-4365/228/1/2}

\bibitem[{{Lusso} \& {Risaliti}(2017)}]{2017A&A...602A..79L}
{Lusso}, E., \& {Risaliti}, G. 2017, \aap, 602, A79, \dodoi{10.1051/0004-6361/201630079}

\bibitem[{{Lyons}(1991)}]{1991pgda.book.....L}
{Lyons}, L. 1991, {A Practical Guide to Data Analysis for Physical Science Students}

\bibitem[{{Lyu} {et~al.}(2017){Lyu}, {Rieke}, \& {Shi}}]{2017ApJ...835..257L}
{Lyu}, J., {Rieke}, G.~H., \& {Shi}, Y. 2017, \apj, 835, 257, \dodoi{10.3847/1538-4357/835/2/257}

\bibitem[{{Lyu} {et~al.}(2023){Lyu}, {Alberts}, {Rieke}, {Shivaei}, {Perez-Gonzalez}, {Sun}, {Hainline}, {Baum}, {Bonaventura}, {Bunker}, {Egami}, {Eisenstein}, {Florian}, {Ji}, {Johnson}, {Morrison}, {Rieke}, {Robertson}, {Rujopakarn}, {Tacchella}, {Scholtz}, \& {Willmer}}]{2023arXiv231012330L}
{Lyu}, J., {Alberts}, S., {Rieke}, G.~H., {et~al.} 2023, arXiv e-prints, arXiv:2310.12330, \dodoi{10.48550/arXiv.2310.12330}

\bibitem[{{MacLeod} {et~al.}(2010){MacLeod}, {Ivezi{\'c}}, {Kochanek}, {Koz{\l}owski}, {Kelly}, {Bullock}, {Kimball}, {Sesar}, {Westman}, {Brooks}, {Gibson}, {Becker}, \& {de Vries}}]{2010ApJ...721.1014M}
{MacLeod}, C.~L., {Ivezi{\'c}}, {\v{Z}}., {Kochanek}, C.~S., {et~al.} 2010, ApJ, 721, 1014, \dodoi{10.1088/0004-637X/721/2/1014}

\bibitem[{{MacLeod} {et~al.}(2019){MacLeod}, {Green}, {Anderson}, {Bruce}, {Eracleous}, {Graham}, {Homan}, {Lawrence}, {LeBleu}, {Ross}, {Ruan}, {Runnoe}, {Stern}, {Burgett}, {Chambers}, {Kaiser}, {Magnier}, \& {Metcalfe}}]{2019ApJ...874....8M}
{MacLeod}, C.~L., {Green}, P.~J., {Anderson}, S.~F., {et~al.} 2019, \apj, 874, 8, \dodoi{10.3847/1538-4357/ab05e2}

\bibitem[{{Mainzer} {et~al.}(2011){Mainzer}, {Bauer}, {Grav}, {Masiero}, {Cutri}, {Dailey}, {Eisenhardt}, {McMillan}, {Wright}, {Walker}, {Jedicke}, {Spahr}, {Tholen}, {Alles}, {Beck}, {Brandenburg}, {Conrow}, {Evans}, {Fowler}, {Jarrett}, {Marsh}, {Masci}, {McCallon}, {Wheelock}, {Wittman}, {Wyatt}, {DeBaun}, {Elliott}, {Elsbury}, {Gautier}, {Gomillion}, {Leisawitz}, {Maleszewski}, {Micheli}, \& {Wilkins}}]{2011ApJ...731...53M}
{Mainzer}, A., {Bauer}, J., {Grav}, T., {et~al.} 2011, \apj, 731, 53, \dodoi{10.1088/0004-637X/731/1/53}

\bibitem[{{Margala} {et~al.}(2016){Margala}, {Kirkby}, {Dawson}, {Bailey}, {Blanton}, \& {Schneider}}]{2016ApJ...831..157M}
{Margala}, D., {Kirkby}, D., {Dawson}, K., {et~al.} 2016, \apj, 831, 157, \dodoi{10.3847/0004-637X/831/2/157}

\bibitem[{{Markowitz} {et~al.}(2014){Markowitz}, {Krumpe}, \& {Nikutta}}]{2014MNRAS.439.1403M}
{Markowitz}, A.~G., {Krumpe}, M., \& {Nikutta}, R. 2014, \mnras, 439, 1403, \dodoi{10.1093/mnras/stt2492}

\bibitem[{{Martin} {et~al.}(2005){Martin}, {Fanson}, {Schiminovich}, {Morrissey}, {Friedman}, {Barlow}, {Conrow}, {Grange}, {Jelinsky}, {Milliard}, {Siegmund}, {Bianchi}, {Byun}, {Donas}, {Forster}, {Heckman}, {Lee}, {Madore}, {Malina}, {Neff}, {Rich}, {Small}, {Surber}, {Szalay}, {Welsh}, \& {Wyder}}]{2005ApJ...619L...1M}
{Martin}, D.~C., {Fanson}, J., {Schiminovich}, D., {et~al.} 2005, \apjl, 619, L1, \dodoi{10.1086/426387}

\bibitem[{{Masci} {et~al.}(2019){Masci}, {Laher}, {Rusholme}, {Shupe}, {Groom}, {Surace}, {Jackson}, {Monkewitz}, {Beck}, {Flynn}, {Terek}, {Landry}, {Hacopians}, {Desai}, {Howell}, {Brooke}, {Imel}, {Wachter}, {Ye}, {Lin}, {Cenko}, {Cunningham}, {Rebbapragada}, {Bue}, {Miller}, {Mahabal}, {Bellm}, {Patterson}, {Juri{\'c}}, {Golkhou}, {Ofek}, {Walters}, {Graham}, {Kasliwal}, {Dekany}, {Kupfer}, {Burdge}, {Cannella}, {Barlow}, {Van Sistine}, {Giomi}, {Fremling}, {Blagorodnova}, {Levitan}, {Riddle}, {Smith}, {Helou}, {Prince}, \& {Kulkarni}}]{2019PASP..131a8003M}
{Masci}, F.~J., {Laher}, R.~R., {Rusholme}, B., {et~al.} 2019, \pasp, 131, 018003, \dodoi{10.1088/1538-3873/aae8ac}

\bibitem[{{Masters} \& {Capak}(2011)}]{2011PASP..123..638M}
{Masters}, D., \& {Capak}, P. 2011, \pasp, 123, 638, \dodoi{10.1086/660023}

\bibitem[{{Matt} {et~al.}(2003){Matt}, {Guainazzi}, \& {Maiolino}}]{2003MNRAS.342..422M}
{Matt}, G., {Guainazzi}, M., \& {Maiolino}, R. 2003, \mnras, 342, 422, \dodoi{10.1046/j.1365-8711.2003.06539.x}

\bibitem[{{Merloni} {et~al.}(2012){Merloni}, {Predehl}, {Becker}, {B{\"o}hringer}, {Boller}, {Brunner}, {Brusa}, {Dennerl}, {Freyberg}, {Friedrich}, {Georgakakis}, {Haberl}, {Hasinger}, {Meidinger}, {Mohr}, {Nandra}, {Rau}, {Reiprich}, {Robrade}, {Salvato}, {Santangelo}, {Sasaki}, {Schwope}, {Wilms}, \& {German eROSITA Consortium}}]{2012arXiv1209.3114M}
{Merloni}, A., {Predehl}, P., {Becker}, W., {et~al.} 2012, arXiv e-prints, arXiv:1209.3114, \dodoi{10.48550/arXiv.1209.3114}

\bibitem[{{Merloni} {et~al.}(2015){Merloni}, {Dwelly}, {Salvato}, {Georgakakis}, {Greiner}, {Krumpe}, {Nandra}, {Ponti}, \& {Rau}}]{2015MNRAS.452...69M}
{Merloni}, A., {Dwelly}, T., {Salvato}, M., {et~al.} 2015, \mnras, 452, 69, \dodoi{10.1093/mnras/stv1095}

\bibitem[{{Miniutti} {et~al.}(2012){Miniutti}, {Brandt}, {Schneider}, {Fabian}, {Gallo}, \& {Boller}}]{2012MNRAS.425.1718M}
{Miniutti}, G., {Brandt}, W.~N., {Schneider}, D.~P., {et~al.} 2012, \mnras, 425, 1718, \dodoi{10.1111/j.1365-2966.2012.21648.x}

\bibitem[{{Netzer}(2015)}]{2015ARA&A..53..365N}
{Netzer}, H. 2015, \araa, 53, 365, \dodoi{10.1146/annurev-astro-082214-122302}

\bibitem[{{Ni} {et~al.}(2020){Ni}, {Brandt}, {Yi}, {Luo}, {Timlin}, {Hall}, {Liu}, {Plotkin}, {Shemmer}, {Vito}, \& {Wu}}]{2020ApJ...889L..37N}
{Ni}, Q., {Brandt}, W.~N., {Yi}, W., {et~al.} 2020, \apjl, 889, L37, \dodoi{10.3847/2041-8213/ab6d78}

\bibitem[{{Ni} {et~al.}(2021){Ni}, {Brandt}, {Chen}, {Luo}, {Nyland}, {Yang}, {Zou}, {Aird}, {Alexander}, {Bauer}, {Lacy}, {Lehmer}, {Mallick}, {Salvato}, {Schneider}, {Tozzi}, {Traulsen}, {Vaccari}, {Vignali}, {Vito}, {Xue}, {Banerji}, {Chow}, {Comastri}, {Del Moro}, {Gilli}, {Mullaney}, {Paolillo}, {Schwope}, {Shemmer}, {Sun}, {Timlin}, \& {Trump}}]{2021ApJS..256...21N}
{Ni}, Q., {Brandt}, W.~N., {Chen}, C.-T., {et~al.} 2021, \apjs, 256, 21, \dodoi{10.3847/1538-4365/ac0dc6}

\bibitem[{{Ni} {et~al.}(2022){Ni}, {Brandt}, {Luo}, {Garmire}, {Hall}, {Plotkin}, {Shemmer}, {Timlin}, {Vito}, {Wu}, \& {Yi}}]{2022MNRAS.511.5251N}
{Ni}, Q., {Brandt}, W.~N., {Luo}, B., {et~al.} 2022, \mnras, 511, 5251, \dodoi{10.1093/mnras/stac394}

\bibitem[{{Paolillo} {et~al.}(2004){Paolillo}, {Schreier}, {Giacconi}, {Koekemoer}, \& {Grogin}}]{2004ApJ...611...93P}
{Paolillo}, M., {Schreier}, E.~J., {Giacconi}, R., {Koekemoer}, A.~M., \& {Grogin}, N.~A. 2004, \apj, 611, 93, \dodoi{10.1086/421967}

\bibitem[{{Pennell} {et~al.}(2017){Pennell}, {Runnoe}, \& {Brotherton}}]{2017MNRAS.468.1433P}
{Pennell}, A., {Runnoe}, J.~C., \& {Brotherton}, M.~S. 2017, \mnras, 468, 1433, \dodoi{10.1093/mnras/stx556}

\bibitem[{{Planck Collaboration} {et~al.}(2020){Planck Collaboration}, {Aghanim}, {Akrami}, {Ashdown}, {Aumont}, {Baccigalupi}, {Ballardini}, {Banday}, {Barreiro}, {Bartolo}, {Basak}, {Battye}, {Benabed}, {Bernard}, {Bersanelli}, {Bielewicz}, {Bock}, {Bond}, {Borrill}, {Bouchet}, {Boulanger}, {Bucher}, {Burigana}, {Butler}, {Calabrese}, {Cardoso}, {Carron}, {Challinor}, {Chiang}, {Chluba}, {Colombo}, {Combet}, {Contreras}, {Crill}, {Cuttaia}, {de Bernardis}, {de Zotti}, {Delabrouille}, {Delouis}, {Di Valentino}, {Diego}, {Dor{\'e}}, {Douspis}, {Ducout}, {Dupac}, {Dusini}, {Efstathiou}, {Elsner}, {En{\ss}lin}, {Eriksen}, {Fantaye}, {Farhang}, {Fergusson}, {Fernandez-Cobos}, {Finelli}, {Forastieri}, {Frailis}, {Fraisse}, {Franceschi}, {Frolov}, {Galeotta}, {Galli}, {Ganga}, {G{\'e}nova-Santos}, {Gerbino}, {Ghosh}, {Gonz{\'a}lez-Nuevo}, {G{\'o}rski}, {Gratton}, {Gruppuso}, {Gudmundsson}, {Hamann}, {Handley}, {Hansen}, {Herranz}, {Hildebrandt}, {Hivon}, {Huang}, {Jaffe}, {Jones}, {Karakci}, {Keih{\"a}nen},
  {Keskitalo}, {Kiiveri}, {Kim}, {Kisner}, {Knox}, {Krachmalnicoff}, {Kunz}, {Kurki-Suonio}, {Lagache}, {Lamarre}, {Lasenby}, {Lattanzi}, {Lawrence}, {Le Jeune}, {Lemos}, {Lesgourgues}, {Levrier}, {Lewis}, {Liguori}, {Lilje}, {Lilley}, {Lindholm}, {L{\'o}pez-Caniego}, {Lubin}, {Ma}, {Mac{\'\i}as-P{\'e}rez}, {Maggio}, {Maino}, {Mandolesi}, {Mangilli}, {Marcos-Caballero}, {Maris}, {Martin}, {Martinelli}, {Mart{\'\i}nez-Gonz{\'a}lez}, {Matarrese}, {Mauri}, {McEwen}, {Meinhold}, {Melchiorri}, {Mennella}, {Migliaccio}, {Millea}, {Mitra}, {Miville-Desch{\^e}nes}, {Molinari}, {Montier}, {Morgante}, {Moss}, {Natoli}, {N{\o}rgaard-Nielsen}, {Pagano}, {Paoletti}, {Partridge}, {Patanchon}, {Peiris}, {Perrotta}, {Pettorino}, {Piacentini}, {Polastri}, {Polenta}, {Puget}, {Rachen}, {Reinecke}, {Remazeilles}, {Renzi}, {Rocha}, {Rosset}, {Roudier}, {Rubi{\~n}o-Mart{\'\i}n}, {Ruiz-Granados}, {Salvati}, {Sandri}, {Savelainen}, {Scott}, {Shellard}, {Sirignano}, {Sirri}, {Spencer}, {Sunyaev}, {Suur-Uski}, {Tauber}, {Tavagnacco},
  {Tenti}, {Toffolatti}, {Tomasi}, {Trombetti}, {Valenziano}, {Valiviita}, {Van Tent}, {Vibert}, {Vielva}, {Villa}, {Vittorio}, {Wandelt}, {Wehus}, {White}, {White}, {Zacchei}, \& {Zonca}}]{2020A&A...641A...6P}
{Planck Collaboration}, {Aghanim}, N., {Akrami}, Y., {et~al.} 2020, \aap, 641, A6, \dodoi{10.1051/0004-6361/201833910}

\bibitem[{{Predehl} {et~al.}(2021){Predehl}, {Andritschke}, {Arefiev}, {Babyshkin}, {Batanov}, {Becker}, {B{\"o}hringer}, {Bogomolov}, {Boller}, {Borm}, {Bornemann}, {Br{\"a}uninger}, {Br{\"u}ggen}, {Brunner}, {Brusa}, {Bulbul}, {Buntov}, {Burwitz}, {Burkert}, {Clerc}, {Churazov}, {Coutinho}, {Dauser}, {Dennerl}, {Doroshenko}, {Eder}, {Emberger}, {Eraerds}, {Finoguenov}, {Freyberg}, {Friedrich}, {Friedrich}, {F{\"u}rmetz}, {Georgakakis}, {Gilfanov}, {Granato}, {Grossberger}, {Gueguen}, {Gureev}, {Haberl}, {H{\"a}lker}, {Hartner}, {Hasinger}, {Huber}, {Ji}, {Kienlin}, {Kink}, {Korotkov}, {Kreykenbohm}, {Lamer}, {Lomakin}, {Lapshov}, {Liu}, {Maitra}, {Meidinger}, {Menz}, {Merloni}, {Mernik}, {Mican}, {Mohr}, {M{\"u}ller}, {Nandra}, {Nazarov}, {Pacaud}, {Pavlinsky}, {Perinati}, {Pfeffermann}, {Pietschner}, {Ramos-Ceja}, {Rau}, {Reiffers}, {Reiprich}, {Robrade}, {Salvato}, {Sanders}, {Santangelo}, {Sasaki}, {Scheuerle}, {Schmid}, {Schmitt}, {Schwope}, {Shirshakov}, {Steinmetz}, {Stewart}, {Str{\"u}der},
  {Sunyaev}, {Tenzer}, {Tiedemann}, {Tr{\"u}mper}, {Voron}, {Weber}, {Wilms}, \& {Yaroshenko}}]{2021A&A...647A...1P}
{Predehl}, P., {Andritschke}, R., {Arefiev}, V., {et~al.} 2021, \aap, 647, A1, \dodoi{10.1051/0004-6361/202039313}

\bibitem[{{Pu} {et~al.}(2020){Pu}, {Luo}, {Brandt}, {Timlin}, {Liu}, {Ni}, \& {Wu}}]{2020ApJ...900..141P}
{Pu}, X., {Luo}, B., {Brandt}, W.~N., {et~al.} 2020, ApJ, 900, 141, \dodoi{10.3847/1538-4357/abacc5}

\bibitem[{{Reis} \& {Miller}(2013)}]{2013ApJ...769L...7R}
{Reis}, R.~C., \& {Miller}, J.~M. 2013, \apjl, 769, L7, \dodoi{10.1088/2041-8205/769/1/L7}

\bibitem[{{Ricci} \& {Trakhtenbrot}(2023)}]{2023NatAs...7.1282R}
{Ricci}, C., \& {Trakhtenbrot}, B. 2023, Nature Astronomy, 7, 1282, \dodoi{10.1038/s41550-023-02108-4}

\bibitem[{{Ricci} {et~al.}(2018){Ricci}, {Ho}, {Fabian}, {Trakhtenbrot}, {Koss}, {Ueda}, {Lohfink}, {Shimizu}, {Bauer}, {Mushotzky}, {Schawinski}, {Paltani}, {Lamperti}, {Treister}, \& {Oh}}]{2018MNRAS.480.1819R}
{Ricci}, C., {Ho}, L.~C., {Fabian}, A.~C., {et~al.} 2018, \mnras, 480, 1819, \dodoi{10.1093/mnras/sty1879}

\bibitem[{{Ricci} {et~al.}(2020){Ricci}, {Kara}, {Loewenstein}, {Trakhtenbrot}, {Arcavi}, {Remillard}, {Fabian}, {Gendreau}, {Arzoumanian}, {Li}, {Ho}, {MacLeod}, {Cackett}, {Altamirano}, {Gandhi}, {Kosec}, {Pasham}, {Steiner}, \& {Chan}}]{2020ApJ...898L...1R}
{Ricci}, C., {Kara}, E., {Loewenstein}, M., {et~al.} 2020, \apjl, 898, L1, \dodoi{10.3847/2041-8213/ab91a1}

\bibitem[{{Richards} {et~al.}(2006){Richards}, {Lacy}, {Storrie-Lombardi}, {Hall}, {Gallagher}, {Hines}, {Fan}, {Papovich}, {Vanden Berk}, {Trammell}, {Schneider}, {Vestergaard}, {York}, {Jester}, {Anderson}, {Budav{\'a}ri}, \& {Szalay}}]{2006ApJS..166..470R}
{Richards}, G.~T., {Lacy}, M., {Storrie-Lombardi}, L.~J., {et~al.} 2006, \apjs, 166, 470, \dodoi{10.1086/506525}

\bibitem[{{Risaliti} {et~al.}(2002){Risaliti}, {Elvis}, \& {Nicastro}}]{2002ApJ...571..234R}
{Risaliti}, G., {Elvis}, M., \& {Nicastro}, F. 2002, \apj, 571, 234, \dodoi{10.1086/324146}

\bibitem[{{Rivers} {et~al.}(2011){Rivers}, {Markowitz}, \& {Rothschild}}]{2011ApJ...742L..29R}
{Rivers}, E., {Markowitz}, A., \& {Rothschild}, R. 2011, \apjl, 742, L29, \dodoi{10.1088/2041-8205/742/2/L29}

\bibitem[{{Roig} {et~al.}(2014){Roig}, {Blanton}, \& {Ross}}]{2014ApJ...781...72R}
{Roig}, B., {Blanton}, M.~R., \& {Ross}, N.~P. 2014, \apj, 781, 72, \dodoi{10.1088/0004-637X/781/2/72}

\bibitem[{{Rosen} {et~al.}(2016){Rosen}, {Webb}, {Watson}, {Ballet}, {Barret}, {Braito}, {Carrera}, {Ceballos}, {Coriat}, {Della Ceca}, {Denkinson}, {Esquej}, {Farrell}, {Freyberg}, {Gris{\'e}}, {Guillout}, {Heil}, {Koliopanos}, {Law-Green}, {Lamer}, {Lin}, {Martino}, {Michel}, {Motch}, {Nebot Gomez-Moran}, {Page}, {Page}, {Page}, {Pakull}, {Pye}, {Read}, {Rodriguez}, {Sakano}, {Saxton}, {Schwope}, {Scott}, {Sturm}, {Traulsen}, {Yershov}, \& {Zolotukhin}}]{2016A&A...590A...1R}
{Rosen}, S.~R., {Webb}, N.~A., {Watson}, M.~G., {et~al.} 2016, \aap, 590, A1, \dodoi{10.1051/0004-6361/201526416}

\bibitem[{{Ruan} {et~al.}(2016){Ruan}, {Anderson}, {Cales}, {Eracleous}, {Green}, {Morganson}, {Runnoe}, {Shen}, {Wilkinson}, {Blanton}, {Dwelly}, {Georgakakis}, {Greene}, {LaMassa}, {Merloni}, \& {Schneider}}]{2016ApJ...826..188R}
{Ruan}, J.~J., {Anderson}, S.~F., {Cales}, S.~L., {et~al.} 2016, ApJ, 826, 188, \dodoi{10.3847/0004-637X/826/2/188}

\bibitem[{{Ruiz} {et~al.}(2022){Ruiz}, {Georgakakis}, {Gerakakis}, {Saxton}, {Kretschmar}, {Akylas}, \& {Georgantopoulos}}]{2022MNRAS.511.4265R}
{Ruiz}, A., {Georgakakis}, A., {Gerakakis}, S., {et~al.} 2022, \mnras, 511, 4265, \dodoi{10.1093/mnras/stac272}

\bibitem[{{Sanders}(2023)}]{2023ascl.soft07017S}
{Sanders}, J. 2023, {Veusz: Scientific plotting package}, Astrophysics Source Code Library, record ascl:2307.017

\bibitem[{{Saxton} {et~al.}(2020){Saxton}, {Komossa}, {Auchettl}, \& {Jonker}}]{2020SSRv..216...85S}
{Saxton}, R., {Komossa}, S., {Auchettl}, K., \& {Jonker}, P.~G. 2020, \ssr, 216, 85, \dodoi{10.1007/s11214-020-00708-4}

\bibitem[{{Schlafly} \& {Finkbeiner}(2011)}]{2011ApJ...737..103S}
{Schlafly}, E.~F., \& {Finkbeiner}, D.~P. 2011, \apj, 737, 103, \dodoi{10.1088/0004-637X/737/2/103}

\bibitem[{{Scott} {et~al.}(2011){Scott}, {Stewart}, {Mateos}, {Alexander}, {Hutton}, \& {Ward}}]{2011MNRAS.417..992S}
{Scott}, A.~E., {Stewart}, G.~C., {Mateos}, S., {et~al.} 2011, \mnras, 417, 992, \dodoi{10.1111/j.1365-2966.2011.19325.x}

\bibitem[{{Shemmer} {et~al.}(2006){Shemmer}, {Brandt}, {Netzer}, {Maiolino}, \& {Kaspi}}]{2006ApJ...646L..29S}
{Shemmer}, O., {Brandt}, W.~N., {Netzer}, H., {Maiolino}, R., \& {Kaspi}, S. 2006, \apjl, 646, L29, \dodoi{10.1086/506911}

\bibitem[{{Shemmer} {et~al.}(2014){Shemmer}, {Brandt}, {Paolillo}, {Kaspi}, {Vignali}, {Stein}, {Lira}, {Schneider}, \& {Gibson}}]{2014ApJ...783..116S}
{Shemmer}, O., {Brandt}, W.~N., {Paolillo}, M., {et~al.} 2014, \apj, 783, 116, \dodoi{10.1088/0004-637X/783/2/116}

\bibitem[{{Shen} {et~al.}(2011){Shen}, {Richards}, {Strauss}, {Hall}, {Schneider}, {Snedden}, {Bizyaev}, {Brewington}, {Malanushenko}, {Malanushenko}, {Oravetz}, {Pan}, \& {Simmons}}]{2011ApJS..194...45S}
{Shen}, Y., {Richards}, G.~T., {Strauss}, M.~A., {et~al.} 2011, \apjs, 194, 45, \dodoi{10.1088/0067-0049/194/2/45}

\bibitem[{{Shen} {et~al.}(2019){Shen}, {Hall}, {Horne}, {Zhu}, {McGreer}, {Simm}, {Trump}, {Kinemuchi}, {Brandt}, {Green}, {Grier}, {Guo}, {Ho}, {Homayouni}, {Jiang}, {I-Hsiu Li}, {Morganson}, {Petitjean}, {Richards}, {Schneider}, {Starkey}, {Wang}, {Chambers}, {Kaiser}, {Kudritzki}, {Magnier}, \& {Waters}}]{2019ApJS..241...34S}
{Shen}, Y., {Hall}, P.~B., {Horne}, K., {et~al.} 2019, \apjs, 241, 34, \dodoi{10.3847/1538-4365/ab074f}

\bibitem[{{Shu} {et~al.}(2020){Shu}, {Zhang}, {Li}, {Jiang}, {Dou}, {Yan}, {Xie}, {Shen}, {Sun}, {Liu}, \& {Wang}}]{2020NatCo..11.5876S}
{Shu}, X., {Zhang}, W., {Li}, S., {et~al.} 2020, Nature Communications, 11, 5876, \dodoi{10.1038/s41467-020-19675-z}

\bibitem[{{Sadowski} {et~al.}(2014){Sadowski}, {Narayan}, {McKinney}, \& {Tchekhovskoy}}]{2014MNRAS.439..503S}
{Sadowski}, A., {Narayan}, R., {McKinney}, J.~C., \& {Tchekhovskoy}, A. 2014, \mnras, 439, 503, \dodoi{10.1093/mnras/stt2479}

\bibitem[{{Sobolewska} \& {Papadakis}(2009)}]{2009MNRAS.399.1597S}
{Sobolewska}, M.~A., \& {Papadakis}, I.~E. 2009, \mnras, 399, 1597, \dodoi{10.1111/j.1365-2966.2009.15382.x}

\bibitem[{{Steffen} {et~al.}(2006){Steffen}, {Strateva}, {Brandt}, {Alexander}, {Koekemoer}, {Lehmer}, {Schneider}, \& {Vignali}}]{2006AJ....131.2826S}
{Steffen}, A.~T., {Strateva}, I., {Brandt}, W.~N., {et~al.} 2006, \aj, 131, 2826, \dodoi{10.1086/503627}

\bibitem[{{Stern} {et~al.}(2018){Stern}, {McKernan}, {Graham}, {Ford}, {Ross}, {Meisner}, {Assef}, {Balokovi{\'c}}, {Brightman}, {Dey}, {Drake}, {Djorgovski}, {Eisenhardt}, \& {Jun}}]{2018ApJ...864...27S}
{Stern}, D., {McKernan}, B., {Graham}, M.~J., {et~al.} 2018, \apj, 864, 27, \dodoi{10.3847/1538-4357/aac726}

\bibitem[{{Stroe} \& {Savu}(2021)}]{2021AJ....161..158S}
{Stroe}, A., \& {Savu}, V.-N. 2021, \aj, 161, 158, \dodoi{10.3847/1538-3881/abe12a}

\bibitem[{{Tanaka} {et~al.}(2004){Tanaka}, {Boller}, {Gallo}, {Keil}, \& {Ueda}}]{2004PASJ...56L...9T}
{Tanaka}, Y., {Boller}, T., {Gallo}, L., {Keil}, R., \& {Ueda}, Y. 2004, PASJ, 56, L9, \dodoi{10.1093/pasj/56.3.L9}

\bibitem[{{Tang} {et~al.}(2021){Tang}, {Silverman}, {Ding}, {Li}, {Lee}, {Strauss}, {Goulding}, {Schramm}, {Kawinwanichakij}, {Xavier Prochaska}, {Hennawi}, {Imanishi}, {Iwasawa}, {Toba}, {Kayo}, {Oguri}, {Matsuoka}, {Onoue}, {Jahnke}, {Ichikawa}, {Hartwig}, {Kashikawa}, {Kawaguchi}, {Kohno}, {Matsuda}, {Nagao}, {Ono}, {Ouchi}, {Shimasaku}, {Suh}, {Suzuki}, {Taniguchi}, {Ueda}, \& {Yasuda}}]{2021ApJ...922...83T}
{Tang}, S., {Silverman}, J.~D., {Ding}, X., {et~al.} 2021, \apj, 922, 83, \dodoi{10.3847/1538-4357/ac1ff0}

\bibitem[{{Taylor}(2005)}]{2005ASPC..347...29T}
{Taylor}, M.~B. 2005, in Astronomical Society of the Pacific Conference Series, Vol. 347, Astronomical Data Analysis Software and Systems XIV, ed. P.~{Shopbell}, M.~{Britton}, \& R.~{Ebert}, 29

\bibitem[{{Timlin} {et~al.}(2020){Timlin}, {Brandt}, {Zhu}, {Liu}, {Luo}, \& {Ni}}]{2020MNRAS.498.4033T}
{Timlin}, John~D., I., {Brandt}, W.~N., {Zhu}, S., {et~al.} 2020, \mnras, 498, 4033, \dodoi{10.1093/mnras/staa2661}

\bibitem[{{Trakhtenbrot} {et~al.}(2019){Trakhtenbrot}, {Arcavi}, {MacLeod}, {Ricci}, {Kara}, {Graham}, {Stern}, {Harrison}, {Burke}, {Hiramatsu}, {Hosseinzadeh}, {Howell}, {Smartt}, {Rest}, {Prieto}, {Shappee}, {Holoien}, {Bersier}, {Filippenko}, {Brink}, {Zheng}, {Li}, {Remillard}, \& {Loewenstein}}]{2019ApJ...883...94T}
{Trakhtenbrot}, B., {Arcavi}, I., {MacLeod}, C.~L., {et~al.} 2019, \apj, 883, 94, \dodoi{10.3847/1538-4357/ab39e4}

\bibitem[{{Traulsen} {et~al.}(2020){Traulsen}, {Schwope}, {Lamer}, {Ballet}, {Carrera}, {Ceballos}, {Coriat}, {Freyberg}, {Koliopanos}, {Kurpas}, {Michel}, {Motch}, {Page}, {Watson}, \& {Webb}}]{2020A&A...641A.137T}
{Traulsen}, I., {Schwope}, A.~D., {Lamer}, G., {et~al.} 2020, \aap, 641, A137, \dodoi{10.1051/0004-6361/202037706}

\bibitem[{{Trouille} \& {Barger}(2010)}]{2010ApJ...722..212T}
{Trouille}, L., \& {Barger}, A.~J. 2010, \apj, 722, 212, \dodoi{10.1088/0004-637X/722/1/212}

\bibitem[{{Ulrich} {et~al.}(1997){Ulrich}, {Maraschi}, \& {Urry}}]{1997ARA&A..35..445U}
{Ulrich}, M.-H., {Maraschi}, L., \& {Urry}, C.~M. 1997, \araa, 35, 445, \dodoi{10.1146/annurev.astro.35.1.445}

\bibitem[{{Vaccari}(2015)}]{2015fers.confE..27V}
{Vaccari}, M. 2015, in The Many Facets of Extragalactic Radio Surveys: Towards New Scientific Challenges, 27, \dodoi{10.22323/1.267.0027}

\bibitem[{{Vanden Berk} {et~al.}(2001){Vanden Berk}, {Richards}, {Bauer}, {Strauss}, {Schneider}, {Heckman}, {York}, {Hall}, {Fan}, {Knapp}, {Anderson}, {Annis}, {Bahcall}, {Bernardi}, {Briggs}, {Brinkmann}, {Brunner}, {Burles}, {Carey}, {Castander}, {Connolly}, {Crocker}, {Csabai}, {Doi}, {Finkbeiner}, {Friedman}, {Frieman}, {Fukugita}, {Gunn}, {Hennessy}, {Ivezi{\'c}}, {Kent}, {Kunszt}, {Lamb}, {Leger}, {Long}, {Loveday}, {Lupton}, {Meiksin}, {Merelli}, {Munn}, {Newberg}, {Newcomb}, {Nichol}, {Owen}, {Pier}, {Pope}, {Rockosi}, {Schlegel}, {Siegmund}, {Smee}, {Snir}, {Stoughton}, {Stubbs}, {SubbaRao}, {Szalay}, {Szokoly}, {Tremonti}, {Uomoto}, {Waddell}, {Yanny}, \& {Zheng}}]{2001AJ....122..549V}
{Vanden Berk}, D.~E., {Richards}, G.~T., {Bauer}, A., {et~al.} 2001, \aj, 122, 549, \dodoi{10.1086/321167}

\bibitem[{{Vasudevan} \& {Fabian}(2007)}]{2007MNRAS.381.1235V}
{Vasudevan}, R.~V., \& {Fabian}, A.~C. 2007, \mnras, 381, 1235, \dodoi{10.1111/j.1365-2966.2007.12328.x}

\bibitem[{{Vestergaard} \& {Peterson}(2006)}]{2006ApJ...641..689V}
{Vestergaard}, M., \& {Peterson}, B.~M. 2006, ApJ, 641, 689, \dodoi{10.1086/500572}

\bibitem[{{Wang} {et~al.}(2022){Wang}, {Luo}, {Brandt}, {Alexander}, {Bauer}, {Gallagher}, {Huang}, {Liu}, \& {Stern}}]{2022ApJ...936...95W}
{Wang}, C., {Luo}, B., {Brandt}, W.~N., {et~al.} 2022, \apj, 936, 95, \dodoi{10.3847/1538-4357/ac886e}

\bibitem[{{Wang} {et~al.}(2024{\natexlab{a}}){Wang}, {Woo}, {Gallo}, {Guo}, {Son}, {Kong}, {Mandal}, {Cho}, {Kim}, \& {Shin}}]{2024ApJ...966..128W}
{Wang}, S., {Woo}, J.-H., {Gallo}, E., {et~al.} 2024{\natexlab{a}}, \apj, 966, 128, \dodoi{10.3847/1538-4357/ad3049}

\bibitem[{{Wang} {et~al.}(2024{\natexlab{b}}){Wang}, {Graham}, {Ford}, {McKernan}, {Ryu}, \& {Stern}}]{2024arXiv240612096W}
{Wang}, Y., {Graham}, M.~J., {Ford}, K.~E.~S., {et~al.} 2024{\natexlab{b}}, arXiv e-prints, arXiv:2406.12096, \dodoi{10.48550/arXiv.2406.12096}

\bibitem[{{Webb} {et~al.}(2020){Webb}, {Coriat}, {Traulsen}, {Ballet}, {Motch}, {Carrera}, {Koliopanos}, {Authier}, {de la Calle}, {Ceballos}, {Colomo}, {Chuard}, {Freyberg}, {Garcia}, {Kolehmainen}, {Lamer}, {Lin}, {Maggi}, {Michel}, {Page}, {Page}, {Perea-Calderon}, {Pineau}, {Rodriguez}, {Rosen}, {Santos Lleo}, {Saxton}, {Schwope}, {Tom{\'a}s}, {Watson}, \& {Zakardjian}}]{2020A&A...641A.136W}
{Webb}, N.~A., {Coriat}, M., {Traulsen}, I., {et~al.} 2020, \aap, 641, A136, \dodoi{10.1051/0004-6361/201937353}

\bibitem[{{Wevers} {et~al.}(2023){Wevers}, {Coughlin}, {Pasham}, {Guolo}, {Sun}, {Wen}, {Jonker}, {Zabludoff}, {Malyali}, {Arcodia}, {Liu}, {Merloni}, {Rau}, {Grotova}, {Short}, \& {Cao}}]{2023ApJ...942L..33W}
{Wevers}, T., {Coughlin}, E.~R., {Pasham}, D.~R., {et~al.} 2023, \apjl, 942, L33, \dodoi{10.3847/2041-8213/ac9f36}

\bibitem[{{Wright} {et~al.}(2010){Wright}, {Eisenhardt}, {Mainzer}, {Ressler}, {Cutri}, {Jarrett}, {Kirkpatrick}, {Padgett}, {McMillan}, {Skrutskie}, {Stanford}, {Cohen}, {Walker}, {Mather}, {Leisawitz}, {Gautier}, {McLean}, {Benford}, {Lonsdale}, {Blain}, {Mendez}, {Irace}, {Duval}, {Liu}, {Royer}, {Heinrichsen}, {Howard}, {Shannon}, {Kendall}, {Walsh}, {Larsen}, {Cardon}, {Schick}, {Schwalm}, {Abid}, {Fabinsky}, {Naes}, \& {Tsai}}]{2010AJ....140.1868W}
{Wright}, E.~L., {Eisenhardt}, P. R.~M., {Mainzer}, A.~K., {et~al.} 2010, \aj, 140, 1868, \dodoi{10.1088/0004-6256/140/6/1868}

\bibitem[{{Wu} \& {Shen}(2022)}]{2022ApJS..263...42W}
{Wu}, Q., \& {Shen}, Y. 2022, \apjs, 263, 42, \dodoi{10.3847/1538-4365/ac9ead}

\bibitem[{{Yan} {et~al.}(2011){Yan}, {Ho}, {Newman}, {Coil}, {Willmer}, {Laird}, {Georgakakis}, {Aird}, {Barmby}, {Bundy}, {Cooper}, {Davis}, {Faber}, {Fang}, {Griffith}, {Koekemoer}, {Koo}, {Nandra}, {Park}, {Sarajedini}, {Weiner}, \& {Willner}}]{2011ApJ...728...38Y}
{Yan}, R., {Ho}, L.~C., {Newman}, J.~A., {et~al.} 2011, \apj, 728, 38, \dodoi{10.1088/0004-637X/728/1/38}

\bibitem[{{Yang} {et~al.}(2016){Yang}, {Brandt}, {Luo}, {Xue}, {Bauer}, {Sun}, {Kim}, {Schulze}, {Zheng}, {Paolillo}, {Shemmer}, {Liu}, {Schneider}, {Vignali}, {Vito}, \& {Wang}}]{2016ApJ...831..145Y}
{Yang}, G., {Brandt}, W.~N., {Luo}, B., {et~al.} 2016, ApJ, 831, 145, \dodoi{10.3847/0004-637X/831/2/145}

\bibitem[{{Yang} {et~al.}(2018){Yang}, {Wu}, {Fan}, {Jiang}, {McGreer}, {Shangguan}, {Yao}, {Wang}, {Joshi}, {Green}, {Wang}, {Feng}, {Fu}, {Yang}, \& {Liu}}]{2018ApJ...862..109Y}
{Yang}, Q., {Wu}, X.-B., {Fan}, X., {et~al.} 2018, ApJ, 862, 109, \dodoi{10.3847/1538-4357/aaca3a}

\bibitem[{{Yang} {et~al.}(2020){Yang}, {Shen}, {Chen}, {Liu}, {Annis}, {Avila}, {Bertin}, {Brooks}, {Buckley-Geer}, {Carnero Rosell}, {Carrasco Kind}, {Carretero}, {da Costa}, {Desai}, {Thomas Diehl}, {Doel}, {Frieman}, {Garcia-Bellido}, {Gaztanaga}, {Gerdes}, {Gruen}, {Gruendl}, {Gschwend}, {Gutierrez}, {Hollowood}, {Honscheid}, {Hoyle}, {James}, {Krause}, {Kuehn}, {Lidman}, {Lima}, {Maia}, {Marshall}, {Martini}, {Menanteau}, {Miquel}, {Plazas Malag{\'o}n}, {Sanchez}, {Scarpine}, {Schindler}, {Schubnell}, {Serrano}, {Sevilla}, {Smith}, {Soares-Santos}, {Sobreira}, {Suchyta}, {Swanson}, {Tarle}, {Vikram}, \& {Walker}}]{2020MNRAS.493.5773Y}
{Yang}, Q., {Shen}, Y., {Chen}, Y.-C., {et~al.} 2020, \mnras, 493, 5773, \dodoi{10.1093/mnras/staa645}

\bibitem[{{Yang} {et~al.}(2023){Yang}, {Green}, {MacLeod}, {Plotkin}, {Anderson}, {Bieryla}, {Civano}, {Eracleous}, {Graham}, {Ruan}, {Runnoe}, \& {Zhao}}]{2023ApJ...953...61Y}
{Yang}, Q., {Green}, P.~J., {MacLeod}, C.~L., {et~al.} 2023, \apj, 953, 61, \dodoi{10.3847/1538-4357/acdedd}

\bibitem[{{Yip} {et~al.}(2004{\natexlab{a}}){Yip}, {Connolly}, {Vanden Berk}, {Ma}, {Frieman}, {SubbaRao}, {Szalay}, {Richards}, {Hall}, {Schneider}, {Hopkins}, {Trump}, \& {Brinkmann}}]{2004AJ....128.2603Y}
{Yip}, C.~W., {Connolly}, A.~J., {Vanden Berk}, D.~E., {et~al.} 2004{\natexlab{a}}, AJ, 128, 2603, \dodoi{10.1086/425626}

\bibitem[{{Yip} {et~al.}(2004{\natexlab{b}}){Yip}, {Connolly}, {Szalay}, {Budav{\'a}ri}, {SubbaRao}, {Frieman}, {Nichol}, {Hopkins}, {York}, {Okamura}, {Brinkmann}, {Csabai}, {Thakar}, {Fukugita}, \& {Ivezi{\'c}}}]{2004AJ....128..585Y}
{Yip}, C.~W., {Connolly}, A.~J., {Szalay}, A.~S., {et~al.} 2004{\natexlab{b}}, AJ, 128, 585, \dodoi{10.1086/422429}

\bibitem[{{York} {et~al.}(2000){York}, {Adelman}, {Anderson}, {Anderson}, {Annis}, {Bahcall}, {Bakken}, {Barkhouser}, {Bastian}, {Berman}, {Boroski}, {Bracker}, {Briegel}, {Briggs}, {Brinkmann}, {Brunner}, {Burles}, {Carey}, {Carr}, {Castander}, {Chen}, {Colestock}, {Connolly}, {Crocker}, {Csabai}, {Czarapata}, {Davis}, {Doi}, {Dombeck}, {Eisenstein}, {Ellman}, {Elms}, {Evans}, {Fan}, {Federwitz}, {Fiscelli}, {Friedman}, {Frieman}, {Fukugita}, {Gillespie}, {Gunn}, {Gurbani}, {de Haas}, {Haldeman}, {Harris}, {Hayes}, {Heckman}, {Hennessy}, {Hindsley}, {Holm}, {Holmgren}, {Huang}, {Hull}, {Husby}, {Ichikawa}, {Ichikawa}, {Ivezi{\'c}}, {Kent}, {Kim}, {Kinney}, {Klaene}, {Kleinman}, {Kleinman}, {Knapp}, {Korienek}, {Kron}, {Kunszt}, {Lamb}, {Lee}, {Leger}, {Limmongkol}, {Lindenmeyer}, {Long}, {Loomis}, {Loveday}, {Lucinio}, {Lupton}, {MacKinnon}, {Mannery}, {Mantsch}, {Margon}, {McGehee}, {McKay}, {Meiksin}, {Merelli}, {Monet}, {Munn}, {Narayanan}, {Nash}, {Neilsen}, {Neswold}, {Newberg}, {Nichol}, {Nicinski},
  {Nonino}, {Okada}, {Okamura}, {Ostriker}, {Owen}, {Pauls}, {Peoples}, {Peterson}, {Petravick}, {Pier}, {Pope}, {Pordes}, {Prosapio}, {Rechenmacher}, {Quinn}, {Richards}, {Richmond}, {Rivetta}, {Rockosi}, {Ruthmansdorfer}, {Sandford}, {Schlegel}, {Schneider}, {Sekiguchi}, {Sergey}, {Shimasaku}, {Siegmund}, {Smee}, {Smith}, {Snedden}, {Stone}, {Stoughton}, {Strauss}, {Stubbs}, {SubbaRao}, {Szalay}, {Szapudi}, {Szokoly}, {Thakar}, {Tremonti}, {Tucker}, {Uomoto}, {Vanden Berk}, {Vogeley}, {Waddell}, {Wang}, {Watanabe}, {Weinberg}, {Yanny}, {Yasuda}, \& {SDSS Collaboration}}]{2000AJ....120.1579Y}
{York}, D.~G., {Adelman}, J., {Anderson}, John~E., J., {et~al.} 2000, \aj, 120, 1579, \dodoi{10.1086/301513}

\bibitem[{{Zabludoff} {et~al.}(2021){Zabludoff}, {Arcavi}, {LaMassa}, {Perets}, {Trakhtenbrot}, {Zauderer}, {Auchettl}, {Dai}, {French}, {Hung}, {Kara}, {Lodato}, {Maksym}, {Qin}, {Ramirez-Ruiz}, {Roth}, {Runnoe}, \& {Wevers}}]{2021SSRv..217...54Z}
{Zabludoff}, A., {Arcavi}, I., {LaMassa}, S., {et~al.} 2021, \ssr, 217, 54, \dodoi{10.1007/s11214-021-00829-4}

\bibitem[{{Zeltyn} {et~al.}(2024){Zeltyn}, {Trakhtenbrot}, {Eracleous}, {Yang}, {Green}, {Anderson}, {LaMassa}, {Runnoe}, {Assef}, {Bauer}, {Brandt}, {Davis}, {Frederick}, {Fries}, {Graham}, {Grogin}, {Guolo}, {Hern{\'a}ndez-Garc{\'\i}a}, {Koekemoer}, {Krumpe}, {Liu}, {Mart{\'\i}nez-Aldama}, {Ricci}, {Schneider}, {Shen}, {{\'S}niegowska}, {Temple}, {Trump}, {Xue}, {Brownstein}, {Dwelly}, {Morrison}, {Bizyaev}, {Pan}, \& {Kollmeier}}]{2024ApJ...966...85Z}
{Zeltyn}, G., {Trakhtenbrot}, B., {Eracleous}, M., {et~al.} 2024, \apj, 966, 85, \dodoi{10.3847/1538-4357/ad2f30}

\bibitem[{Zhan(2021)}]{zhan2021wide}
Zhan, H. 2021, Chinese Science Bulletin, 66, 1290

\bibitem[{{Zhang} {et~al.}(2019){Zhang}, {Bao}, \& {Yuan}}]{2019MNRAS.490L..81Z}
{Zhang}, X.-G., {Bao}, M., \& {Yuan}, Q. 2019, \mnras, 490, L81, \dodoi{10.1093/mnrasl/slz151}

\bibitem[{{Zhang} {et~al.}(2023){Zhang}, {Luo}, {Brandt}, {Du}, {Hu}, {Huang}, {Pu}, {Wang}, \& {Yi}}]{2023ApJ...954..159Z}
{Zhang}, Z., {Luo}, B., {Brandt}, W.~N., {et~al.} 2023, \apj, 954, 159, \dodoi{10.3847/1538-4357/ace7c2}

\bibitem[{{Zhu} {et~al.}(2023){Zhu}, {Brandt}, {Zou}, {Luo}, {Ni}, {Xue}, \& {Yan}}]{2023MNRAS.522.3506Z}
{Zhu}, S., {Brandt}, W.~N., {Zou}, F., {et~al.} 2023, \mnras, 522, 3506, \dodoi{10.1093/mnras/stad1178}

\bibitem[{{Zhu} {et~al.}(2020){Zhu}, {Brandt}, {Luo}, {Wu}, {Xue}, \& {Yang}}]{2020MNRAS.496..245Z}
{Zhu}, S.~F., {Brandt}, W.~N., {Luo}, B., {et~al.} 2020, \mnras, 496, 245, \dodoi{10.1093/mnras/staa1411}

\bibitem[{{Zhu} {et~al.}(2021){Zhu}, {Timlin}, \& {Brandt}}]{2021MNRAS.505.1954Z}
{Zhu}, S.~F., {Timlin}, J.~D., \& {Brandt}, W.~N. 2021, \mnras, 505, 1954, \dodoi{10.1093/mnras/stab1406}

\bibitem[{{Zhuang} {et~al.}(2024){Zhuang}, {Yang}, {Shen}, {Adamow}, {Friedel}, {Gruendl}, {Liu}, {Martini}, {Abbott}, {Anderson}, {Assef}, {Bauer}, {Bielby}, {Brandt}, {Burke}, {Casares}, {Chen}, {De Rosa}, {Drlica-Wagner}, {Dwelly}, {Eltvedt}, {Fonseca Alvarez}, {Fu}, {Fuentes}, {Graham}, {Grier}, {Golovich}, {Hall}, {Hartigan}, {Horne}, {Koekemoer}, {Krumpe}, {Li}, {Lidman}, {Malik}, {Mangian}, {Merloni}, {Ricci}, {Salvato}, {Sharp}, {Stone}, {Trilling}, {Tucker}, {Wen}, {Wideman}, {Xue}, {Yu}, \& {Zucker}}]{2024arXiv240206052Z}
{Zhuang}, M.-Y., {Yang}, Q., {Shen}, Y., {et~al.} 2024, arXiv e-prints, arXiv:2402.06052.
\newblock \doarXiv{2402.06052}

\bibitem[{{Zou} {et~al.}(2022){Zou}, {Brandt}, {Chen}, {Leja}, {Ni}, {Yan}, {Yang}, {Zhu}, {Luo}, {Nyland}, {Vito}, \& {Xue}}]{2022ApJS..262...15Z}
{Zou}, F., {Brandt}, W.~N., {Chen}, C.-T., {et~al.} 2022, \apjs, 262, 15, \dodoi{10.3847/1538-4365/ac7bdf}

\bibitem[{{Zu} {et~al.}(2011){Zu}, {Kochanek}, \& {Peterson}}]{2011ApJ...735...80Z}
{Zu}, Y., {Kochanek}, C.~S., \& {Peterson}, B.~M. 2011, \apj, 735, 80, \dodoi{10.1088/0004-637X/735/2/80}

\end{thebibliography}

\end{document}